\documentclass[10pt]{article}

\usepackage{amsmath}
\usepackage{amsfonts}
\usepackage{amssymb}
\usepackage{amscd}
\newcounter{theorem}
\renewcommand{\thetheorem}{\arabic{section}.\arabic{theorem}}

\newenvironment{thm}[1]{\par
\begin{sloppypar}\refstepcounter{theorem}%
\noindent{\bf #1 \thetheorem.}\it{}}{\end{sloppypar}}

\newenvironment{theorem}{\begin{thm}{Theorem}}{\end{thm}}
\newenvironment{proposition}{\begin{thm}{Proposition}}{\end{thm}}
\newenvironment{corollary}{\begin{thm}{Corollary}}{\end{thm}}
\newenvironment{lemma}{\begin{thm}{Lemma}}{\end{thm}}

\newenvironment{defi}[1]{\par
\begin{sloppypar}\refstepcounter{theorem}%
\noindent{\bf #1 \thetheorem.}\rm{}}{\end{sloppypar}}
\newenvironment{definition}{\begin{defi}{Definition}}{\end{defi}}

\newenvironment{remark}{\begin{defi}{Remark}}{\end{defi}}
\newenvironment{hypothesis}{\begin{defi}{Hypothesis}}{\end{defi}}

\newcommand{\eh}{\hfill}\newlength{\sperr}
\newenvironment{proof}{{\settowidth{\sperr}{\rm Proof}
\par\addvspace{0.3cm}\noindent\parbox[t]{1.3\sperr}{\rm P\eh r\eh o\eh o\eh
f\eh.}}}{\nopagebreak\mbox{}\hfill $\blacksquare $\par\addvspace{0.25cm}}

\newenvironment{proofT}{{\settowidth{\sperr}{\rm Proof of the Theorem}
\par\addvspace{0.3cm}\noindent\parbox[t]{1.3\sperr}{\rm P\eh r\eh o\eh o\eh
f\eh\eh o\eh f\eh\eh t\eh h\eh e\eh\eh T\eh h\eh e\eh o\eh r\eh e\eh m\eh.}}}{\nopagebreak\mbox{}\hfill $\blacksquare $\par\addvspace{0.25cm}}
\def\dB{\,\sharp^B\,}

\def\dB{\,\sharp^B\,}

\def\e{\mathfrak{e}_}
\def\l{\mathfrak{l}_}
\def\ad{\mathfrak{ad}}
\def\aut{\mathbb{A}{\rm ut}}
 
\def\R{{\rm I\kern-.2em R}}   
  \def\H{{\cal H}} 
   \def\Op{\mathfrak{Op}}  
\def\X{\mathcal X}

\begin{document}

\title{Commutator Criteria for Magnetic Pseudodifferential Operators}

\date{\today}

\author{Viorel Iftimie\footnote{Institute
of Mathematics Simion Stoilow of the Romanian Academy, P.O.  Box
1-764, Bucharest, RO-70700, Romania.} ,
Marius M\u{a}ntoiu\footnote{Institute
of Mathematics Simion Stoilow of the Romanian Academy, P.O.  Box
1-764, Bucharest, RO-70700, Romania.}
\footnote{Universidad de Chile, Las
Palmeras 3425, Casilla 653, Santiago Chile.}\ \ and Radu
Purice\footnote{Institute
of Mathematics Simion Stoilow of the Romanian Academy, P.O.  Box
1-764, Bucharest, RO-70700, Romania.}
\footnote{Laboratoire Europ\'een Associ\'e CNRS Franco-Roumain {\it Math-Mode}}}

\maketitle

\begin{abstract}

The gauge covariant magnetic Weyl calculus has been introduced and
studied in previous works. We prove criteria in terms of commutators for operators to be magnetic
pseudodifferential operators of suitable symbol classes. The approach is completely intrinsic; neither the
statements nor the proofs depend on a choice of a vector potential. We apply
this criteria to inversion problems, functional calculus, affiliation results and to the study of the evolution
group generated by a magnetic pseudodifferential operator.

\end{abstract}

\section{Introduction}
This paper is devoted to the study of some commutator techniques in the frame of the twisted pseudodifferential calculus associated to Quantum Hamilonian sytems in a bounded, smooth magnetic field. In order to take advantage of the gauge invariance we formulate our statements about commutators in an algebraic frame using the twisted Moyal algebra (Section 2). In Section 3 we study the magnetic translations and their generators in order to formulate our main result, an analogue of Beals' Criterion \cite{Be}, in an algebraic setting. This theorem is proved in Section 4 for the case of symbols of type $S^0_0(\Xi)$ (\ref{symb-classes}). In the 4-th Section we extend our main Theorem to symbols of a general class $S^m_\rho(\Xi)$ (\ref{symb-classes}) and also prove a Bony type Criterion \cite{Bo1}. We apply these criteria in Section 5 for inverses and fractional powers of some twisted pseudodifferential operators and to the functional calculus they generate. In a last chapter we use the idea of Bony to define Fourier Integral Operators and introduce a class of \textit{Twisted Fourier Integral Operators}. We prove that for a large class of first order elliptic symbols, the unitary groups that they generate (in any Hilbert representation) are such operators.
\subsection{Beals criterion in a classical setting}

For a linear operator $T$ acting in $L^2(\mathbb R^n)$ or in some other related function space,
it is often useful to know if it is a pseudodifferential operator with the symbol in a certain class.
A possible answer (\cite{Be}, \cite{Bo1}, \cite{He}) can be given in terms of commutators: very roughly,
a necessary and sufficient condition would be that the successive commutators of $T$ with an infinite
family of specified simple operators has a specified behavior.
Very often this involves boundedeness of the commutators seen as
operators between some Sobolev spaces. We indicate a particular case, relevant for our purposes.

The Weyl calculus is a systematic procedure to associate to suitable functions $f$ (classical observables) on
$\mathbb R^{2n}$ (the phase space), operators $\Op(f)$ acting on functions $u:\mathbb R^n\rightarrow \mathbb C$.
Formally this is given by
\begin{equation}\label{Op}
\left[\mathfrak{Op}(f)u\right](x)\,:=\,(2\pi)^{-n}\int_{\mathbb R^n}dy\int_{\mathbb R^n}d\eta\,
\exp\{i<x-y,\eta>\}\,f\left(\frac{x+y}{2},\eta\right)\,u(y).
\end{equation}
This formula has various interpretations for various elements $f$ and $u$. Frequently it is assumed that $f$ belongs to one
of H\"ormander's symbol classes $S^m_{\rho,\delta}(\mathbb R^n)$ and (\ref{Op}) is regarded as an oscillatory integral.
Then the operators $\Op(f)$ have nice multiplication properties,
they act continuously in the Schwartz space $\mathcal S(\mathbb R^n)$ and, in the case $m=0,\rho=\delta=0$, they define
bounded operators in the Hilbert space $\H:=L^2(\mathbb R^n)$.

In Quantum Mechanics $\H$ is considered to be the state space of a quantum particle moving
in $\mathbb R^n$; the basic observables of position and momenta are, respectively, the operators $Q_j$ of multiplication
by the coordinate $x_j$ and the operators $D_j:=-i\partial_{x_j}$, $j=1,\dots,n$.
The correspondence $f\mapsto\Op(f)\equiv f(Q,D)$, often called "a quantization", is interpreted as a functional
calculus for the family of unbounded self-adjoint non-commuting operators $(Q;D)=(Q_1,\dots,Q_n;D_1,\dots,D_n)$.

Independent of any interpretation, the operators $Q$ and $D$ are involved in Beals' characterization of those
operators that are pseudodifferential (here in the Weyl sense) with symbols in the
classes $S^m_{\rho,0}(\mathbb R^n)$. We describe only the result for $S^0_{0,0}(\mathbb R^{2n})$. Let us denote by
$\mathfrak{ad}_{Q_j}$, $\mathfrak{ad}_{D_j}$, the commutators with $Q_j$ and $D_j$, $j=1,\dots,n$; in a general notation,
$\mathfrak{ad}_S[T]:=[S,T]=ST-TS$, for convenient operators $S,T$
acting in $\H$.

{\it Then the operator $T$ is of the form $\Op(f)$ for
some $f\in S^0_{0,0}(\mathbb R^{2n})$ if and only if for all
multi-indices $(a_1,\dots,a_n)$ and $(\alpha_1,\dots,\alpha_n)$ in $\mathbb
N^{n}$, the commutators
\begin{equation}
\mathfrak{ad}^{a_1}_{Q_1}\dots\mathfrak{ad}^{a_n}_{Q_n}\mathfrak{ad}^{\alpha_1}_{D_1}\dots
\mathfrak{ad}^{\alpha_n}_{D_n}[T]
\end{equation}
define bounded operators in} $\H$.

Aside the obvious direct interest of having criteria for an operator to be pseudodifferential in purely hilbertian
terms, such a result is also very handy for deciding under which conditions
inverses or functions of Weyl operators are still of the same type.

\subsection{Magnetic pseudodifferential operators}

The main goal of the present article is to prove analogous results for a generalization of the Weyl calculus adapted
to the situation in which a variable magnetic field is also present. We recall very
briefly some facts concerning the {\it magnetic pseudodifferential calculus} that we have developed in
\cite{MP} and \cite{IMP}.
Other references are \cite{Mu}, \cite{KO1}, \cite{KO2}, \cite{MP1} and \cite{MPR1}.

First some notations. We denote $\mathbb R^n$ by $\mathcal X$, with elements $x,y,z$. $\mathcal X'$ will be the dual of
$\mathcal X$, with elements $\xi,\eta,\zeta$. We also denote by $<\cdot,\cdot>$ the duality form
$<\xi,x>=\xi(x)=<x,\xi>$. The phase space will be $\Xi=\mathbb R^{2n}=\mathcal X\oplus\mathcal
X'$, with elements $X=(x,\xi),\,Y=(y,\eta),\,Z=(z,\zeta)$. In fact these notations will be used in a rigid manner: if the
contrary is not explicitly stated, when one encounters $X\in \Xi$,
one should think that its components in $\mathcal X$ and $\mathcal
X'$, respectively, are called $x$ and $\xi$; the same for $Y$ and $Z$.
$\mathcal S$ stands for the Schwartz space, $\mathcal S'$ for its dual, formed of tempered distributions, and
$\mathbb B(\mathcal R;\mathcal T)$ is the vector space of all linear continuous operators acting between the locally
convex spaces $\mathcal R$ and $\mathcal T$. For any real euclidean space $\mathcal{Y}$ we consider on $\mathcal{S}(\mathcal{Y})$ the family of norms indexed by $M\in\mathbb{N}$ (defining its localy convex Topology):
$$
\||\varphi|\|_M:=\underset{Y\in\mathcal{Y}}{\sup}\sum_{|\mathfrak{a}|+|\mathfrak{b}|\leq M}\left|Y^{\mathfrak{a}}\big(\partial^{\mathfrak{b}}\varphi\big)(Y)\right|.
$$

Suppose given a magnetic field $B$, i.e. a closed $2$-form on $\mathcal X$ with components of class $C^\infty(\mathcal{X})$. Since $dB=0$,
the magnetic field can be written as the differential $B=dA$ of a $1$-form $A$ on $\X$ with components of class $C^\infty(\mathcal{X})$, called {\it vector
potential}. In such a situation, aside the position operators $Q_1,\dots,Q_n$, one works with {\it the magnetic momenta}
$\Pi^A_1:=D_1-A_1,\dots,\Pi^A_n:=D_n-A_n$. By analogy with the Weyl calculus, one would like to construct a quantization
assigning to phase-space functions $f$ operators $\Op^A(f)$ which admit the interpretation $\Op^A(f)=f(Q;\Pi^A)$, within a
functional calculus. The commutation relations satisfied by the $2n$ operators $(Q;\Pi^A)$ are more involved that those for
$B=0$ (especially when $B$ is not a polynomial), so a new pseudodifferential calculus is required. The solution
$\Op^A(f)=\Op(f^A)$ was offered in the literature, with $f^A(x,\xi):=f(x,\xi-A(x))$. It fails, because {\it it is not gauge
covariant}: two vector potentials $A,A'$ which differ by an exact $1$-form $A'=A+d\varphi$ define the same magnetic field, but in
general there is no reasonable connection between the operators $\Op(f^A)$ and $\Op(f^{A'})$.

The solution is to introduce in (\ref{Op}) an extra phase factor $\exp\left\{-i\Gamma^A([x,y])\right\}$, where
$\Gamma^A([x,y]):=\int_{[x,y]}A$ is the circulation of the $1$-form $A$ through the segment $[x,y]:=\{tx+(1-t)y\mid
t\in[0,1]\}$. So, for any test function $f\in\mathcal{S}(\Xi)$, we
define the following operator, which is not equivalent in any sense with $\Op(f^A)$:
\begin{equation}\label{OpA}
\left[\mathfrak{Op}^A(f)u\right](x)\,:=\,(2\pi)^{-n}\int_{\mathcal{X}}dy\int_{\mathcal{X}^\prime}d\eta\,
\exp\{i<x-y,\eta>\}\,\exp\left\{-i\Gamma^A([x,y)]\right\}\,f\left(\frac{x+y}{2},\eta\right)\,u(y).
\end{equation}
A thorough justification of this formula, properties and applications can be found in the references cited above. We note
that {\it gauge covariance} is recovered: if $dA=dA'$, then
$\mathfrak{Op}^A(f)$ and $\mathfrak{Op}^{A'}(f)$ are unitarily equivalent.

Although for some developments this is not necessary, let us assume that the components $A_j$ of the vector potential are in
$C^\infty_{\rm{pol}}(\X)$, the space of all $C^\infty$ functions on $\X$ with each
derivative dominated by an (arbitrary) polynomial. This can always be achieved if the components $B_{jk}$ are in $C^\infty_{\rm{pol}}(\X)$, $\forall j,k=1,\dots,n$. We have proved
in \cite{MP} that the application $\mathfrak{Op}^A$ is an isomorphism
\begin{equation}
\mathfrak{Op}^A:\mathcal{S}(\Xi)\rightarrow\mathbb{B}(\mathcal{S}^{\prime}(\mathcal{X});\mathcal{S}(\mathcal{X})),
\end{equation}
and may be extended to an isomorphism
\begin{equation}
\mathfrak{Op}^A:\mathcal{S}^\prime(\Xi)\rightarrow\mathbb{B}(\mathcal{S}(\mathcal{X});\mathcal{S}^\prime(\mathcal{X})).
\end{equation}
Thus there exists a vector subspace $\mathfrak{M}^B(\Xi)$ of
$\mathcal{S}^\prime(\Xi)$ sent bijectively by $\Op^A$ onto
$\mathbb{B}(\mathcal{S}\left(\mathcal{X})\right):=\mathbb{B}(\mathcal{S}(\mathcal{X});\mathcal S(\X))$.
The fact that it only depends on the magnetic field is an easy consequence of
gauge covariance. In \cite{IMP} and \cite{MP} it is shown that
$\mathfrak{M}^B(\Xi)$ contains all H\"ormander's classes of
symbols
$$
S^m_{\rho,\delta}(\Xi):=\left\{f\in
C^\infty(\Xi)\mid\forall(a,\alpha)\in\mathbb{N}^n\times\mathbb{N}^n,\,\exists
C_{a\alpha}>0, \ |(\partial^a_x\partial^\alpha_\xi f)(x,\xi)|\leq
C_{a\alpha}<\xi>^{m-\rho\vert\alpha\vert+\delta\vert
a\vert}\right\}
$$
for $m\in\mathbb R$, $\rho\ge 0,\delta<1$, as well as the space $C^\infty_{\rm{pol,u}}(\Xi)$ composed of those
$f\in C^\infty_{\rm{pol}}(\Xi)$ for which all the derivatives are dominated by a {\it fixed} polynomial (depending on $f$).

\subsection{The magnetic Beals criterion; the represented version of a particular case}

Now we come to the Hilbert space $\H:=L^2(\X)$. For the hilbertian
framework we shall need a stronger assumption on the magnetic field, so we recall a function space
that will be used very often subsequently. For any $m$-dimensional euclidean space $\mathcal
Y$ (the cases $\mathcal Y=\X$ and $\mathcal Y=\Xi$ will be relevant) we set $BC^\infty(\mathcal Y)=\{f\in C^\infty(\mathcal Y)\mid
\partial^\alpha f \ \text{is bounded for any\ }\alpha\in\mathbb N^m\}$. A particular case of a result of \cite{IMP} says
that $\Op^A\left[S^0_{0,0}(\Xi)\right]\subset \mathbb B(\H)$ if $B_{jk}\in BC^\infty(\X)$.

The next statement
is an extension of the result of Beals described in the first subsection, which can be recovered for $B=0$.
It is one of our main results.

\medskip
\begin{theorem}\label{trep}
Assume that the components of the magnetic field $B$ belong to $BC^\infty(\X)$. Choose a vector potential $A$
defining $B$ (i.e. $B=dA$) which belongs to $C^\infty_{\rm{pol}}$. A linear continuous operator $T:\mathcal
S(\mathcal X)\rightarrow \mathcal S'(\mathcal X)$ is a magnetic
pseudodifferential operator with symbol of class $S^0_{0,0}(\Xi)$ if and only if the commutators
$$\ad_{Q_1}^{a_1}\dots\ad_{Q_n}^{a_n}\ad_{\Pi^A_1}^{\alpha_1}\dots
\ad^{\alpha_n}_{\Pi^A_n}T$$
are bounded operators on $\H$ for all multi-indices
$(a,\alpha)=(a_1,\dots,a_n,\alpha_1,\dots,\alpha_n)$.
\end{theorem}

\medskip
This Theorem will be recast in a more tractable setting in the next section. It will be proved in Section 4,
after some preparations involving magnetic commutators and phase-space translations, object of Section 3.
More general results, including a treatment of the class $S^m_{\rho,0}(\Xi)$, will be given in Sections 6 and 7.
The final two sections will contain applications, mainly investigating the functional calculus applied to a
magnetic pseudodifferential operator.

\section{The main results in an intrinsic setting}

\subsection{The need of an intrinsic approach}

The goal of this Section is to rephrase our problem in a more elegant and tractable intrinsic language.

Many of the drawbacks of the mathematical theory of systems placed in magnetic fields come
from the following fact: Although the single physically relevant object is the magnetic field, in most cases the objects
one studies involve the choice of a vector potential. Not only is this vector potential highly non-unique, but it is also
worse behaved than the magnetic field.

On one hand, obtaining gauge-invariant assertions is a difficult matter,
both concerning the assumptions and the conclusions.
Most often, the underlying hypothesis says that a certain result holds for magnetic fields admitting a vector potential
with some specified properties, although one suspects that some simple condition imposed directly on $B$ would suffice.
And it happens sometimes that the output is not obviously a gauge-covariant assertion.

On the other hand, rather nice magnetic fields admit as a best choice a vector potential which is a more
"singular" function than $B$ itself.
Within the class of bounded magnetic fields, for instance, a large subclass only corresponds to unbounded vector potentials.

While trying to prove Theorem \ref{trep} we had to overcome these obstacles. The way out is actually built in the
formalism itself. Beyond the magnetic pseudodifferential operators, one also disposes of algebraic structures, defined only in terms of $B$. The main notion is a symbol composition $\dB$ which extends the usual
Weyl-Moyal multiplication law, for which $\Op^A(f)\Op^A(g)=\Op^A(f\dB g)$. Thus $\left(\Op^A\right)_{dA=B}$ is seen
as a family of equivalent representations of some algebra, the choice of this algebra being at our disposition and depending
on the type of symbols (classical observables) one would like to treat.
Some choices are well-suited to various practical problems (spectral analysis, as in \cite{MPR2}, quantization, as in
\cite{MP1}). This will also be the case for our commutator characterization of magnetic pseudodifferential operators,
as shown below.

We note that all these are consistent with the $C^*$-algebraic approach to quantization, cf. \cite{La} and references therein.
Presentations of the general algebraic formalism for systems in magnetic fields, including $C^*$-algebras generated
by twisted dynamical systems, can be found in \cite{MPR1} and \cite{MP1}.

\subsection{The magnetic composition law}

First we note that
$\Xi$ is a symplectic space with the canonical symplectic form
$$\sigma(X,Y)=\sigma[(y,\eta), (z,\zeta)]=<\eta,z>-<\zeta,y>.$$
As said before, the magnetic field is described by a closed 2-form $B$ of class
$BC^\infty(\mathcal{X})$. It has a natural raising to a closed
2-form on $\Xi$ and it is easy to verify that the sum
$$\big(\sigma_B\big)_Z(X,Y):=\sigma(X,Y)+B(z)(x,y)$$
defines a symplectic form on $\Xi$.

For any $k$-form $C$ in $\X$, given a compact $k$-manifold $\mathcal{K}\subset\mathcal{X}$, we set
\begin{equation}\label{be}
 \Gamma^C[\mathcal K]:=\int_\mathcal K C
\end{equation}
and
\begin{equation}\label{gin}
\Omega^C[\mathcal{K}]:=\exp\left\{-i\int_{\mathcal{K}}C\right\}=\exp\left\{-i\,\Gamma^C[\mathcal
K]\right\},
\end{equation}
involving the invariant integral of the $k$-form along the compact $k$-manifold.

On the Schwartz space of test functions $\mathcal{S}(\Xi)$ we
introduce the {\it magnetic composition}:
\begin{equation}\label{dB}
 (f\dB g)(X)\,:=\,\pi^{-2n}\,\int_{\Xi}\int_{\Xi}\,dY\,dZ\,e^{-2i\sigma(X-Y,X-Z)}\Omega^B[\mathcal{T}(x,y,z)]f(Y)g(Z),
\end{equation}
where $\mathcal{T}(x,y,z)$ is the triangle having the vertices:
$x-y+z,\,y-z+x,\,z-x+y$. Under the assumption that $B_{jk}\in C^\infty_{\rm{pol}}(\X)$, it is easy to show that
$\mathcal{S}(\Xi)$ is a $^*$-algebra, the involution being the usual complex
conjugation. The point in introducing $\dB$ is that one has
$\mathfrak{Op}^A(f)\mathfrak{Op}^A(g)=\mathfrak{Op}^A(f\dB g)$ for any $f,g\in\mathcal S(\Xi)$.

Once we have verified (see \cite{MP}) that for any three functions $f,g,h$ from $\mathcal{S}(\Xi)$ one has the following equality of the two $L^2(\Xi)$-scalar products:
\begin{equation*}\label{dual}
 \left(f,g\dB h \right)\,=\,\left(f\dB\overline{g},h\right),
\end{equation*}
we can extend the magnetic composition $\dB$ by duality and define the {\it magnetic Moyal $^*-$algebra}
as being the unital associative algebra $\mathfrak{M}^B(\Xi)$ of tempered
distributions $F\in\mathcal{S}^\prime(\Xi)$ satisfying $g\dB
F\in\mathcal{S}(\Xi)$ and $F\dB g\in\mathcal{S}(\Xi)$ for any test function $g\in\mathcal{S}(\Xi)$.
Actually it is the same space defined in the previous section, and if $A_j\in C^\infty_{\rm{pol}}$ then
$\Op^A:\mathfrak{M}^B(\Xi)\rightarrow \mathbb B(\mathcal S(\X))$ is an isomorphism of involutive algebras.
The magnetic composition extends to composition laws $\dB:\mathcal
S'(\Xi)\times\mathfrak{M}^B(\Xi)\rightarrow\mathcal S'(\Xi)$ and $\dB:\mathfrak M^B(\Xi)\times\mathcal S'(\Xi)
\rightarrow\mathcal S'(\Xi)$. We shall denote by $f^-$ the inverse of $f\in\mathfrak{M}^B(\Xi)$ when it exists.

An important matter is the behavior of the Weyl calculus with respect to H\"ormander's symbol classes
$S^{m}_{\rho,\delta}(\Xi)$. In the present article we are only interested in the case $\delta=0$, for which
we use the simplified notation
\begin{equation}\label{symb-classes}
S^m_{\rho}\equiv S^m_{\rho,0}:=\{f\in C^\infty(\Xi)\mid
|\left(\partial_x^a\partial_\xi^\alpha f\right)(x,\xi)| \le
C_{a\alpha}<\xi>^{m-\rho|\alpha|}\}.
\end{equation}
By Theorem 2.2 in \cite{IMP}, if $B\in BC^\infty$, $m_1,m_2\in\mathbb R$ and $\rho\in(0,1]$, then
\begin{equation}\label{hor}
S^{m_1}_{\rho}(\Xi)\dB S^{m_2}_{\rho}(\Xi)\subset S^{m_1+m_2}_{\rho}(\Xi).
\end{equation}
Actually the case $\rho=0$ is a consequence of our Proposition in Appendix \ref{comp-symb}, but the asymptotic development
contained in \cite{IMP}, Theorem 2.2 will no longer hold for this case. In dealing with symbols we shall very often make
use without explicitly marking it, of the regularization procedure described in Appendix \ref{reg-proc}.

\medskip
Let us also define $\mathfrak C^B(\Xi):=\left(\mathfrak{Op}^A\right)^{-1}[\mathbb B(\mathcal H)]$. It is
obviously a vector subspace of $\mathcal S'(\Xi)$ and a
$^*$-algebra for the magnetic composition. We transport the norm of $\mathbb B(\mathcal H)$: $\parallel
f\parallel_{\mathfrak C^B}:=\parallel\mathfrak{Op}^A(f)\parallel_{\mathbb B(\mathcal
H)}$ will be a $C^*$-norm on $\mathfrak C^B(\Xi)$. Gauge covariance shows that the $C^*$-algebra $\mathfrak C^B(\Xi)$
is independent of the vector potential $A$.
For a nicer point of view on the norm $\parallel \cdot\parallel_B$, involving twisted crossed products, we refer to \cite{MPR1}.
As already noticed above, by the magnetic version of the Calderon-Vaillancourt theorem proved in \cite{IMP}, one has
 $S^0_{0}(\Xi)\subset \mathfrak C^B(\Xi)$.

\subsection{Statement of the result}

Let us consider real linear
functions belonging to $\mathfrak{M}^B(\Xi)$ of the form $\mathfrak{l}_X(Y):=\sigma(X,Y)$, for some $X\in\Xi$.
By quantization, they produce the basic operators of our theory, as explained below.

Let us denote by $(e_1,\dots,e_n;\epsilon_1,\dots,\epsilon_n)$ the canonical base of $\,\Xi=\mathbb R^{2n}$.
We choose a vector potential $A$ associated to the magnetic field $B$ (i.e. such that $B=dA$), and consider the
representation $\mathfrak{Op}^A:\mathfrak{C}^B(\Xi)\rightarrow
\mathbb{B}[L^2(\mathcal{X})]$ associated to it. We observe that we can then associate to our linear functions
$\mathfrak{l}_{e_j}$, the operators of multiplication with the variables, as operators in
$\mathbb{B}(\mathcal{S}(\mathcal{X});\mathcal{S}^\prime(\mathcal{X}))$
$$
\mathfrak{Op}^A(\mathfrak{l}_{e_j})=Q_j,\quad\text{ such that}\quad(Q_jf)(x):=x_jf(x)
$$
and to the elements $\mathfrak{l}_{\epsilon_j}$ the magnetic momentum operators
$$
\mathfrak{Op}^A(\mathfrak{l}_{\epsilon_j})=\Pi^A_j,\quad\text{
such that }\ (\Pi^A_ju)(x):=-i(\partial_{x_j}u)(x)-A_j(x)u(x).
$$

We introduce now a family of derivations, which play at an intrinsic level the role of basic commutators.

\medskip
\begin{definition}
For any $X\in\Xi$ we set
\begin{equation}\label{md}
 \ad_X^B[F]\,:=\,\mathfrak{l}_X\dB F-F\dB\mathfrak{l}_X,\quad\forall F\in\mathcal S'(\Xi),\quad\forall X\in\Xi.
\end{equation}
\end{definition}
Obviously we have on $\mathcal S(X)$
$$\ad_{\Op^A(\mathfrak l_X)}\left(\Op^A(F)\right)=\Op^A\left(\ad^B_X(F)\right),\ \ \forall F\in\mathfrak M^B(\Xi).$$

One of the main results in this paper, the intrinsic version of Theorem \ref{trep}, is

\medskip
\begin{theorem}\label{intrinseca}
If $B$ is of class $BC^\infty(\X)$, then $f\in S^0_{0}(\Xi)$ if and only if for all $N\in\mathbb N$ and
all $U_1,\dots,U_N\in\Xi$ with $|U_1|=\dots=|U_N|=1$
\begin{equation*}
\ad^B_{U_1}\dots\ad^B_{U_N}[f]\in\mathfrak C^B(\Xi).
\end{equation*}
\end{theorem}

The statement above can be understood in terms of $C^\infty$-vectors and this reinterpretation seems to us
interesting even for the case $B=0$. To this end we shall use an action of the linear space $\Xi$
on the algebra $\mathfrak{M}^B(\Xi)$, defined
by conjugation with exponentials of linear functions.  The family of exponentials
\begin{equation}\label{e}
\e X:=\exp\{-i\mathfrak{l}_X\},\ \ X\in\Xi
\end{equation}
could be called the {\it algebraic Weyl system}. The functions $\e X$ are unitary
elements in $\mathfrak{M}^B(\Xi)$ for any $X\in\Xi$; the corresponding operators $W^A(X):=\Op^A(\e X)$ were
studied and used in \cite{MP} as a sort of building blocks for the magnetic Weyl calculus.
Then we define a family of automorphisms, indexed by $\Xi$, of the magnetic Moyal algebra:
$$
\Xi\ni X\mapsto\mathfrak T^B_X\in\aut\left[\mathfrak{M}^B(\Xi)\right],
$$
\begin{equation}\label{E1}
\mathfrak T^B_X[F]\,:=\,\e{-X}\dB F\dB\e X,\quad\forall F\in\mathfrak{M}^B(\Xi).
\end{equation}
Each $\mathfrak T^B_X$ also acts on other spaces, as $\mathcal S(\Xi),\mathcal S'(\Xi),L^2(\Xi)$ or $\mathfrak C^B(\Xi)$. In
particular, since $\Op^A(\e{\pm X})$ are bounded (unitary) operators,
$\mathfrak T^B_X$ is an automorphism of the $C^*$-algebra $\mathfrak C^B(\Xi)$. Generically, we call
$\left(\mathfrak T^B_X\right)_{X\in\Xi}$ {\it the family of magnetic phase-space translations}. For $B=0$ this
reduces to the usual phase-space translations $(\mathfrak T_X)_{X\in\Xi}$.

\medskip
\begin{definition}
For a magnetic field $B$ with components of class $BC^\infty(\mathcal{X})$, we define
the following linear space:
$$
C^\infty\left(\mathfrak T^B;\mathfrak C^B\right):=\left\{f\in \mathfrak
C^B(\Xi)\mid X\mapsto \mathfrak T^B_X[f]\in \mathfrak C^B(\Xi) \rm\ \ {is} \
C^\infty\ \rm{in}\ X=0\right\}
$$
and endow it with the following family of seminorms:
\begin{equation}\label{treica}
\left\{\parallel\cdot\parallel^{\mathfrak T^B,\mathfrak
C^B}_{U_1,\dots,U_N}\ \mid\ |U_1|=\dots=|U_N|=1,\  N\in\mathbb
N\right\},\ \ \ \parallel f\parallel^{\mathfrak T^B,\mathfrak
C^B}_{U_1,\dots,U_N}:=\parallel \ad^B_{U_1}\dots\ad^B_{U_N}[f]\parallel_{\mathfrak C^B}.
\end{equation}
\end{definition}

At a more elementary level, let us observe that the symbol space $S^0_{0}(\Xi)$ is nothing
but $BC^\infty(\Xi)=C^\infty(\mathfrak T;BC(\Xi))$. Here we considered automorphisms defined by translations
$$\left[\mathfrak T_Y(f)\right](X):=f(X-Y),\ \ X,Y\in\Xi$$
on the $C^*$-algebra
$$BC(\Xi):=\{f:\Xi\rightarrow \mathbb C\mid f\ \text{is bounded and continuous}\},$$
endowed with the sup-norm $\parallel\cdot\parallel_\infty.$
$BC^\infty(\Xi)$ is also a Fr\'echet space with the family of seminorms
\begin{equation}\label{treico}
\left\{\parallel\cdot\parallel^{\mathfrak T,BC}_{(a,\alpha)}\mid (a,\alpha)\in\mathbb N^{2n}\right\},
\ \ \ \parallel f\parallel^{\mathfrak T,BC}_{(a,\alpha)}:=\parallel \partial_x^a\partial_\xi^\alpha f
\parallel_\infty,
\end{equation}
obviously equivalent with the family
\begin{equation}\label{treicu}
\left\{\parallel\cdot\parallel^{\mathfrak T,BC}_{U_1,\dots,U_N}\ \mid\ |U_1|=\dots=|U_N|=1,\ N\in\mathbb
N\right\},\ \ \ \parallel f\parallel^{\mathfrak T,BC}_{U_1,\dots,U_N}:=\parallel \partial_{U_1}\dots\partial_{U_N}
f\parallel_{\infty}.
\end{equation}

At first sight the two spaces $BC^\infty(\Xi)$ and $C^\infty(\mathfrak T^B,\mathfrak C^B)$ seem to be very different.
They involve different families of derivations acting in different $C^*$-algebras ($BC(\Xi)$ and
$\mathfrak C^B(\Xi)$ are quite different even when $B=0$).
But we shall prove in Section 4 the following result, that implies Theorem \ref{intrinseca}, which in turn implies
Theorem \ref{trep}.

\medskip
\begin{theorem}\label{main}
If $B$ is a magnetic field with components of class $BC^\infty(\mathcal{X})$, then the spaces
$C^\infty(\mathfrak T;BC)$ and $C^\infty(\mathfrak T^B,\mathfrak C^B)$ coincide and have
isomorphic Fr\'echet structures.
\end{theorem}

\section{Magnetic phase-space translations and commutators}\label{mag-deriv}

We shall introduce  notations that will allow us to put into evidence some algebraic and topologic structures appearing in a rather natural way when dealing with the magnetic translations of symbols.

\paragraph*{Notations}
\begin{itemize}
 \item For a distribution $F\in\mathcal{S}^\prime(\Xi)$ and a test function $g\in\mathcal{S}(\Xi)$, we define
 the following commutative {\it mixed product} (this is a mixture between pointwise multiplication in the first variable
 and convolution in the second):
\begin{equation}\label{star}
 (F\star g)(x,\xi)\,:=\,\int_{\mathcal{X}^\prime}d\eta\,F(x,\xi-\eta)\,g(x,\eta)\qquad\text{with }F\star g\in\mathcal{S}^\prime(\Xi).
\end{equation}
\item For any $p\in[1,\infty]$ we shall consider
the complex linear space $BC(\mathcal{X};L^p(\mathcal{X}^\prime))$ of bounded continuous functions $f:\mathcal{X}\rightarrow L^p(\mathcal{X}^\prime)$ endowed with the norm
\begin{equation}
 \|f\|_{\infty,p}:=\underset{x\in\mathcal{X}}{\sup}\left\{\int_{\mathcal{X}^\prime}d\xi\,|f(x,\xi)|^p\right\}^{1/p}.
\end{equation}
\item For $m\in\mathbb R$ we define the weight function
$\mathfrak{w}_m(v):=<v>^m\equiv\left(1+|v|^2\right)^{m/2}$ and the following functions on phase space: $\mathfrak
q_m(x,\xi):=(\mathfrak{w}_m\otimes 1)(x,\xi)=<x>^m$ and $\mathfrak
p_m(x,\xi):=(1\otimes\mathfrak{w}_m)(x,\xi)=<\xi>^m$.
\item Let us define the complex linear space $\mathfrak{A}(\Xi)$ as the space of functions $a\in
BC^\infty(\mathcal{X};L^1(\mathcal{X}^\prime))\subset BC(\mathcal{X};L^1(\mathcal{X}^\prime))$ such that
$\mathfrak{p}_p\cdot\big(\partial_x^\alpha a\big)\in BC(\mathcal{X};L^1(\mathcal{X}^\prime))$ for any $p>0$ and any multi-index $\alpha\in\mathbb{N}^n$ (i.e. having rapid decay in the variable $\xi\in\mathcal{X}^\prime$ together with all its derivatives with respect to the $x\in\mathcal{X}$).
\end{itemize}

Let us point out that $\|f\|_{\infty,\infty}=\|f\|_\infty$.
Taking into account that the $\infty$-norm is a cross-norm for the usual multiplication as well as the
Hausdorff-Young inequality, we see that the above norms behave well with respect to our mixt product $\star$.
In fact the mixt product in (\ref{star}) also defines a bilinear abelian composition law on
$L^1(\mathcal{X}^\prime;BC(\mathcal{X}))$ and we have the following results:

\medskip
\begin{proposition}\label{L-inf-p}
 \begin{enumerate}
  \item The space $BC(\mathcal{X};L^1(\mathcal{X}^\prime))$ is a Banach algebra for the $\star$-product and $\mathfrak{A}(\Xi)$ is closed for the operation $\star$.
  \item For any $p\in[1,\infty]$, the $\star$-product defines bicontinuous bilinear maps:
$$
BC(\mathcal{X};L^1(\mathcal{X}^\prime))\times BC(\mathcal{X};L^p(\mathcal{X}^\prime))\ni(f,F)\mapsto f\star F
\in BC(\mathcal{X};L^p(\mathcal{X}^\prime)),
$$
$$
BC(\mathcal{X};L^1(\mathcal{X}^\prime))\times L^p(\Xi)\ni(f,\Phi)\mapsto f\star\Phi\in L^p(\Xi),
$$
with the estimations: $\|f\star F\|_{\infty,p}\,\leq\,\|f\|_{\infty,1}\|F\|_{\infty,p}$, respectively
$\|f\star\Phi\|_p\,\leq\,\|f\|_{\infty,1}\|\Phi\|_p$.
  \item If $f\in\mathfrak{A}(\Xi)$, then for any symbol $\phi\in S^m_{\rho}(\Xi)$, we have $f\star\phi\in S^m_{\rho}(\Xi)$, the map being continuous.
\item For $f\in\mathfrak{A}(\Xi)$ and for any $M\in\mathbb{N}$ there exists a constant $C(M,n;f)\in\mathbb{R}_+$ such that
$$
\||f\star\psi|\|_M\leq C(M,n;f)\||\psi|\|_M,\quad\forall\psi\in\mathcal{S}(\Xi).
$$
 \end{enumerate}
\end{proposition}

The proof is defered to an Appendix.

Since translations in the $\X$-variable will occur very often, we use a special notation:
$\tau_x:=\mathfrak T_{(x.0)}$.
\subsection{The magnetic phase-space translations}\label{mag-transl}

To study the families (\ref{e}) and (\ref{E1}), we introduce first a 2-cocycle associated to the symplectic form
$\sigma$ and the magnetic field $B$. Its cohomological
and analytical importance was outlined in our previous works. The space
$$
C(\mathcal{X},U(1)):=\{\varphi:\X\rightarrow\mathbb C\mid \varphi\ \text{is continuous}, \ |\varphi(x)|=1,\ \forall x\in\X\}
$$
can be seen as the group of all unitary elements of the $C^*$-algebra $BC(\X)$.

\medskip
\begin{definition}
$$
\omega^B:\Xi\times\Xi\rightarrow C(\mathcal{X},U(1)),
$$
\begin{equation}
\left[\omega^B(X,Y)\right](z):=\exp\{(i/2)\sigma(X,Y)\}\Omega^B[\mathcal{T}(z,z-y/2,z+x/2)].
\end{equation}
\end{definition}
The multiplication properties of the functions $\e X$, $X\in\Xi$ will be essential in the sequel.
By straightforward computations we obtain
\medskip
\begin{lemma}\label{chiau}
$$
\e X\dB\e Y=\omega^B(X,Y)\e{X+Y}=
$$
$$
=\left\{\tau_{-(x+y)/2}\left[\omega^B(X,Y)\right]\right\}\dB\e{X+Y}=\e{X+Y}\dB\left\{\tau_{(x+y)/2}\left[\omega^B(X,Y)\right]\right\}.
$$
\end{lemma}

The next step is an explicit formula for the phase-space magnetic translation
$\mathfrak T^B_U$, with $U=(u,\mu)\in\Xi$.

\begin{proposition}\label{magn-transl}
For any 3 points $q,x,y\in \mathcal{X}$ let us define the parallelogram
\begin{equation}
 \mathcal{P}(q;x,y)\,:=\,\{q+sx+ty\mid s\in[-1/2,1/2],\,t\in[-1,0]\},
\end{equation}
having edges parallel to the vectors $x$ and $y$, respectively. We consider the distribution $\Omega^B[\mathcal{P}(x;y,u)]$ (see (\ref{be}) and (\ref{gin})) and its Fourier transform
 with respect to
 the second variable:
\begin{equation}
 \widetilde{\Omega^B_\mathcal{P}}[u](x,\xi)\,:=(2\pi)^{-n}\,\int_{\mathcal{X}}dy\, e^{-i<y,\xi>}\Omega^B[\mathcal{P}
 (x;y,u)].
\end{equation}
\begin{itemize}
\item We have the following explicit formula:
\begin{equation}
 \Omega^B[\mathcal{P}(x;y,u)]=\exp\left\{-i\Gamma^B[\mathcal{P}(x;y,u)]\right\}=
 \exp\left\{-i\sum_{j,k=1}^ny_ju_k\int_{-1/2}^{1/2}ds\int_{-1}^0dtB_{jk}(x+sy+tu)\right\}.
\end{equation}
\item For $U=(u,\mu)\in\Xi$ and $f\in\mathcal{S}(\Xi)$, we have
\begin{equation}\label{aut}
\mathfrak T^B_{U}[f]\,=\,\widetilde{\Omega^B_\mathcal{P}}[u]\star\mathfrak T_U[f].
\end{equation}
\end{itemize}
\end{proposition}
\begin{proof}
 Straightforward computations give
$$
(\e{-U}\dB f)(Y)\,=\,\pi^{-2n}\int_{\Xi}\,dY_1\int_{\Xi}\,dY_2\,e^{-2i\sigma(Y-Y_1,Y-Y_2)}\Omega^B[\mathcal{T}
(y,y_1,y_2)]e^{i\sigma(U,Y_1)}f(Y_2)\,=
$$
(by integrating upon $\eta_1\in\mathcal{X}^*$, and using Fourier inversion formula for $\delta_0(y-y_2-u/2)$)
$$
=\,\pi^{-n}\,e^{2i<\eta,y-u/2>}\,\int_{\Xi}\,dY^\prime\,e^{2i<y^\prime,\eta^\prime>-2i<\eta^\prime,y>-2i<\eta-\mu/2,
y^\prime>}\Omega^B[\mathcal{T}(y,y^\prime,y-u/2)] f(y-u/2,\eta^\prime).
$$
Thus
$$
\mathfrak T^B_U[f](X)\,=\,\pi^{-2n}\int_{\Xi}\,dY\int_{\Xi}\,dZ\,e^{-2i\sigma(X-Y,X-Z)}\Omega^B[\mathcal{T}(x,y,z)]
\left(\e {-U}\dB f\right)(Y)e^{-i\sigma(U,Z)}\,=
$$
$$
=\,\pi^{-3n}\int_{\Xi}\,dY\int_{\Xi}\,dZ\,\int_{\Xi}\,dY^\prime\,e^{-2i<\xi-\eta,x-z>+2i<\xi-\zeta,x-y>-
i<\mu,z>+i<\zeta,u>}\times
$$
$$
\times e^{2i<\eta,y-u/2>-2i<\eta-\mu/2,y^\prime>-2i<\eta^\prime,y-y^\prime>}\Omega^B[\mathcal{T}(x,y,z)]
\Omega^B[\mathcal{T}(y,y^\prime,y-u/2)]f(y-u/2,\eta^\prime)\,=
$$
(we integrate with respect to $\eta$ and $\zeta$ by using the Fourier inversion formula for $\delta_0(y-x+u/2)$ and for $\delta_0(z-x-y+u/2+y^\prime)$ respectively)
$$
=\,\pi^{-n}\int_{\mathcal{X}}dy^\prime\,e^{-2i<\xi-\mu,y^\prime-x+u/2>}\Omega^B[\mathcal{T}(x,x-u/2,2x-u-y^\prime)]
\Omega^B[\mathcal{T}(x-u/2,y^\prime,x-u)]\times
$$
$$
\times\left\{\int_{\mathcal{X}^*}d\eta^\prime\,e^{2i<\eta^\prime,y^\prime-x+u/2>}f(x-u,\eta^\prime)\right\}\,=
$$
(changing variables from $y^\prime$ to $y=2y^\prime-2x+u$)
$$
=\, (2\pi)^{-n}\int_{\mathcal{X}}dy\,e^{-i<\xi-\mu,y>}\Omega^B[\mathcal{P}(x;y,u)]
\left\{\int_{\mathcal{X}^*}d\eta^\prime\,e^{i<\eta^\prime,y>}f(x-u,\eta^\prime)\right\}\,=
\,\left(\widetilde{\Omega^B_\mathcal{P}}[u]\star\mathfrak T_U[f]\right)(X).
$$
\end{proof}
\begin{remark}\label{notgrup}
Using Stokes Theorem for the closed 2-form $B$ and the formula $\mathfrak T_{(u,\mu)}(g\star h)=
\mathfrak T_{(u,0)}g\star \mathfrak T_{(u,\mu)}h$, we
get by a straightforward computation the following formula for composing magnetic translations:
$$
\left(\mathfrak T^B_U\circ\mathfrak T^B_V\right)[f]=\Sigma^B_{u,v}\star\mathfrak T^B_{U+V}[f],
$$
where
$$
\Sigma^B_{u,v}(x,\xi):=(2\pi)^{-n}\int_{\mathcal{X}}dy\,e^{-i<\xi,y>}\Omega^B\left[<x+\frac{y}{2},x+\frac{y}{2}-u,
x+\frac{y}{2}-u-v>\right]\times
$$
$$
\times\Omega^B\left[<x-\frac{y}{2},x-\frac{y}{2}-u,x-\frac{y}{2}-u-v>\right]
$$
and $\,<a,b,c>\ \subset\mathcal X$ is the triangle defined by the vertices $a,b,c$. Although $U\mapsto \mathfrak T^B_U$ is
not a representation of $\Xi$ by automorphisms, it is however clear that
$$
\mathfrak T^B_{sU}\circ\mathfrak T^B_{tU}=\mathfrak T^B_{(s+t)U},\ \ \ \forall U\in\Xi,\ \ \ \forall s,t\in\mathbb R.
$$
\end{remark}

\subsection{The magnetic derivations}

The mappings $\left(\ad^B_X\right)_{X\in\Xi}$, defined in (\ref{md}), will be called {\it magnetic derivations}.
They are the infinitesimal objects associated with the family of automorphisms $\left(\mathfrak T^B_X\right)_{X\in\Xi}$.
Our notation stresses their interpretation as commutators, but another natural one would be
$\ad ^B_U\equiv D_U^B=-i\partial^B_U$. Anyhow, for $B=0$ they reduce to usual derivations.

\medskip
\begin{proposition}
For any $X\in\Xi$ and $t_0\in\mathbb{R}$, we have the equalities
\begin{equation}
i\left.\left(\partial_t\e {tX}\right)\right|_{t=t_0}\,=\,\l X\dB\e {t_0X}\,=\,\e {t_0X}\dB\l X
\end{equation}
and
\begin{equation}\label{parti}
i\left.\partial_t\mathfrak T^B_{tX}[F]\right|_{t=t_0}\,=\,-\ad^B_X\left[\mathfrak T^B_{t_0X}[F]\right].
\end{equation}
\end{proposition}
\begin{proof}
Due to the fact that $\omega^B$ is obviously unitary in $\mathfrak{C}^B(\Xi)$,  we have
$$
\omega^B(tX,t_0X)(z)\,=\,\exp\{(i/2)tt_0\sigma(X,X)\}\Omega^B[\mathcal{T}(z,z-t_0x/2,z+tx/2)]\,=
$$
$$
=\,\Omega^B[\mathcal{T}(z,z-t_0x/2,z+tx/2)].
$$
Remark that the three vertices: $z-(z-t_0x/2)+(z+tx/2)$, $(z-t_0x/2)-(z+tx/2)+z$,
$(z+tx/2)-z+(z-t_0x/2)$ of the above triangle
are in fact equal to $z+(t+t_0)x/2$, $z-(t+t_0)x/2$, $z+(t-t_0)x/2$ and thus are colinear. It follows that
the flux of $B$ through the given triangle is $0$ and we get
$$
\omega^B(tX,t_0X)(z)\,=\,1
$$
and by Lemma \ref{chiau}
$$
t^{-1}\left[ \e {(t+t_0)X}-\e {t_0X} \right]\,=\,t^{-1}\left[\omega^B(tX,t_0X)^{-1}\e {tX}\dB\e {t_0X} - \e {t_0X}\right]\,=
$$
$$
=\,t^{-1}\left[\e {tX}-1\right]\dB\e {t_0X}\,=\,t^{-1}\left[\exp\{-it\l X\}-1\right]\dB\e {t_0X}\,
\underset{t\rightarrow0}{\longrightarrow}\,(-i)\l X\dB\e {t_0X}.
$$
Similarly
$$
t^{-1}\left[ \e {(t+t_0)X}-\e {t_0X} \right]\,\,\underset{t\rightarrow0}{\longrightarrow}\,(-i)\e {t_0X}\dB\l X.
$$
For the second equality we observe that
$$
t^{-1}\left\{\mathfrak T^B_{(t+t_0)X}[F]\,-\,\mathfrak T^B_{t_0X}[F]\right\}\,=
$$
$$
=\,t^{-1}\left\{\e {-(t+t_0)X}\dB\,F\dB\,\e {(t+t_0)X}\,-\,\e {-t_0X}\dB\,F\dB\,\e {t_0X}\right\}\,=
$$
$$
=\,t^{-1}\left\{\left[\e {-(t+t_0)X}\,-\,\e {-t_0X}\right]\dB\,F\dB\,\e {(t+t_0)X}\right\}\,+\,t^{-1}
\left\{\e {-t_0X}\dB\,F\dB\left[\e {(t+t_0)X}\,-\,\e {t_0X}\right]\right\}.
$$
\end{proof}
It is extremely useful to express the magnetic derivations in terms of usual ones.

\medskip
\begin{proposition}\label{delta-B}
 For $U=(u,\mu)\in\Xi$ and $f\in\mathcal{S}(\Xi)$ we have
\begin{equation}\label{deriv}
\ad_U^B[f]\,=\,(-i)\,\,\underset{\epsilon\rightarrow0}{\lim}\,\epsilon^{-1}\left(\mathfrak T^B_{\epsilon U}
[f]\,-\,f\right)\,=\,u\cdot(D_x+\delta^B)f\,+\,\mu\cdot D_\xi f,
\end{equation}
where
$$
\delta^B_jf:=\sum_{k=1}^nc_{jk}^B\star D_{\xi_k}f,\quad\text{with}\quad c_{jk}^B(x,\xi):=(2\pi)^{-n}
\int_{\mathcal X}dy\,e^{-i<\xi,y>}b_{jk}^B(x,y),
$$
and $\,b_{jk}^B(x,y):=\int_{-1/2}^{1/2}ds\,B_{jk}(x+sy)\,$ are functions belonging to $BC^\infty(\mathcal X
\times\mathcal X)$.
\end{proposition}
\begin{proof}
The first equality is a particular case of (\ref{parti}).

The last equality in (\ref{deriv}) is easily obtained from (\ref{aut}) by differentiation. For $j\in\{1,\ldots,n\}$
we have
$$
-i\partial_{x_j}f+\delta^B_j f:=\mathfrak{ad}^B_{e_j}[f]=(-i)\left.\frac{\partial}{\partial u_j}\right|_{U=0}\left\{\mathfrak T^B_U[f]\right\}=
\left.(-i)\frac{\partial}{\partial u_j}\left\{\widetilde{\Omega^B_\mathcal{P}}[u]\star\mathfrak T_U[f]\right\}
\right|_{U=0},
$$
so that, taking into account the bilinearity of the mixed product $\star$, we get
$$
\left(\delta^B_j f\right)(X)=(-i)\left\{\left.\left(\frac{\partial}{\partial u_j}
\widetilde{\Omega^B_\mathcal{P}}[u]\right)\right|_{u=0}\star f\right\}(X).
$$
But
$$
\left.\left(\frac{\partial}{\partial u_j}\widetilde{\Omega^B_\mathcal{P}}[u]\right)\right|_{u=0}(X)=(2\pi)^{-n}\,
\int_{\mathcal{X}}dy\, e^{-i<y,\xi>}\left.\left(\frac{\partial}{\partial u_j}\Omega^B[\mathcal{P}(x;y,u)]
\right)\right|_{u=0}=
$$
$$
=-i(2\pi)^{-n}\,\int_{\mathcal{X}}dy\, e^{-i<y,\xi>}\sum_{k=1}^ny_k\int_{-1/2}^{1/2}ds\,B_{kj}(x+sy)=
$$
$$
=\sum_{k=1}^n\frac{\partial}{\partial \xi_k}\left((2\pi)^{-n}\,\int_{\mathcal{X}}dy\, e^{-i<y,\xi>}
\int_{-1/2}^{1/2}dsB_{kj}(x+sy) \right),
$$
and using the properties of the usual convolution we get
$$
\delta^B_j f=(-i)\sum_{k=1}^n\left((2\pi)^{-n}\,\int_{\mathcal{X}}dy\, e^{-i<y,\xi>}\int_{-1/2}^{1/2}dsB_{kj}
(x+sy) \right)\star\left(\frac{\partial}{\partial \xi_k}f \right).
$$
\end{proof}

Let us notice that $D_{\xi_l}(c\star f)=c\star D_{\xi_l}f$ and
$\,D_{x_l}(c\star f)=D_{x_l}c\star f+c\star D_{x_l}f$.
Using then the associativity and commutativity properties of $\star$, it is easy to verify that the
different components of the vectorial operator $\delta^B$ commute with each other and we have
\begin{equation}\label{unca}
\delta^B_j\delta^B_k[f]=\delta^B_k\delta^B_j[f]=\sum_{l=1}^n
\sum_{m=1}^nc_{jl}^B\star
c_{km}^B\star[D_{\xi_l}D_{\xi_m} f],
\end{equation}

\noindent Moreover we obtain
\begin{equation}\label{doica}
\left[D_{\xi_l},\delta^B_j\right]=0,\ \
\left[D_{x_l},\delta^B_j\right]=\delta_j^{D_lB},
\end{equation}
and thus
\begin{equation}\label{lie}
\left[\ad^B_{(u,\mu)},\ad^B_{(v,\nu)}\right]=-\sum_{j,k=1}^n(u_kv_j-u_jv_k)\delta^{D_jB}_k=-\sum_{j,k=1}^n(u\wedge
v)_{kj}\delta^{D_jB}_k.
\end{equation}

\subsection{Estimations for the magnetic derivatives}

\begin{proposition}\label{alg-coef}
 For $B$ with components of class $BC^\infty(\mathcal{X})$ and for any $j\in\{1,\ldots,n\}$ we have
$$
i\delta^B_jf:=\sum_{1\leq|\alpha|\leq2([n/2]+1)+1}c_{j\alpha}^B\star\partial^\alpha_{\xi}f\qquad\forall f\in\mathcal{S}(\Xi).
$$
with the coefficients $c_{j\alpha}^B$ belonging to the space
$\mathfrak{A}(\Xi)$.
\end{proposition}
\begin{proof}
 If we denote by $\mathcal{F}_2$ the Fourier transform in the second variable (the $\xi$-variable), then we can
 write for any natural numbers $m$ and $M$
$$
i\delta^B_j
f=\sum_{k=1}^nc_{jk}^B\star\partial_{\xi_k}f=\sum_{k=1}^n\mathcal{F}_2\left[b_{jk}^B\right]\star\partial_{\xi_k}f=
\sum_{k=1}^n\mathcal{F}_2\left[(1\otimes\mathfrak{w}_{(-2M)})b_{jk}^B\right]\star(1\otimes<D>^{2M})\partial_{\xi_k}f=
$$
$$
=\sum_{k=1}^n(1\otimes\mathfrak{w}_{(-2m)})\mathcal{F}_2\left[(1\otimes<D>^{2m})(1\otimes\mathfrak{w}_{(-2M)})
b_{jk}^B\right]\star(1\otimes<D>^{2M})\partial_{\xi_k}f=
$$
$$
=\sum_{k=1}^n\mathfrak{p}_{(-2m)}\mathcal{F}_2\left[(1\otimes<D>^{2m})(1\otimes\mathfrak{w}_{(-2M)})b_{jk}^B\right]
\star\left[\sum\limits_{0\leq|\beta|\leq2M}C^{(2M)}_\beta\partial_{\xi}^{\beta}\partial_{\xi_k}f\right]=
$$
$$
=:\sum\limits_{1\leq|\beta|\leq2M+1}\mathfrak{p}_{(-2m)}\mathcal{F}_2\left[(1\otimes<D>^{2m})
(1\otimes\mathfrak{w}_{(-2M)})\tilde{b}_{j\beta}^B\right]\star\left[\partial_{\xi}^{\beta}f\right],
$$
with $\tilde{b}_{j\beta}^B\in BC^\infty(\mathcal{X}\times\mathcal{X})$. Thus we have the following expression
for the coefficient functions:
$$
c^B_{j\alpha}=\mathfrak{p}_{(-2m)}\mathcal{F}_2\left[(1\otimes<D>^{2m})
(1\otimes\mathfrak{w}_{(-2M)})\tilde{b}_{j\beta}^B\right].
$$
Now we can chose
$m$ such that $\mathfrak{w}_{-2m}\in L^1(\mathcal{X})$; take $2m>n$.
By also choosing $2M>n$ we get that
$$
\|c^B_{j\alpha}\|_{\infty,1}\leq\|\mathfrak{p}_{-2m}\|_{L^1}\underset{X\in\Xi}{\sup}
\left|\mathcal{F}_2\left[(1\otimes<D>^{2m})(1\otimes\mathfrak{w}_{(-2M)})\tilde{b}_{j\beta}^B\right](X)\right|\leq
$$
$$
\leq\|\mathfrak{w}_{(-2m)}\|_{L^1}\underset{x\in\mathcal{X}}{\sup}
\int_{\mathcal{X}^\prime}\left|[(1\otimes<D>^{2m})(1\otimes\mathfrak{w}_{(-2M)})\tilde{b}_{j\beta}^B](x,y)\right|\leq
$$
$$
\leq C\|\mathfrak{w}_{(-2m)}\|_{L^1}\|\mathfrak{w}_{(-2M)}\|_{L^1}\left[\underset{|\beta|\leq2m}{\max}\
\
\underset{(x,y)\in\mathcal{X}\times\mathcal{X}}{\sup}|[\partial_y^\beta
\tilde{b}_{j\beta}^B](x,y)|\right].
$$

For the second conclusion we repeat the proof of Proposition \ref{alg-coef} above, taking now $2m>n+p$.
\end{proof}

\medskip
\begin{remark}\label{stable}
Among others, the results above imply that (together with Prop.3.6, pct.3)
\begin{equation}\label{stabli}
\ad^B_U\left[S^m_{\rho}(\Xi)\right]\subset S^m_{\rho}(\Xi),\ \ \forall U\in\Xi, \,m\in\mathbb R,\,\rho\in[0,1].
\end{equation}
\end{remark}

\medskip
\begin{remark}\label{x-der-c}
Using the explicit form of the functions $b^B_{jk}$ appearing in Proposition \ref{delta-B}, one easily proves that
$$
\partial_{x_l}c^B_{j\alpha}=
\mathfrak{p}_{(-2m)}\mathcal{F}_2
\left[(1\otimes<D>^{2m})(1\otimes\mathfrak{w}_{(-2M)})\left(\partial_{x_l}\tilde{b}_{j\beta}^B\right)\right]=
$$
$$
=\mathfrak{p}_{(-2m)}\mathcal{F}_2
\left[(1\otimes<D>^{2m})(1\otimes\mathfrak{w}_{(-2M)})\tilde{b}_{j\beta}^{\partial_{l}B}\right]=c^{\partial_{l}B}_{j\alpha}
$$
\end{remark}
\medskip

\begin{corollary}\label{cor-alg-coef}
For $B$ with components of class $BC^\infty(\mathcal{X})$ and any $j\in\{1,\ldots,n\}$ we have
\begin{itemize}
 \item $\|\delta^B_j[f]\|_\infty\leq C_\infty\sum_{|\alpha|\leq2[n/2]+3}\|\partial_\xi^\alpha f\|_\infty,\quad\forall f\in\mathcal{S}(\Xi)$
\item $\mathfrak{ad}^B_U$ sends $S^m_\rho(\Xi)$ in $S^m_\rho(\Xi)$ for any $m\in \mathbb R$ and any $\rho\in[0,1]$.
\item $\delta^B_j, \mathfrak{ad}^B_U:BC^{\infty}(\Xi)\rightarrow BC^{\infty}(\Xi)$ are continuous operators.
\item $g\in\mathcal{S}(\Xi)$ implies $\delta^B_j[g]\in\mathcal{S}(\Xi)$, the map $\delta^B_j$ being continuous
for the topology of $\mathcal{S}(\Xi)$.
\end{itemize}
\end{corollary}

Taking into account this result, the explicit formula (\ref{deriv}) relying usual derivatives to the magnetic ones
and the commutation relations (\ref{doica}) and (\ref{lie}), we get easily

\medskip
\begin{corollary}\label{scurtu}
On $BC^\infty(\Xi)=S^0_{0}(\Xi)=C^\infty(\mathfrak T;BC(\Xi))$, the families of seminorms $\left\{\parallel\cdot
\parallel^{\mathfrak T,BC}_{(a,\alpha)}\mid (a,\alpha)\in\mathbb N^{2n}\right\}$ and
\begin{equation}
\left\{\parallel\cdot\parallel^{\mathfrak T^B,BC}_{(U_1,\dots,U_N)}\mid |U_1|=\dots=|U_N|=1\right\},\ \ \ {\rm with}\ \ \
\parallel F\parallel^{\mathfrak T^B,BC}_{(U_1,\dots,U_N)}:=\parallel \mathfrak{ad}^B_{U_1}\dots\mathfrak{ad}^B_{U_N}[F]
\parallel_\infty
\end{equation}
are equivalent.
\end{corollary}

\section{Proof of Theorem \ref{main}}
We shall consider a fixed an euclidean basis $\{e_1,\ldots,e_n\}$ of our configuration space $\mathcal{X}$ and the dual basis $\{\epsilon_1,\ldots,\epsilon_n\}$ in $\mathcal{X}^\prime$. We shall constantly use the multi-index type notations
$$
\big(\mathfrak{ad}^B_{\boldsymbol{e}}\big)^a:=\big(\mathfrak{ad}^B_{e_1}\big)^{a_1}\circ\cdots\circ\big(\mathfrak{ad}^B_{e_n}\big)^{a_n}
$$
$$
\big(\mathfrak{ad}^B_{\boldsymbol{\epsilon}}\big)^\alpha:=\big(\mathfrak{ad}^B_{\epsilon_1}\big)^{\alpha_1}\circ\cdots\circ\big(\mathfrak{ad}^B_{\epsilon_n}\big)^{\alpha_n}
$$
\subsection{Proof of the embedding $BC^\infty(\Xi)\subset C^\infty(\mathfrak T^B;\mathfrak{C}^B)$}

Recalling that $BC^\infty(\Xi)=S^0_{0}(\Xi)$, this inclusion
will be a consequence of the following Calderon-Vaillancourt type theorem, that we proved in \cite{IMP}:

\medskip
\begin{theorem}\label{CV}
Assume that the magnetic field $B$ has components of class
$BC^\infty$. Let $G\in S^0_{\rho,\rho}(\Xi)$ for some $\rho\in[0,1)$. Then $G\in\mathfrak C^B(\Xi)$
(i.e. $\mathfrak{Op}^A(G)\in\mathbb{B}(L^2(\mathcal{X}))$) for $B=dA$) and we have the inequality
\begin{equation}\label{inequality}
\left\|G\right\|_{\mathfrak C^B}\,\leq\,c(n)\,\underset{|a|\leq p(n)}{\sup}\;\underset{|\alpha|\leq
p(n)}{\sup}\;\underset{X\in\Xi}{\sup}\left[<\xi>^{\rho(|\alpha|-|a|)}\left|\partial^a_x\partial^\alpha_\xi
G(X)\right|\right],
\end{equation}
where $c(n),p(n)$ are constants depending only on $n$, that can be determined explicitly.
\end{theorem}

\medskip
Let $F$ be an element of $BC^\infty(\Xi)$. By Corollary \ref{cor-alg-coef}, $G:=\mathfrak{ad}^B_{U_1}\ldots\mathfrak{ad}^B_{U_N}[F]$
also belongs to $BC^\infty(\Xi)$, so we can use Theorem \ref{CV} with $\rho=0$ and get
$$
\parallel \mathfrak{ad}^B_{U_1}\ldots\mathfrak{ad}^B_{U_N}[F]\parallel_{\mathfrak C^B}\le
c(n)\,\underset{|a|\leq p(n)}{\sup}\;\underset{|\alpha|\leq
p(n)}{\sup}\;\parallel\partial^a_x\partial^\alpha_\xi\mathfrak{ad}^B_{U_1}\ldots\mathfrak{ad}^B_{U_N}[F]\parallel_\infty
.
$$
Then applying Corollary \ref{scurtu} gives the result.

\subsection{Proof of the embedding $C^\infty(\mathfrak T^B;\mathfrak{C}^B)\subset BC^\infty(\Xi)$}

The basic step is

\medskip
\begin{theorem}\label{aciu}
There exists $N\in\mathbb{N}$, dependig only on the dimension $n$ and on the magnetic field $B$, such that any distribution $f\in\mathcal{S}^\prime(\Xi)$ satisfying $\|\big(\mathfrak{ad}^B_{\boldsymbol{e}}\big)^a\big(\mathfrak{ad}^B_{\boldsymbol{\epsilon}}\big)^\alpha [f]\|_{\mathfrak C^B}<\infty$ for all $|a|+|\alpha|\leq N$ is in fact a bounded measurable function and there exist a constant $C<\infty$ such that
$$
\|f\|_{\infty}\leq C \sum_{|a|+|\alpha|=
N}\|\big(\mathfrak{ad}^B_{\boldsymbol{e}}\big)^a\big(\mathfrak{ad}^B_{\boldsymbol{\epsilon}}\big)^\alpha [f]\|_{\mathfrak C^B}.
$$
\end{theorem}

\medskip
Then for any $F\in C^\infty(\mathfrak T^B;\mathfrak C^B)$, the function $f:=\mathfrak{ad}^B_{U_1}\ldots\mathfrak{ad}^B_{U_N}[F]$
can be plugged in Theorem \ref{aciu}. The Fr\'echet spaces embedding
$C^\infty(\mathfrak T^B;\mathfrak{C}^B)\subset BC^\infty(\Xi)$ follows by applying Corollary \ref{scurtu}.

So we only need to prove Theorem \ref{aciu}. For this we use a strategy inspired by \cite{Bo1} (Lemma 2.2);
some changes are needed to implement magnetic translations and derivatives.

\begin{proof}
{\it 1.} Since the algebraic tensor product $\mathcal{S}(\mathcal{X})\odot\mathcal{S}(\mathcal{X})$ is dense
in $\mathcal{S}(\mathcal{X}\times\mathcal{X})$ and the inclusion $\mathcal{S}(\mathcal{X})\hookrightarrow L^2(\mathcal{X})$
is continuous, taking into account the explicit formula for the kernel of a magnetic pseudodifferential operator, we deduce
that the inclusion map
$\mathfrak{C}^B(\Xi)\hookrightarrow\mathcal{S}^\prime(\Xi)$ is continuous. It follows that there exist constants
$C_1>0$ and $N_1\in\mathbb{N}$ such that for any $f\in\mathfrak{C}^B(\Xi)$ and any $\varphi\in\mathcal{S}(\Xi)$
$$
|<f,\varphi>|\leq C \|f\|_{\mathfrak C^B}\,\left\{\underset{|\mathfrak a|+|\mathfrak b|\leq N_1}{\max}\
\underset{X\in\Xi}{\sup}\left|X^{\mathfrak a}\big(\partial^{\mathfrak b}\varphi\big)\right|\right\}
\equiv C \|f\|_{\mathfrak C^B}\,\||\varphi|\|_{N_1},
$$
where $\mathfrak a$ and $\mathfrak b$ are multi-indices in $\mathbb{N}^{2n}$ and $\||\cdot|\|_{N_1}$ is one of the semi-norms of
$\mathcal S(\Xi)$.

{\it 2.} If $\ast$ denotes the usual convolution and $\check f(Y):=\overline{f(-Y)}$, we evidently have for any $X\in\Xi$
\begin{equation}\label{relation}
(\check f\ast\phi)(X)=<f,\mathfrak T_{-X}[\phi]>.
\end{equation}
Now we choose a function $\chi\in \mathcal{S}(\Xi)$ such that its Fourier transform $\hat{\chi}\in
C^\infty_0(\Xi)$ with $\hat{\chi}(X)=1$ on a neighborhood of the
origin of $\Xi$. For any $\epsilon\in(0,\epsilon_0]$ and $X\in\Xi$ we set $\chi_{_{X,\epsilon}}(Y):=\epsilon^{-2n}\chi((Y-X)/\epsilon)$, $\chi_X(Y):=\chi_{X,1}(Y)$. We have
$$
\frac{\partial\chi_{_{X,\epsilon}}}{\partial\epsilon}(Y)=-2n\epsilon^{-1}\chi_{_{X,\epsilon}}(Y)
-\epsilon^{-1}\epsilon^{-2n}\sum_{j=1}^{2n}\left(\frac{Y_j-X_j}{\epsilon}\right)\partial_j\chi((Y-X)/\epsilon)=:\epsilon^{-1}\psi_{_{X,\epsilon}}(Y),
$$
where the function $\widehat\psi$ vanishes on a neighborhood of $0$.
Therefore, for any $N\in\mathbb{N}$ we can find $C^\infty_0$
functions $\left\{\widehat\theta_{\mathfrak a}\mid |\mathfrak a|=N\right\}$ satisfying
$\widehat\psi(Z)=\sum_{|\mathfrak a|=N}(iZ)^{\mathfrak a}\widehat\theta_{\mathfrak a}(Z)$.
Thus for any  $f\in\mathfrak{C}^B(\Xi)$ and for $\delta>0$
$$
\check f*\chi_{_{X,\delta}}=\check f*\chi_{_{X}}-\int_\delta^1\check f*\left(\frac{\partial\chi_{_{X,\epsilon}}}{\partial\epsilon}\right)d\epsilon=
\check f*\chi_{_{X}}-
\int_\delta^1\epsilon^{N-1}\underset{|\mathfrak a|=N}{\sum}\check f*\left[\partial^{\mathfrak a}\left(
{\theta_{\mathfrak a}}\right)_{_{X,\epsilon}}\right]d\epsilon=
$$
\begin{equation}\label{importanta}
=\check f*\chi_{_{X}}-\int_\delta^1\epsilon^{N-1}\underset{|\mathfrak a|=N}{\sum}\big(\partial^{\mathfrak a}\check f\big)*\left(
{\theta_{\mathfrak a}}\right)_{_{X,\epsilon}} d\epsilon.
\end{equation}
Since $\underset{\delta\rightarrow 0}{\lim}\check{f}*\chi_{_{X,\delta}}=\check{f}(X)$ weakly in
$\mathcal{S}^\prime(\Xi)$, if we can find a finite bound for the right hand-side member we deduce that $f$ belongs to $L^\infty(\Xi)$ and we obtain
$$
\|f\|_\infty\leq
C\|f\|_{\mathfrak C^B}\,\||\chi|\|_{N_1}+C\underset{|\mathfrak a|=N}
{\sum}\|\big(\partial^{\mathfrak a}f\big)\|_{\mathfrak C^B}\,\left(\underset{\delta\rightarrow0}
{\lim}\int_\delta^1\epsilon^{N-1}\epsilon^{-N_1-2n}\left\|\left|\theta_{\mathfrak a}\right|\right\|_{N_1} d\epsilon\right).
$$
The problem now is to replace in the last term above the usual derivatives of $f$ with
magnetic derivatives.

{\it 3.}  Let us first notice that
$$
\partial^{\mathfrak a}f=\partial_x^a\partial_\xi^\alpha f=\left[\prod_{1\leq j\leq n}{\left(\mathfrak{ad}^B_{e_j}-\sum_{1\leq|
\beta|\leq2([n/2]+1)}\boldsymbol{c}_{j,\beta}^B\star(\mathfrak{ad}^B_{\boldsymbol{\epsilon}})^\beta \right)^{\!\!a_j}}\right](\mathfrak{ad}^B_{\boldsymbol{\epsilon}})^\alpha[f].
$$
By direct computations it is easy to show that
\begin{equation}\label{1}
\mathfrak{ad}^B_{e_j}\left[c_{k,\beta}^B\star g
\right]=c_{k,\beta}^B\star(\mathfrak{ad}^B_{e_j}g)-ic_{k,\beta}^{\partial_jB}\star g
\end{equation}
and
\begin{equation}\label{2}
c_{k,\beta}^B\star(\mathfrak{ad}^B_{\boldsymbol{\epsilon}})^{\beta}\left[
c_{l,\gamma}^B\star(\mathfrak{ad}^B_{\boldsymbol{\epsilon}})^{\gamma}g\right]=\left(c_{k,\beta}^B\star
c_{l,\gamma}^B\right)\star\left[(\mathfrak{ad}^B_{\boldsymbol{\epsilon}})^{\beta+\gamma}g \right].
\end{equation}
Thus we conclude by a simple induction procedure that for any multi-index $\mathfrak a\in\mathbb{N}^{2n}$, the derivative
$\partial^{\mathfrak a}f$ is a finite sum of terms of the form $r\star(\mathfrak{ad}^B_{\boldsymbol{e}})^{a}(\mathfrak{ad}^B_{\boldsymbol{\epsilon}})^{\gamma}f$ with
$r\in\mathfrak{A}(\Xi)$ (the algebra introduced in the Notations at the begining of the previous section), $\gamma\in\mathbb{N}^n$ and $(\mathfrak{ad}^B_{\boldsymbol{e}})^{a}(\mathfrak{ad}^B_{\boldsymbol{\epsilon}})^{\gamma}f\in\mathfrak{C}^B(\Xi)$.

{\it 4.}
Let us choose $r\in\mathfrak{A}(\Xi)$, $g\in\mathfrak{C}^B(\Xi)$ and $\psi\in\mathcal{S}(\Xi)$; then $r\star
g\in\mathcal{S}^\prime(\Xi)$. The following triple integral is absolutely convergent and
$$
\left[(r\star g)*\psi\right](X)=\int_{\mathcal{X}}dy\int_{\mathcal{X}^\prime}d\eta\int_{\mathcal{X}^\prime}d\zeta\
r(y,\zeta) g(y,\eta-\zeta)\psi(x-y,\xi-\eta)=
$$
$$
=\int_{\mathcal{X}}dy\int_{\mathcal{X}^\prime}d\eta\int_{\mathcal{X}^\prime}d\zeta\
r(y,\xi-\eta-\zeta) g(y,\zeta)\psi(x-y,\eta)=
$$
$$
=\int_{\mathcal{X}}dy\int_{\mathcal{X}^\prime}d\eta\int_{\mathcal{X}^\prime}d\zeta\
g(y,\zeta)\tau_{-x}[r](y-x,\xi-\eta-\zeta)\psi(x-y,\eta)=
$$
$$
=\left\{g*\left[\left(\tau_{-x}[r]\right)^{\vee_1}\star\psi\right]\right\}(x,\xi),
$$
where $\left(\tau_{-x}[r]\right)^{\vee_1}(y,\eta):=\left(\tau_{-x}[r]\right)(-y,\eta)=r(x-y,\eta)$. Thus, using the first two points of the proof we get
$$
\underset{X\in\Xi}{\sup}\left|\left[(r\star
g)*\psi\right](X)\right|=\underset{(x,\xi)\in\Xi}{\sup}\left|\left\{g*\left[\left(\tau_{-x}[r]\right)^{\vee_1}\star\psi\right]\right\}(x,\xi)\right|\leq
$$
$$
\leq\underset{y\in\mathcal{X}}{\sup}\,\underset{(x,\xi)\in\Xi}{\sup}\left|\left\{g*\left[\left(\tau_{-x}[r]\right)^{\vee_1}\star\psi\right]\right\}(x,\xi)\right|\leq
\underset{y\in\mathcal{X}}{\sup}\,C\|g\|_{\mathfrak C^B}\,\||\tau_y[r]^{\vee_1}\star\psi|\|_{N_1}\leq
$$
$$
\leq C(N_1,n;r)\|g\|_{\mathfrak C^B}\,\||\psi|\|_{N_1},
$$
using point 4 of Proposition \ref{L-inf-p} in the last step.

{\it 5.} By writing the last term of the right hand side of (\ref{importanta}) as a linear combination of terms of the form $(r\star g)*\psi$ with $g$ replaced by a suitable sequence of magnetic derivatives
applied to $f$, $\psi=\left(\theta_{\mathfrak a}\right)_{_{X,\epsilon}}$ and taking $N$ large, we finish the proof of our Theorem.
\end{proof}

\section{Commutator Criteria for Magnetic Pseudodifferential Operators}

\subsection{The class of symbols $\boldsymbol{S^m_{0}(\Xi)}$}

For $\lambda>0$ and $m>0$ we define
\begin{equation}\label{pma}
\mathfrak{p}_{m,\lambda}(X):=\mathfrak{p}_m(X)+\lambda=<\xi>^m+\lambda,
\end{equation}
that is clearly an elliptic element of $S^m_1(\Xi)\subset S^m_0(\Xi)\subset\mathfrak{M}^B(\Xi)$.
In a previous paper \cite{MPR2} we have shown that for $\lambda>0$ large enough, $\mathfrak{p}_{m,\lambda}$
is invertible in $\mathfrak{M}^B(\Xi)$ with a quasi-explicit inverse
\begin{equation}\label{tit}
\mathfrak{p}_{m,\lambda}^-\,=\,\mathfrak{p}_{m,\lambda}^{-1}\dB\sum_{k\in\mathbb{N}}\mathfrak{r}_\lambda^{(\dB
k)}\,=\,\mathfrak{p}_{m,\lambda}^{-1}\dB(1-\mathfrak r_\lambda)^-,
\end{equation}
where $\mathfrak{p}_{m,\lambda}^{-1}(x,\xi)=(<\xi>^m+\lambda)^{-1}$ is the usual point-wise inverse and
$$
\mathfrak{r}_\lambda=\mathfrak{p}_{m,\lambda}\dB\mathfrak{p}_{m,\lambda}^{-1}-1\in S^0_1(\Xi)\subset S^0_0(\Xi)\subset\mathfrak C^B(\Xi)
$$
is a reminder that can be controlled as in \cite{MPR2}. The series converges in $\mathfrak C^B(\Xi)$.

\noindent For any $m>0$ we fix $\lambda>0$ such that $\mathfrak{p}_{m,\lambda}$ is invertible. We shall use the weight function
\begin{equation}\label{pmam}
\mathfrak{s}_m:=\left\{
\begin{array}{c}
\mathfrak{p}_{m,\lambda},\quad\text{for}\quad m>0\\
\mathfrak{p}^-_{|m|,\lambda},\quad\text{for}\quad m<0,
\end{array}
\right.
\end{equation}
such that $\mathfrak s_m^-=\mathfrak s_{-m}$. For $m=0$ we set simply $\mathfrak s_0:=1$.

\medskip
\begin{theorem}\label{main-m}
A distribution $F\in\mathcal{S}^\prime(\Xi)$ is a symbol of type
$S^m_{0}(\Xi)$ if and only if for any $N\in\mathbb{N}$ and any
family of $N$ vectors $\{X_1,\ldots,X_N\}\subset\Xi$ the following is true:
$$
\mathfrak{s}_{m}^-\dB\left(\ad^B_{X_1}\cdot\ldots\cdot\ad^B_{X_N}[F]\right)\in\mathfrak{C}^B(\Xi)
\qquad\forall m\in\mathbb{R}.
$$
The families of semi-norms $\|\mathfrak{s}_{m}^-\partial_x^a\partial_\xi^\alpha F\|_\infty$,
with $(a,\alpha)\in\mathbb{N}^{2n}$, and
$\|\mathfrak{s}_{m}^-\dB\left(\ad^B_{X_1}\cdot\ldots\cdot\ad^B_{X_N}[F]\right)\|_{\mathfrak C^B}$,
indexed by $N\in\mathbb{N}$ and $N$-tuples of vectors in $\Xi$, define equivalent topologies on $S^m_0(\Xi)$.
\end{theorem}

\begin{proof}
{\sf Step 1.} The Theorem is true for $m=0$ (it is just our Theorem \ref{intrinseca}).

{\sf Step 2.}
For $m<0$, $\mathfrak{s}^-_m=\mathfrak{s}_{|m|}=\mathfrak{p}_{|m|,\lambda}\in S^{|m|}_{1}(\Xi)\subset S^{|m|}_{0}(\Xi)$.
If $F\in S^m_{0}(\Xi)$, by Corollary \ref{cor-alg-coef}
$$
\ad^B_{X_1}\ldots\ad^B_{X_N}[F]\in S^m_0(\Xi),\ \ \ \forall X_1,\dots,X_n\in\Xi,\ \forall n\in\mathbb N.
$$
By (\ref{hor})  $\mathfrak{s}^-_m\dB (\ad^B_{X_1}\ldots\ad^B_{X_N}[F])\in S^0_{0}(\Xi)$
and thus it belongs to $\mathfrak{C}^B(\Xi)$, as we already know from Theorem
\ref{CV}. Thus the direct implication in the statement of the Theorem \ref{main-m} is proved for $m<0$.

{\sf Step 3.} For $m<0$ we shall prove that for any $F\in\mathcal{S}^\prime(\Xi)$:
\begin{equation}\label{step3}
\forall N\in\mathbb{N},\ \forall\{X_1,\ldots,X_N\}\subset\Xi,\
\mathfrak{s}_m^-\dB\left(\ad^B_{X_1}\cdot\ldots\cdot\ad^B_{X_N}[F]\right)\in\mathfrak{C}^B(\Xi)\ \Rightarrow\ F\in S^0_{0}(\Xi).
\end{equation}
Since $\mathfrak{s}_m=\mathfrak{p}_{|m|,\lambda}^-\in\mathfrak{C}^B(\Xi)$, for any $X_1,\ldots,X_N\in\Xi$ we have
$$
\ad^B_{X_1}\ldots\ad^B_{X_N}[F]=\mathfrak{s}_m\dB\left(\mathfrak{s}_m^-
\dB\ad^B_{X_1}\ldots\ad^B_{X_N}[F]\right)\in\mathfrak{C}^B(\Xi),
$$
and using Theorem \ref{intrinseca} we conclude that $F\in S^0_{0}(\Xi)$.

{\sf Step 4.} For $m<0$ we shall prove that for any $F\in\mathcal{S}^\prime(\Xi)$:
\begin{equation}\label{step4}
\forall N\in\mathbb{N},\ \forall\{X_1,\ldots,X_N\}\subset\Xi,\
\mathfrak{s}_m^-\dB\left(\ad^B_{X_1}\cdot\ldots\cdot\ad^B_{X_N}[F]\right)\in\mathfrak{C}^B(\Xi)\
\Rightarrow \mathfrak{s}_m^-\dB F\in S^0_{0}(\Xi).
\end{equation}
Due to the hypothesis of (\ref{step4}) with $N=0$,
$\mathfrak{s}_m^-\dB F$ belongs to $\mathfrak{C}^B(\Xi)$ and we compute
\begin{equation}\label{step4-1}
\ad^B_{X_1}\ldots\ad^B_{X_M}\left[\mathfrak{s}_m^-\dB F\right]=
\end{equation}
$$
=\sum_{k=0}^M\sum_{j_1,\ldots,j_k}\left(\ad^B_{X_{j_1}}\ldots\ad^B_{X_{j_k}}\left[\mathfrak{p}_{|m|,\lambda}\right]\right)
\dB\mathfrak{s}_m\dB\mathfrak{s}_m^-\dB\left(\ad^B_{X_{j^\prime_1}}\ldots\ad^B_{X_{j^\prime_{M-k}}}[F]\right),
$$
where $j_1<\dots<j_k$, $j'_1<\dots<j'_{M-k}$ and
$\{j_1,\ldots,j_k\}\bigcup\{j^\prime_1,\ldots,j^\prime_{M-k}\}=\{1,\ldots,M\}$.

The factors $\mathfrak{s}_m^-\dB\left(\ad^B_{X_{j^\prime_1}}\ldots\ad^B_{X_{j^\prime_{M-k}}}[F]\right)$
belong to $\mathfrak{C}^B(\Xi)$, by the hypothesis in (\ref{step4}).

Using (\ref{tit}) for $m<0$ and the fact that $\mathfrak{ad}^B_X$ preserves the symbol space $S^m_0(\Xi)$, we can write
\begin{equation}\label{step4-2}
\left(\ad^B_{X_{j_1}}\ldots\ad^B_{X_{j_k}}\left[\mathfrak{p}_{|m|,\lambda}\right]\right)\dB\mathfrak{s}_m=
\left(\ad^B_{X_{j_1}}\ldots\ad^B_{X_{j_k}}\left[\mathfrak{p}_{|m|,\lambda}\right]\right)\dB\mathfrak{p}_{|m|,\lambda}^{-1}\dB
(1-\mathfrak{r}_{\lambda})^-\in
\end{equation}
$$
\in S^{|m|}_{0}(\Xi)\dB
S^{-|m|}_{0}(\Xi)\dB\mathfrak C^B(\Xi)\subset\mathfrak{C}^B(\Xi).
$$
We conclude by induction on $M$ that $\ad^B_{X_1}\ldots\ad^B_{X_M}\left[\mathfrak{s}_m^-\dB F\right]\in\mathfrak{C}^B(\Xi)$ for any
$X_1,\dots,X_M$, thus by Theorem \ref{intrinseca} we get $\mathfrak{s}_m^-\dB F\in S^0_{0}(\Xi)$.

{\sf Step 5.} For $m<0$ we shall prove that for any $F\in\mathcal{S}^\prime(\Xi)$
\begin{equation}\label{step5}
\forall N\in\mathbb{N},\ \forall\{X_1,\ldots,X_N\}\subset\Xi,\
\mathfrak{s}_m^-\dB\left(\ad^B_{X_1}\cdot\ldots\cdot\ad^B_{X_N}[F]\right)\in\mathfrak{C}^B(\Xi)\
\Rightarrow\ F\in S^m_{0}(\Xi).
\end{equation}
Using the results in \cite{MPR2} (see formula (2.6) in the proof of Theorem 1.8 in section 2.1 of the paper) for the symbol
$\mathfrak{s}_m^-=\mathfrak{p}_{|m|,\lambda}\in S^{|m|}_{1}(\Xi)$, there exists $\mathfrak{u}\in
S^r_{0}(\Xi)$ with $r<0$ such that for any $M\in\mathbb{N}$
\begin{equation}
\left[\sum_{k=0}^M \mathfrak{u}^{(\dB
k)}\right]\dB\mathfrak{p}_{|m|,\lambda}^{-1}\dB\mathfrak{p}_{|m|,\lambda}=1-\mathfrak{u}^{(\dB(M+1))}.
\end{equation}
Thus we can write
$$
F=\left[\sum_{k=0}^M\mathfrak{u}^{(\dB
k)}\right]\dB\mathfrak{p}_{|m|,\lambda}^{-1}\dB\left[\mathfrak{p}_{|m|,\lambda}\dB
F\right]+\mathfrak{u}^{(\dB (M+1))}\dB F.
$$
Now $F$ and $\mathfrak{p}_{|m|,\lambda}\dB F$ belong to $S^0_{0}(\Xi)$ by Step 3 and Step 4, $\left[\sum_{0\leq k\leq
M}\mathfrak{u}^{(\dB k)}\right]$ belongs to $S^0_{0}(\Xi)$ due to the properties of $\mathfrak{u}$ and
$\mathfrak{p}_{|m|,\lambda}^{-1}\in S^m_{0}(\Xi)$. Thus if we choose $M>m/r$ and use the Theorem of magnetic
composition of symbols from \cite{IMP}, we get the desired conclusion $F\in S^m_{0}(\Xi)$.

{\it Thus we have proved the Theorem for $m\leq0$.}

{\sf Step 6.} We show that for any $p>0$ the distribution $\mathfrak{s}_p^-\in\mathfrak{C}^B(\Xi)$ is a symbol of type
$S^{-p}_{0}(\Xi)$. In fact we apply the result of
the Theorem with $m=-p<0$ for $F=\mathfrak{s}_p^-$. To the obvious relation
$$
\mathfrak{s}_p\dB\mathfrak{s}^-_p=1
$$
one applies the operator
$\ad^B_{X_1}\cdot\ldots\cdot\ad^B_{X_N}$, using the Leibnitz rule for derivations to obtain
\begin{equation}
\sum_{0\leq k\leq N}\
\sum_{j_1,\ldots,j_k}\left(\ad^B_{X_{j_1}}\ldots\ad^B_{X_{j_k}}\left[\mathfrak{s}_p\right]\right)
\dB\left(\ad^B_{X_{j^\prime_1}}\ldots\ad^B_{X_{j^\prime_{N-k}}}\left[\mathfrak{s}_p^-\right]\right)=0,
\end{equation}
where $j_1<\dots<j_k$, $j'_1<\dots<j'_{N-k}$ and
$\{j_1,\ldots,j_k\}\bigcup\{j^\prime_1,\ldots,j^\prime_{N-k}\}=\{1,\ldots,N\}$.
We can rewrite it as
\begin{equation}
\mathfrak{s}_p\dB\left(\ad^B_{X_{1}}\ldots\ad^B_{X_{N}}\left[\mathfrak{s}_p^-\right]\right)
=-\sum_{k=0}^{N-1}\sum_{j_1,\ldots,j_k}
\left(\ad^B_{X_{j^\prime_1}}\ldots\ad^B_{X_{j^\prime_{N-k}}}\left[\mathfrak{s}_p\right]\right)\dB
\left(\ad^B_{X_{j_1}}\ldots\ad^B_{X_{j_k}}\left[\mathfrak{s}_p^-\right]\right),
\end{equation}
or
\begin{equation}
\ad^B_{X_{1}}\ldots\ad^B_{X_{N}}\left[\mathfrak{s}_p^-\right]
=-\mathfrak{s}^-_p\dB\sum_{k=0}^{N-1}\sum_{j_1,\ldots,j_k}\left(\ad^B_{X_{j^\prime_1}}\ldots\ad^B_{X_{j^\prime_{N-k}}}
\left[\mathfrak{s}_p\right]\right)\dB\left(\ad^B_{X_{j_1}}\ldots\ad^B_{X_{j_k}}\left[\mathfrak{s}_p^-\right]\right).
\end{equation}
Taking $m=-p<0$ we obtain
\begin{equation}
\mathfrak{s}_m^-\dB\left(\ad^B_{X_{1}}\ldots\ad^B_{X_{N}}\left[\mathfrak{s}_p^-\right]\right)=
\mathfrak{s}_{-p}^-\dB\left(\ad^B_{X_{1}}\ldots\ad^B_{X_{N}}\left[\mathfrak{s}_p^-\right]\right)=
\end{equation}
$$
=-\mathfrak{s}_{p}\dB\mathfrak{s}^-_p\dB\sum_{k=0}^{N-1}\sum_{j_1,\ldots,j_k}\left(\ad^B_{X_{j^\prime_1}}
\ldots\ad^B_{X_{j^\prime_{N-k}}}\left[\mathfrak{s}_p\right]\right)
\dB\left(\ad^B_{X_{j_1}}\ldots\ad^B_{X_{j_k}}\left[\mathfrak{s}_p^-\right]\right)=
$$
$$
=-\sum_{k=0}^{N-1}\sum_{j_1,\ldots,j_k}\left(\ad^B_{X_{j^\prime_1}}\ldots\ad^B_{X_{j^\prime_{N-k}}}\left[\mathfrak{s}_p\right]\right)
\dB\mathfrak{s}_p^-\dB\mathfrak{s}_p\dB\left(\ad^B_{X_{j_1}}\ldots\ad^B_{X_{j_k}}\left[\mathfrak{s}_p^-\right]\right)=
$$
$$
=-\left\{\sum_{0\leq k\leq
(N-1)}\sum_{j_1,\ldots,j_k}\left(\ad^B_{X_{j^\prime_1}}\ldots\ad^B_{X_{j^\prime_{N-k}}}\left[\mathfrak{s}_p\right]\right)
\dB\mathfrak{s}_p^{-1}\dB(1-\mathfrak{r})^-\right\}\dB\left\{\mathfrak{s}_m^-\dB
\left(\ad^B_{X_{j_1}}\ldots\ad^B_{X_{j_k}}\left[\mathfrak{s}_p^-\right]\right)\right\}.
$$
Thus starting with the known relations $\mathfrak{s}_p^-\in\mathfrak{C}^B(\Xi)$ and
$(1-\mathfrak{r})^-\in\mathfrak{C}^B(\Xi)$ (proved in \cite{MPR2}) and
$$
\left(\ad^B_{X_{j^\prime_1}}\ldots\ad^B_{X_{j^\prime_{N-k}}}\left[\mathfrak{s}_p\right]\right)
\dB\mathfrak{s}_p^{-1}\in \mathfrak{C}^B(\Xi)
$$
(shown in (\ref{step4-2})), and using induction, we see that all the conditions
$$
\mathfrak{s}_m^{-}\sharp^B\left(\ad^B_{X_{1}}\ldots\ad^B_{X_{N}}\left[\mathfrak{s}_p^-\right]\right)\in\mathfrak{C}^B(\Xi)
$$
are satisfied and thus $\mathfrak{s}_p^-\in S^{-p}_{0}(\Xi)$.

{\sf Step 7.} We shall consider now the case $m>0$ and prove first the direct implication. Assume that $F\in
S^m_{0}(\Xi)$, which implies
$\ad^B_{X_{1}}\ldots\ad^B_{X_{N}}[F]\in S^m_{0}(\Xi)$. Step 6 implies $\mathfrak{s}_m^-\in
S^{-m}_{0}(\Xi)$. By the Theorem of magnetic composition of symbols in \cite{IMP} we get
$\mathfrak{s}_m^-\dB\left(\ad^B_{X_{1}}\ldots\ad^B_{X_{N}}[F]\right)\in S^0_{0}(\Xi)\subset\mathfrak{C}^B(\Xi)$.

{\sf Step 8.} For $m>0$ we prove now the inverse implication. For $F\in\mathcal{S}^\prime(\Xi)$ we show
\begin{equation}\label{step8}
\forall N\in\mathbb{N},\ \forall\{X_1,\ldots,X_N\}\subset\Xi,\
\mathfrak{s}_m^-\dB\left(\ad^B_{X_1}\cdot\ldots\cdot\ad^B_{X_N}[F]\right)\in\mathfrak{C}^B(\Xi)\
\Rightarrow \mathfrak{s}_m^-\dB F\in S^0_{0}(\Xi).
\end{equation}
Due to the hypothesis of (\ref{step8}) with $N=0$,
$\mathfrak{s}_m^-\dB F$ belongs to $\mathfrak{C}^B(\Xi)$ and we compute
\begin{equation}\label{step8-1}
\ad^B_{X_1}\ldots\ad^B_{X_M}\left[\mathfrak{s}_m^-\dB F\right]=
\end{equation}
$$
=\sum_{k=0}^M\sum_{j_1,\ldots,j_k}\left(\ad^B_{X_{j_1}}\ldots\ad^B_{X_{j_k}}\left[\mathfrak{s}_m^-\right]\right)
\dB\left(\ad^B_{X_{j^\prime_1}}\ldots\ad^B_{X_{j^\prime_{M-k}}}[F]\right)=
$$
$$
=\sum_{k=0}^{M}\sum_{j_1,\ldots,j_k}\left(\ad^B_{X_{j_1}}\ldots\ad^B_{X_{j_k}}\left[\mathfrak{s}_{m}^-\right]\right)
\dB\mathfrak{s}_m\dB\mathfrak{s}_m^-\dB\left(\ad^B_{X_{j^\prime_1}}\ldots\ad^B_{X_{j^\prime_{M-k}}}[F]\right),
$$
where $j_1<\dots<j_k$, $j'_1,\dots,j'_{M-k}$ and
$\{j_1,\ldots,j_k\}\bigcup\{j^\prime_1,\ldots,j^\prime_{M-k}\}=\{1,\ldots,M\}$.

\noindent
The factors
$\mathfrak{s}_m^-\dB\left(\ad^B_{X_{j^\prime_1}}\ldots\ad^B_{X_{j^\prime_{M-k}}}[F]\right)$
belong to $\mathfrak{C}^B(\Xi)$ by the hypothesis in (\ref{step8}).
Using the result of Step 6 we know that $\mathfrak{s}_{m}^-\in S^{-m}_{0}(\Xi)$ and thus we have
\begin{equation}\label{step8-3}
\left(\ad^B_{X_{j_1}}\ldots\ad^B_{X_{j_k}}\left[\mathfrak{s}_{m}^-\right]\right)\dB
\mathfrak{s}_{m}\in S^{-m}_{0}(\Xi)\dB S^{m}_{0}(\Xi)\subset S^0_{0}(\Xi)\subset\mathfrak{C}^B(\Xi).
\end{equation}
By the hypothesis in (\ref{step8}) we conclude that $\ad^B_{X_1}\ldots\ad^B_{X_M}\left[\mathfrak{s}_m^-\dB
F\right]\in\mathfrak{C}^B(\Xi)$ for any family of vectors of $\Xi$ and thus, by Theorem \ref{main}, we conclude that
$\mathfrak{s}_m^-\dB F$ belongs to $S^0_{0}(\Xi)$.\\
Then
$$
F\ =\ \mathfrak{s}_m\dB\mathfrak{s}_m^-\dB F\in S^{m}_{0}(\Xi)\dB S^0_{0}(\Xi)\subset S^{m}_{0}(\Xi).
$$
\end{proof}

\subsection{The class of symbols $\boldsymbol{S^m_{\rho}(\Xi)}$}

\begin{theorem}\label{main-m_rho}
A distribution $F\in\mathcal{S}^\prime(\Xi)$ is a symbol of type $S^m_{\rho}(\Xi)$ (with $0\leq\rho\leq1$)
if and only if for any $p,q\in\mathbb{N}$ and for any $u_1,\ldots,u_p\in\mathcal{X}$ and any
$\mu_1,\ldots,\mu_q\in\mathcal{X}^\prime$ the following is true:
\begin{equation}\label{cond-m_rho-0}
\mathfrak{s}_{m-q\rho}^-\dB\left(\ad^B_{u_1}\cdot\ldots\cdot\ad^B_{u_p}
\ad^B_{\mu_1}\cdot\ldots\cdot\ad^B_{\mu_q}[F]\right)\in\mathfrak{C}^B(\Xi).
\end{equation}
The two families of norms:
$$
\|\mathfrak{s}_{m-|\alpha|\rho}^-\partial_\xi^\alpha\partial_x^aF\|_\infty,
$$
indexed by $(a,\alpha)\in\mathbb{N}^{2n}$, and
$$
\|\mathfrak{s}_{m-q\rho}^-\dB\left(\ad^B_{u_1}\cdot\ldots\cdot\ad^B_{u_p}
\ad^B_{\mu_1}\cdot\ldots\cdot\ad^B_{\mu_q}[F]\right)\|_{\mathfrak C^B},
$$
indexed
by $(p,q)\in\mathbb{N}^2$ and sets of vectors in $\Xi$, define equivalent topologies on $S^m_\rho(\Xi)$.
\end{theorem}

\begin{proof}
Obviously $F\in S^m_{\rho}(\Xi)$ is equivalent to
\begin{equation}\label{cond-m_rho}
\partial_\xi^\alpha F\in S^{m-|\alpha|\rho}_{0}(\Xi),\quad\forall\alpha\in\mathbb{N}^n,
\end{equation}
that due to our Theorem \ref{main-m} is equivalent to
\begin{equation}\label{cond-m_rho-2}
\mathfrak{s}_{(m-|\alpha|\rho)}^-\dB\left[\ad^B_{X_1}\ldots\ad^B_{X_N}\partial_\xi^\alpha
F\right]\in \mathfrak{C}^B(\Xi),\quad\forall\alpha\in\mathbb{N}^n,\
\forall N\in\mathbb{N},\ \forall X_1,\ldots,X_N\in\Xi,
\end{equation}
i.e.
\begin{equation}\label{cond-m_rho-3}
\mathfrak{s}_{(m-|\alpha|\rho)}^-\dB\left[\ad^B_{X_1}\ldots\ad^B_{X_N}
(\mathfrak{ad}^B_{\epsilon_1})^{\alpha_1}\ldots(\mathfrak{ad}^B_{\epsilon_n})^{\alpha_n}
F\right]\in \mathfrak{C}^B(\Xi),\quad\forall\alpha\in\mathbb{N}^n,\
\forall N\in\mathbb{N},\ \forall X_1,\ldots,X_N\in\Xi.
\end{equation}
Choosing $|\alpha|=q$, $N=p$ and $X_j=u_j$ for $j\in\{1,\ldots,p\}$, we see that (\ref{cond-m_rho-3}) implies
the condition (\ref{cond-m_rho-0}) in the Theorem and thus the direct implication is proved.\\
Now suppose that (\ref{cond-m_rho-0}) is true for any $p,q\in\mathbb{N}$ and for any family of $p$ vectors
$\{u_1,\ldots,u_p\}\subset\mathcal{X}$ and any family of $q$ vectors $\{\mu_1,\ldots,\mu_q\}\subset\mathcal{X}^\prime$.
Let us take $N\in\mathbb{N}$, $\{X_1,\ldots,X_N\}\subset\Xi$ and
$\alpha\in\mathbb{N}^n$. Due to the commutation relations (\ref{lie}), we can rearrange the operator
$\ad^B_{X_1}\ldots\ad^B_{X_N}$ as a sum of operators $\ad^B_{u_1}\cdot\ldots\cdot\ad^B_{u_p}
\ad^B_{\mu_1}\cdot\ldots\cdot\ad^B_{\mu_s}$ with $p+s=N$. Then (\ref{cond-m_rho-0}) implies that
\begin{equation}\label{cond-m_rho-4}
\mathfrak{s}_{[m-(s+|\alpha|)\rho]}^-\dB\left(\ad^B_{u_1}\cdot\ldots\cdot\ad^B_{u_p}
\ad^B_{\mu_1}\cdot\ldots\cdot\ad^B_{\mu_s}
(\mathfrak{ad}^B_{\epsilon_1})^{\alpha_1}\ldots(\mathfrak{ad}^B_{\epsilon_n})^{\alpha_n}
[F]\right)\in \mathfrak{C}^B(\Xi).
\end{equation}
Thus, we conclude that for any $\alpha\in\mathbb{N}^n$, $\forall
N\in\mathbb{N}$ and $\forall\ X_1,\ldots,X_N \in\Xi$:
$$
\mathfrak{s}_{m-|\alpha|\rho}^-\dB\left(\ad^B_{X_1}\ldots\ad^B_{X_N}
(\mathfrak{ad}^B_{\epsilon_1})^{\alpha_1}\ldots(\mathfrak{ad}^B_{\epsilon_n})^{\alpha_n}
[F]\right)=
$$
$$
=\mathfrak{s}_{m-|\alpha|\rho}^-\dB\mathfrak{s}_{m-(s+|\alpha|)\rho}
\dB\mathfrak{s}_{m-(s+|\alpha|)\rho}^-\dB\left(\ad^B_{X_1}\ldots\ad^B_{X_N}
(\mathfrak{ad}^B_{\epsilon_1})^{\alpha_1}\ldots(\mathfrak{ad}^B_{\epsilon_n})^{\alpha_n}
[F]\right)\in\mathfrak{C}^B(\Xi)
$$
due to relation (\ref{cond-m_rho-4}) and using
the result of Step 6 of the Proof of Theorem \ref{main-m}
$$
\mathfrak{s}_{m-|\alpha|\rho}^-\dB\mathfrak{s}_{m-(s+|\alpha|)\rho}\in S^{-(m-|\alpha|\rho)}_{0}(\Xi)\dB
S^{m-(s+|\alpha|)\rho}_{0}(\Xi)\subset S^{-s\rho}_{0}(\Xi)\subset S^0_{0}(\Xi)\subset\mathfrak{C}^B(\Xi).
$$
Thus (\ref{cond-m_rho}) implies (\ref{cond-m_rho-3}) and we get also the inverse implication.
\end{proof}
Since $F\in S^m_\rho(\Xi)$ if and only if $F^*\in S^m_\rho(\Xi)$, by taking the adjoints we prove

\medskip
\begin{corollary}
A distribution $F\in\mathcal{S}^\prime(\Xi)$ is a symbol of type $S^m_{\rho}(\Xi)$ (with $0\leq\rho\leq1$)
if and only if for any $p$ and $q$ in $\mathbb{N}$ and for any family of $p$ vectors
$\{u_1,\ldots,u_p\}\subset\mathcal{X}$ and any family of $q$
vectors $\{\mu_1,\ldots,\mu_q\}\subset\mathcal{X}^\prime$ the following is true:
\begin{equation}\label{cond-m_rho-5}
\left(\ad^B_{u_1}\cdot\ldots\cdot\ad^B_{u_p}
\ad^B_{\mu_1}\cdot\ldots\cdot\ad^B_{\mu_q}[F]\right)\dB\mathfrak{s}_{m-q\rho}^-\in\mathfrak{C}^B(\Xi).
\end{equation}
\end{corollary}

\subsection{The Bony criterion}

Following the work of J-M. Bony \cite{Bo1} we shall reformulate our main theorem by replacing the commutators with linear
distributions (of the type $\mathfrak{l}_X$) by symbols of class $S^+_\rho(\Xi)$.
\begin{definition}
 Let $\rho\in[0,1]$; we define the class of symbols $S^+_\rho(\Xi)$ as
$$
S^+_\rho(\Xi):=\left\{\varphi\in C^\infty(\Xi)\ \mid\ \left|\left(\partial^a_x\partial^\alpha_\xi\varphi\right)(X)\right|\leq C_{a\alpha}<\xi>^{\rho(1-|\alpha|)},\ \text{for}\ |a|+|\alpha|\geq1\right\}.
$$
\end{definition}

For any $\varphi\in S^+_\rho(\Xi)\subset\mathfrak{M}^B(\Xi)$ we can define the derivation
\begin{equation}
\mathfrak{ad}^B_\varphi[F]:=\varphi\dB F-F\dB\varphi,\qquad\forall
F\in\mathfrak{M}^B(\Xi).
\end{equation}

\begin{theorem}\label{Bony}
A distribution $F\in\mathcal{S}^\prime(\Xi)$ is a symbol of type
$S^m_{\rho}(\Xi)$ (with $0\leq\rho\leq1$) if and only if for any $N\in\mathbb{N}$ and any family of $N$ symbols
$\{\varphi_1,\ldots,\varphi_N\}\subset S^+_\rho(\Xi)$ the following is true:
\begin{equation}\label{cond-Bony}
\mathfrak{s}_m^-\dB\mathfrak{ad}^B_{\varphi_1}\ldots\mathfrak{ad}^B_{\varphi_N}[F]\in\mathfrak{C}^B(\Xi).
\end{equation}
\end{theorem}

\begin{proof}

\noindent
{\it Step 1.}
First let us consider that $F\in S^m_{\rho}(\Xi)$ and let us compute its commutator with symbols $\varphi\in S^+_\rho(\Xi)$.
For any distribution $F\in\mathcal{S}^\prime(\Xi)$ we shall introduce the
notations
$$
\nabla_{_{\!\tiny{\mathcal{X}}}}F:=\partial_x F,\quad
\nabla_{_{\!\tiny{\mathcal{X^\prime}}}}F:=\partial_\xi F,
$$
$$
(F)_{X,s}(Y):=F(X+s(Y-X)),\qquad\text{for }s\in\mathbb{R},
$$
and remark that for any $\varphi\in S^+_\rho(\Xi)$ we have that $\nabla_{_{\!\tiny{\mathcal{X}}}}\varphi\in S^\rho_\rho(\Xi)$
and $\nabla_{_{\!\tiny{\mathcal{X^\prime}}}}\varphi\in S^0_\rho(\Xi)$.
Then
$$
\mathfrak{ad}^B_\varphi[F](X)=(\varphi\dB F)(X)-(F\dB\varphi)(X)=
$$
$$
=\pi^{-2n}\int_\Xi\int_\Xi dY\,dZ\,e^{-2i\sigma(X-Y,X-Z)}\Omega^B[\mathcal{T}(x,y,z)]
\left\{\varphi(Y)F(Z)-\varphi(Z)F(Y)\right\}=
$$
$$
=\pi^{-2n}\varphi(X)\int_\Xi\int_\Xi
dY\,dZ\,e^{-2i\sigma(X-Y,X-Z)}\Omega^B[\mathcal{T}(x,y,z)]\left\{F(Z)-F(Y)\right\}-
$$
$$
-\pi^{-2n}\int_0^1ds\,\int_\Xi\int_\Xi
dY\,dZ\,e^{-2i\sigma(X-Y,X-Z)}\Omega^B[\mathcal{T}(x,y,z)]
\left\{[(X-Y)\cdot(\nabla\varphi)((1-s)X+sY)]F(Z)\right\}+
$$
$$
+\pi^{-2n}\int_0^1ds\,\int_\Xi\int_\Xi
dY\,dZ\,e^{-2i\sigma(X-Y,X-Z)}\Omega^B[\mathcal{T}(x,y,z)]
\left\{[(X-Z)\cdot(\nabla\varphi)((1-s)X+sZ)]F(Y)\right\}.
$$
The first term is evidently vanishing and we obtain by usual integration by parts techniques that
\begin{equation}\label{com-phi-F}
\mathfrak{ad}^B_\varphi[F](X)=-\frac{\pi^{-2n}}{2i}\sum_{j=1}^n\int_0^1ds\,\int_\Xi\int_\Xi
dY\,dZ\,e^{-2i\sigma(X-Y,X-Z)}\Omega^B[\mathcal{T}(x,y,z)]\times
\end{equation}
$$
\times\left\{[(\nabla_{_{\!\tiny{\mathcal{X}}}}\varphi)_{X,s}(Y)]_j[(\partial_{\zeta_j}
F)(Z)]+
[(\nabla_{_{\!\tiny{\mathcal{X}}}}\varphi)_{X,s}(Z)]_j[(\partial_{\eta_j}
F)(Y)] -\right.
$$
$$
-[(\nabla_{_{\!\tiny{\mathcal{X^\prime}}}}\varphi)_{X,s}(Y)]_j\left[(\partial_{z_j}
F)(Z)-i\left(\partial_{z_j}\Gamma^B[\mathcal{T}(x,y,z)]\right)F(Z) \right]-
$$
$$
\left.-[(\nabla_{_{\!\tiny{\mathcal{X^\prime}}}}\varphi)_{X,s}(Z)]_j\left[(\partial_{y_j}
F)(Y)-i\left(\partial_{y_j}\Gamma^B[\mathcal{T}(x,y,z)]\right)F(Y) \right] \right\}.
$$
Observing that the functions $\Omega^B[\mathcal{T}(x,y,z)]$ and $\Gamma^B[\mathcal{T}(x,y,z)]$ belong to $BC^\infty(\mathcal{X};C^\infty_{\sf pol}(\mathcal{X}\times\mathcal{X}))$, due to our hypothesis on the magnetic field $B$ and using the results of Appendix \ref{comp-symb} we conclude that $\mathfrak{ad}^B_\varphi[F]$ is a symbol of type $S^m_{\rho}(\Xi)$
and thus, using also the results of the last section, $\mathfrak{s}_m^-\dB\mathfrak{ad}^B_\varphi[F]$ belongs to
$\mathfrak{C}^B(\Xi)$. Replacing now $F$ with $\mathfrak{ad}^B_\varphi[F]$ we may iterate the above argument and
obtain the condition (\ref{cond-Bony}) of the Theorem.

\medskip
\noindent
{\it Step 2.}
Let us prove now the inverse implication. Thus let us consider a tempered distribution $F$ on $\Xi$ that satisfies
(\ref{cond-Bony}) in the statement of the Theorem for some $m\in\mathbb{R}$ and $\rho\in[0,1]$. For any vector
$u\in\mathcal{X}$ and any vector $\mu\in\mathcal{X}^\prime$ we observe that:
\begin{eqnarray}
&&\partial_{\xi_j}\mathfrak{l}_u=\partial_{\xi_j}[\sigma((u,0),(x,\xi))]=-u_j,\\
&&\partial_{x_j}\mathfrak{l}_u=\partial_{x_j}[\sigma((u,0),(x,\xi))]=0,\\
&&\partial_{\xi_j}\mathfrak{l}_\mu=\partial_{\xi_j}[\sigma((0,\mu),(x,\xi))]=0,\\
&&\partial_{x_j}\mathfrak{l}_\mu=\partial_{x_j}[\sigma((0,\mu),(x,\xi))]=\mu_j.
\end{eqnarray}
Thus, for any $u\in\mathcal{X}$ and any
$\mu\in\mathcal{X}^\prime$ the distributions $\mathfrak{l}_u$ and $\mathfrak{l}_\mu$ belong to $S^+_\rho(\Xi)$ (for any $\rho\in[0,1]$). Using (\ref{cond-Bony}), that we suppose to hold, we deduce that
\begin{equation}\label{concl-1}
\mathfrak{s}_{m}^-\dB\left(\ad^B_{u_1}\ldots\ad^B_{u_p}
\ad^B_{\mu_1}\ldots\ad^B_{\mu_q}\left[
\mathfrak{ad}^B_{\varphi_1}\ldots\mathfrak{ad}^B_{\varphi_N}[F]\right]\right)\in\mathfrak{C}^B(\Xi).
\end{equation}

\noindent
{\it Conclusion 1.}
Using our Theorem \ref{main-m} we conclude that $\mathfrak{ad}^B_{\varphi_1}\ldots\mathfrak{ad}^B_{\varphi_N}[F]$ is a
symbol of class $S^m_0(\Xi)\subset\mathfrak{M}^B[\Xi]$ for any $N\in\mathbb{N}$ and any family of $S^+_\rho(\Xi)$-symbols.

Now, for $0<\rho\leq1$ we shall verify the hypothesis (\ref{cond-m_rho-0}) of Theorem \ref{main-m_rho}. The idea is to
use a special symbol of the type $\mathfrak{f}_{\mu,\rho}:=\mathfrak{l}_\mu\sharp^B\mathfrak{s}_\rho$.
Unfortunately, this is not an element of $S^+_\rho(\Xi)$ (for
$\rho>0$) so that we shall need a localization procedure in order
to control the dependence on $x\in\mathcal{X}$.
We follow the procedure elaborated by J-M. Bony and J-Y. Chemin \cite{BC},
but dealing only with a specific class of metrics we shall avoid the use of the general {\it confinement norms} and prove some confinement results in Appendix \ref{confinement}.
We shall use repeatedly the following observations
\begin{remark}\label{produse}
\begin{enumerate}
\item For any two $C^\infty(\mathcal{X})$ functions $f$ and $g$, considering them as functions on $\Xi$ constant with respect to $\xi\in\mathcal{X}^\prime$, we have $f\sharp^Bg=f\sharp g=f\cdot g$.
\item For any $C^\infty(\mathcal{X})$ function $f$, considering it as function on $\Xi$ constant with respect to $\xi\in\mathcal{X}^\prime$, and any symbol $F$ we have $f\sharp^BF=f\sharp F$ and $F\sharp^B f=F\sharp f$.
\item For $\mathfrak{f}\in C^\infty(\mathcal{X}^\prime)\cap S^s_\rho(\Xi)$ and $g\in C^\infty(\mathcal{X})\cap S^+_0(\Xi)$ we have that $\mathfrak{f}\sharp^Bg=\mathfrak{f}\cdot g+\gamma(\mathfrak{f},g)$, where
$\gamma(\mathfrak{f},g)\in S^{s-\rho}_0(\Xi)$.
\item For $F\in S^s_0(\Xi)$ and $\phi\in S^0_0(\Xi)$ or $\mathfrak{g}\in C^\infty(\mathcal{X}^\prime)\cap S^\rho_\rho(\Xi)$, we have that
$$
\mathfrak{ad}^B_\phi[F]\in S^s_0(\Xi),\qquad \mathfrak{ad}^B_{\mathfrak{g}}[F]\in S^s_0(\Xi).
$$
\end{enumerate}
\end{remark}

\noindent
While the first point of the above remark is evident, the second follows easily by integration by parts, and the last two can be easily proved by a very similar procedure to the proof of (\ref{com-phi-F}) and using once again the results of Appendix \ref{comp-symb}.

Suppose chosen a positive function $\chi\in C^\infty_0(\mathcal{X})$ such that $\int_{\mathcal{X}}dx\,\chi(x)=1$ and $\textsf{supp}\chi\subset B_R(0)$. Then we have
$$
\mathfrak T^B_{(x,0)}[\chi]=\tau_x[\chi]=:\chi_x\in C^\infty_0(\mathcal{X}),\qquad\forall x\in\mathcal{X}
$$
defining a continuous map $\mathcal{X}\ni x\mapsto\tau_x[\chi]\in C^\infty_0(\mathcal{X})$, so that we can define the following integrals in the weak sense as elements of $\mathcal{S}^\prime(\Xi)$ (the translations acting continuously on the Fr\'{e}chet space $\mathcal{S}(\mathcal{X})$), and we have in the sense of distributions
$$
\int_{\mathcal{X}}dx\,\mathfrak T^B_{(x,0)}[\chi]=\int_{\mathcal{X}}dx\,\tau_{x}[\chi]=1.
$$
Thus, due to the continuity of the magnetic Moyal product as map $\mathcal{S}^\prime(\Xi)\times\mathcal{S}^\prime(\Xi)\rightarrow\mathcal{S}^\prime(\Xi)$, we can write for any $G\in\mathfrak{M}^B(\Xi)$, in the sense of distributions:
$$
G=\int_{\mathcal{X}}dxG_x:=\int_{\mathcal{X}}dx\left[\int_{\mathcal{X}}dy\,\tau_{x}[\chi]\sharp G\sharp\tau_{y}[\chi]\right].
$$
Remark that for any $F\in\mathfrak{M}^B(\Xi)$ we have (for any $\phi\in\mathcal{S}(\mathcal{X})$)
$$
\left<F\dB G,\phi\right>=\left<G,\phi\dB F\right>=\left<\int_{\mathcal{X}}dxG_x,\phi\dB F\right>=\left<\int_{\mathcal{X}}dxF\dB G_x,\phi\right>.
$$
For any fixed $x\in\mathcal{X}$ we can write
$$
\int_{\mathcal{X}}dy\,\tau_{x}[\chi]\sharp G\sharp\tau_{y}[\chi]=\int_{B_{2R}(x)}dy\,\left\{\tau_{x}[\chi]\sharp
G\sharp\tau_{y}[\chi]\right\}+\int_{B^c_{2R}(x)}dy\,\left\{\tau_{x}[\chi]\sharp
G\sharp\tau_{y}[\chi]\right\}
$$
and we introduce the notations
$$
\dot{G}_x:=\int_{B_{2R}(x)}dy\,\left\{\tau_{x}[\chi]\sharp
G\sharp\tau_{y}[\chi]\right\},\qquad
\tilde{G}_x:=\int_{B^c_{2R}(x)}dy\,\left\{\tau_{x}[\chi]\sharp
G\sharp\tau_{y}[\chi]\right\}.
$$
Let us also denote by
$$
\theta_x:=\int_{B_{2R}(x)}dy\,\tau_{y}[\chi]=\tau_{x}\left[\int_{B_{2R}(0)}dy\,\tau_{y}[\chi]
\right]=:\tau_x[\theta]\in C^\infty_0(\mathcal{X}),
$$
and observe that $\theta(z)=1$ for $z\in B_{R}(0)$ so that
$$
\theta\chi=\chi, \qquad\dot{G}_x=\chi_x\sharp
G\sharp\theta_x
$$
$$
\tilde{G}_x=\int_{|x-y|\geq2R}\left[<x-y>^N\left(\tau_{x}[\chi]\sharp
G\sharp\tau_{y}[\chi]\right) \right]<x-y>^{-N}dy
$$

First let us consider $G=\mathfrak{ad}^B_{\varphi_1}\ldots\mathfrak{ad}^B_{\varphi_J}[F]$.

Using Conclusion 1 above for our $F$, and the Theorem for the magnetic composition of symbols, it is easy to see that
$\dot{G}_x\in S^m_0(\Xi)$ uniformly with respect to $x\in\mathcal{X}$. In fact we shall use several times the observation that the symbol type norms (on any $S^m_\rho(\Xi)$) are left invariant by translations with $\mathcal{X}$-variables. 

For $\tilde{G}_x$ we use Lemma \ref{confinare} in Appendix \ref{confinement} to prove that the integral is finite and defines an element in $S^m_0(\Xi)$ uniformly with respect to $x\in\mathcal{X}$. 

Let us prove that
$$
\mathfrak{s}^-_{m-\rho}\sharp^B\mathfrak{ad}^B_\mu[G]\in \mathfrak{C}^B(\Xi).
$$
In fact, we can write
$$
\mathfrak{s}^-_{m-\rho}\sharp^B\mathfrak{ad}^B_\mu[G]=\left[\mathfrak{s}^-_{m-\rho}\sharp^B\mathfrak{s}^-_\rho\sharp^B\mathfrak{s}_m\right]\sharp^B\mathfrak{s}^-_{m}\sharp^B\mathfrak{s}_{\rho}\sharp^B\mathfrak{ad}^B_\mu[G].
$$
and due to the fact that (by the magnetic composition of symbols) $\mathfrak{s}^-_{m-\rho}\sharp^B\mathfrak{s}^-_\rho\sharp^B\mathfrak{s}_m\in S^0_0(\Xi)$ we only have to estimate
\begin{equation}\label{localization-F}
\mathfrak{s}^-_{m}\sharp^B\mathfrak{s}_{\rho}\sharp^B\mathfrak{ad}^B_\mu[G]=\int_{\mathcal{X}}dx\left\{\mathfrak{s}^-_{m}\sharp^B\mathfrak{s}_{\rho}\sharp^B\mathfrak{ad}^B_\mu\left[\dot{G}_x\right]+\mathfrak{s}^-_{m}\sharp^B\mathfrak{s}_{\rho}\sharp^B\mathfrak{ad}^B_\mu\left[\tilde{G}_x\right] \right\}.
\end{equation}
In order to control the integral appearing in (\ref{localization-F}) we shall apply the usual Cotlar-Stein argument together with the confinement result in Lemma \ref{CS-confinare} of Appendix \ref{confinement}. In
fact we shall use the following form of the 'integral Cotlar-Stein Lemma' proved in \cite{BL} (Lemma 4.2.3'):
\begin{proposition}\label{Cotlar-Stein}
 Let $\{A_x\}_{x\in\mathcal{X}}$ be a family of bounded operators on a Hilbert space $\mathcal{H}$
 such that the following two estimations are satisfied:
\begin{itemize}
 \item $\underset{y\in\mathcal{X}}{\sup}\left(\int_{\mathcal{X}}dz\|A_y^*A_z\|^{1/2}_{\mathbb{B}
 (\mathcal{H})}\right)\leq M$,
 \item $\underset{y\in\mathcal{X}}{\sup}\left(\int_{\mathcal{X}}dz\|A_yA_z^*\|^{1/2}_{\mathbb{B}
 (\mathcal{H})}\right)\leq M$.
\end{itemize}
Then $\int_{\mathcal{X}}dx\,A_x$ converges for the strong topology and we have
$$
\left\|\int_{\mathcal{X}}dx\,A_x\right\|\leq M.
$$
\end{proposition}
\noindent
We have to verify that all the terms appearing in \ref{localization-F} verify the conditions of Lemma \ref{CS-confinare} in Appendix \ref{confinement}.

Let us consider the first contribution to the integral in \ref{localization-F} and remark that
$$
\mathfrak{ad}^B_\mu[\dot{G}_x]\equiv\mathfrak{ad}^B_{\mathfrak{l}_\mu}[\dot{G}_x]=\mathfrak{ad}^B_{\mathfrak{l}_{\mu,x}}[\dot{G}_x]
$$
for any $x\in\mathcal{X}$, where $\mathfrak{l}_{\mu,x}:=\tau_x[\mathfrak{l}_\mu]$, i.e. $\mathfrak{l}_{\mu,x}(Z)=<\mu,z-x>$ with $<\mu,x>$ a constant term (with respect to the variable $z$). Moreover, if we fix a $C^\infty_0(\mathcal{X})$ function $\psi$ such that $\psi\cdot\theta=\theta$ (then we also have $\psi\cdot\chi=\chi$) and denote by $\psi_x:=\tau_x\psi$ and by $\psi_{\mu,x}:=\mathfrak{l}_{\mu,x}\psi_x$; we get (remark that $\psi^2_x\chi_x=\psi_x(\psi_x\chi_x)=\psi_x\chi_x=\chi_x$)
$$
\mathfrak{s}_{\rho}\sharp^B\mathfrak{ad}^B_\mu\left[\dot{G}_x\right]=\mathfrak{s}_{\rho}\sharp^B\mathfrak{ad}^B_{\mathfrak{l}_{\mu,x}}\left[\dot{G}_x\right]=
$$
$$
=\mathfrak{s}_{\rho}\sharp^B\left[\mathfrak{l}_{\mu,x}\sharp^B\dot{G}_x-\dot{G}_x\sharp^B\mathfrak{l}_{\mu,x}\right]=
\mathfrak{s}_{\rho}\sharp^B\left[\mathfrak{l}_{\mu,x}\sharp^B\left(\chi_x\sharp G\sharp\theta_x\right)-\left(\chi_x\sharp G\sharp\theta_x\right)\sharp^B\mathfrak{l}_{\mu,x}\right]=
$$
$$
=
\mathfrak{s}_{\rho}\sharp^B\left[\mathfrak{l}_{\mu,x}\sharp^B\left(\chi_x\sharp^B G\sharp^B\theta_x\right)-\left(\chi_x\sharp^B G\sharp^B\theta_x\right)\sharp^B\mathfrak{l}_{\mu,x}\right]=
$$
$$
=\mathfrak{s}_{\rho}\sharp^B\left[\left(\mathfrak{l}_{\mu,x}\sharp^B\chi_x\right)\sharp^B G\sharp^B\theta_x\sharp^B\psi_x-\psi_x\sharp^B\chi_x\sharp^B G\sharp^B\left(\theta_x\sharp^B\mathfrak{l}_{\mu,x}\right)\right]=
$$
$$
=\mathfrak{s}_{\rho}\sharp^B\psi_x\sharp^B\left[\left(\psi_{\mu,x}\sharp^B\chi_x\right)\sharp^B G\sharp^B\theta_x-\chi_x\sharp^B G\sharp^B\left(\theta_x\sharp^B\psi_{\mu,x}\right)\right]\sharp^B\psi_x=
$$
\begin{equation}\label{Fdot-tot}
=\mathfrak{s}_{\rho}\sharp^B\psi_x\sharp^B\mathfrak{ad}^B_{\psi_{\mu,x}}\left[\dot{G}_x\right]\sharp^B\psi_x=\psi_x\sharp^B\mathfrak{s}_{\rho}\sharp^B\mathfrak{ad}^B_{\psi_{\mu,x}}\left[\dot{G}_x\right]\sharp^B\psi_x+\mathfrak{ad}^B_{\mathfrak{s}_\rho}[\psi_x]\sharp^B\mathfrak{ad}^B_{\psi_{\mu,x}}\left[\dot{G}_x\right]\sharp^B\psi_x
\end{equation}
and (see Remark \ref{produse})
\begin{equation}\label{Fdot}
\mathfrak{s}_{\rho}\sharp^B\mathfrak{ad}^B_{\psi_{\mu,x}}\left[\dot{G}_x\right]=\left(\mathfrak{ad}^B_{\mathfrak{s}_\rho\cdot\psi_{\mu,x}}\left[\dot{G}_x\right]\right)+ \left(\mathfrak{ad}^B_{\gamma(\mathfrak{s}_\rho,\psi_{\mu,x})}\left[\dot{G}_x\right] \right)- \left(\mathfrak{ad}^B_{\mathfrak{s}_\rho}\left[\dot{G}_x\right]\sharp^B\psi_{\mu,x} \right).
\end{equation}
First we remark that $\psi_x=\mathfrak{T}^B_{(x,0)}[\psi]$ and
$$
\mathfrak{ad}^B_{\mathfrak{s}_\rho}[\psi_x]=\mathfrak{ad}^B_{\mathfrak{s}_\rho}\left[\mathfrak{T}^B_{(x,0)}[\psi]\right]=\mathfrak{T}^B_{(x,0)}\left[\mathfrak{ad}^B_{\mathfrak{s}_\rho}[\psi]\right]
$$
define translations of symbols of class $S^0_1(\Xi)$ with rapid decay in the $\mathcal{X}$ variable. Then, the family $\left\{\mathfrak{ad}^B_{\psi_{\mu,x}}\left[\dot{G}_x\right]\right\}_{_{x\in\mathcal{X}}}$ defines a family of symbols in $S^m_0(\Xi)$ uniformly with respect to $x\in\mathcal{X}$. Observing that
$$
\big(\partial_{y_j}(\mathfrak{s}_\rho\cdot\psi_{\mu,x})\big)(Y)=(<\eta>^\rho+\lambda)\big[\mu_j\psi(y-x)+<\mu,(y-x)>(\partial_{y_j}\psi)(y-x)\big],
$$
$$
\big(\partial_{\eta_j}(\mathfrak{s}_\rho\cdot\psi_{\mu,x})\big)(Y)=\rho\frac{\eta_j}{<\eta>}<\eta>^{\rho-1}<\mu,(y-x)>\psi(y-x),
$$
we deduce that $\mathfrak{s}_\rho\cdot\psi_{\mu,x}\in S^+_\rho(\Xi)$ uniformly for $x\in\mathcal{X}$ and due to our Conclusion 1, the first term in \ref{Fdot} is uniformly bounded in $S^m_0(\Xi)$. Using the above Remark \ref{produse} for $\gamma(\mathfrak{s}_\rho,\psi_{\mu,x})$ we conclude that $\mathfrak{ad}^B_{\gamma(\mathfrak{s}_\rho,\psi_{\mu,x})}\left[\dot{G}_x\right]\in S^m_0(\Xi)$ uniformly with respect to $x\in\mathcal{X}$.
Now for the last term in \ref{Fdot} we use once again the Remark \ref{produse}. We conclude thus that we can apply Lemma \ref{CS-confinare} in Appendix \ref{confinement} and thus Proposition \ref {Cotlar-Stein} above.

Let us study now the second contribution to the integral in \ref{localization-F}:
\begin{equation}\label{Ftilde-1}
\mathfrak{s}_{\rho}\sharp^B\mathfrak{ad}^B_\mu\left[\tilde{G}_x\right]=\mathfrak{ad}^B_\mu\left[\mathfrak{s}_{\rho}\sharp^B\tilde{G}_x\right]-\mathfrak{ad}^B_\mu\left[\mathfrak{s}_{\rho}\right]\sharp^B \tilde{G}_x
\end{equation}
and the second term defines clearly an element of $S^m_0(\Xi)$ uniformly in $x\in\mathcal{X}$. For the first term,denoting by $\theta^c:=1-\theta$ we remark that $\chi_x\cdot\theta^c_x=0$. Let us fix a function $\phi\in C^\infty_0(\mathcal{X})$ such that $\phi\cdot\chi=\chi$ and $\phi\cdot\theta^c=0$. Then
$$
\mathfrak{s}_{\rho}\sharp^B\tilde{G}_x=(\mathfrak{s}_{\rho}\cdot\phi^2_x)\sharp^B\chi_x\sharp^BG\sharp^B\theta^c_x+\gamma(\mathfrak{s}_{\rho},\phi^2_x)\sharp^B\tilde{G}_x-\chi_x\sharp^BG\sharp^B(\mathfrak{s}_{\rho}\cdot\phi_x\cdot\theta^c_x)=
$$
$$
=(\phi_x\cdot\mathfrak{s}_{\rho}\cdot\phi_x)\sharp^B\tilde{G}_x-\tilde{G}_x\sharp^B(\mathfrak{s}_{\rho}\cdot\phi_x)+\gamma(\mathfrak{s}_{\rho},\phi^2_x)\sharp^B\tilde{G}_x+\chi_x\sharp^BG\sharp^B\gamma(\theta^c_x,\mathfrak{s}_{\rho}\cdot\phi_x)=
$$
\begin{equation}\label{Ftilde-2}
=\phi_x\sharp^B\mathfrak{ad}^B_{\mathfrak{s}_{\rho}\cdot\phi_x}[\tilde{G}_x]+\left[\gamma(\mathfrak{s}_{\rho},\phi^2_x)-\gamma(\phi_x,\mathfrak{s}_{\rho}\cdot\phi_x)\right]\sharp^B\tilde{G}_x+\chi_x\sharp^BG\sharp^B\gamma(\theta^c_x,\mathfrak{s}_{\rho}\cdot\phi_x)
\end{equation}
where $\mathfrak{s}_{\rho}\cdot\phi_x\in S^+_\rho(\Xi)$ uniformly for $x\in\mathcal{X}$. It is easy to verify that $\gamma^B(\theta^c_x,\mathfrak{s}_{\rho}\cdot\phi_x)$ is a symbol of class $S^0_1(\Xi)$ with rapid decay in the $\mathcal{X}$ variable and uniformly with respect to $x\in\mathcal{X}$ (just repeat the arguments of the proof of \ref{com-phi-F} snd use once again the results of Appendix \ref{comp-symb}). Thus the last term can once again be treated by applying Lemma \ref{CS-confinare} in Appendix \ref{confinement} (with $G_x=G$ constant). For the first two terms above we use Lemma \ref{CS-s-confinare} in Appendix \ref{confinement} and thus end the proof.

\noindent
{\it Conclusion 2.}
Thus we proved that for $F$ verifying the condition (\ref{cond-Bony}) of the Theorem, for any $N\in\mathbb{N}$, $\{\varphi_1,\ldots,\varphi_N\}\subset S^+_\rho(\Xi)$, denoting $G:=\mathfrak{ad}^B_{\varphi_1}\cdots\mathfrak{ad}^B_{\varphi_N}[F]$ we have for any $\mu\in\mathcal{X}^\prime$
$$
\mathfrak{s}^-_{m-\rho}\sharp^B\mathfrak{ad}^B_\mu[G]\in \mathfrak{C}^B(\Xi).
$$

Taking any family $\{X_1,\ldots,X_M\}\subset\Xi$ and taking into account the commutation of the $\mathfrak{ad}^B_X$ operators (\ref{lie}) we obtain that
$$
\mathfrak{s}^-_{m-\rho}\sharp^B\mathfrak{ad}^B_{X_1}\cdots\mathfrak{ad}^B_{X_M}[\mathfrak{ad}^B_\mu G]=\mathfrak{s}^-_{m-\rho}\sharp^B\mathfrak{ad}^B_\mu\left[\mathfrak{ad}^B_{X_1}\cdots\mathfrak{ad}^B_{X_M}[G] \right]=\mathfrak{s}^-_{m-\rho}\sharp^B\mathfrak{ad}^B_\mu[\widetilde{G}]
$$
with
$$
\widetilde{G}:=\mathfrak{ad}^B_{X_1}\cdots\mathfrak{ad}^B_{X_M}\cdot\mathfrak{ad}^B_{\varphi_1}\cdots\mathfrak{ad}^B_{\varphi_N}[F]
$$
to which we apply once again Conclusion 1 and Conclusion 2. Finally we use Theorem \ref{main-m} to conclude that $\mathfrak{ad}^B_\mu\left[\mathfrak{ad}^B_{\varphi_1}\ldots\mathfrak{ad}^B_{\varphi_J}[F]\right]\in S^{m-\rho}_0(\Xi)$.

Taking now $G=\mathfrak{ad}^B_\mu\left[\mathfrak{ad}^B_{\varphi_1}\ldots\mathfrak{ad}^B_{\varphi_J}[F]\right]\in S^{m-\rho}_0(\Xi)$ and repeating exactly the above procedure we shall obtain that
$$
\mathfrak{s}^-_{m-2\rho}\sharp^B\mathfrak{ad}^B_{\mu_1}\left[\mathfrak{ad}^B_\mu\left[\mathfrak{ad}^B_{\varphi_1}\ldots\mathfrak{ad}^B_{\varphi_J}[F]\right]\right]\in \mathfrak{C}^B(\Xi).
$$
Iterating these arguments (and taking into account the commutation properties of the $\mathfrak{ad}^B_X$ operators) one clearly finishes the proof.
\end{proof}

\section{Applications}

\subsection{Inversion}

\begin{proposition}\label{th}
If $F\in S^0_\rho(\Xi)$ is invertible in the $C^*$-algebra $\mathfrak{C}^B(\Xi)$,
then the inverse $F^-$ also belongs to $S^0_\rho(\Xi)$.
\end{proposition}

\begin{proof}
By Theorem \ref{Bony}, we need to show that for arbitrary $N\in\mathbb{N}\smallsetminus{0}$ and any family
$\varphi_1,\dots,\varphi_N$ in $S^+_\rho(\Xi)$ we have $\ad^B_{\varphi_1}\dots\ad^B_{\varphi_N}[F^-]\in\mathfrak C^B(\Xi)$.

For any subset $K:=\{k_1,\dots,k_m\}$ of the ordered set $J:=\{1,\dots,N\}$ we write
$\mathfrak{D}^B_K:=\ad^B_{\varphi_{k_1}}\dots\ad^B_{\varphi_{k_m}}$.

\noindent
It is known and easy to prove by induction that
\begin{equation}\label{mult-ad-inv}
\mathfrak{D}^B_J[F^-]=\sum C_{J_1,\dots,J_p}\,F^-\sharp^B\mathfrak{D}^B_{J_1}[F]\sharp^B F^-\sharp^B\ldots\sharp^BF^-\sharp^B
\mathfrak{D}^B_{J_p}[F]\sharp^B F^-.
\end{equation}
The sum is over all partitions $J=\sqcup_{i=1}^pJ_i$ where, for example, the partition $(J_1,J_2)$ is considered different
from $(J_2,J_1)$.
The coefficients $C_{J_1,\dots,J_p}$ take only the values $\pm 1$, but this is not important.

Once again by Theorem \ref{Bony} we know that each $\mathfrak{D}^B_{J_i}[F]$ belongs to $\mathfrak C^B(\Xi)$,
while $F^-\in\mathfrak C^B(\Xi)$ by assumption. It follows that $\mathfrak{D}^B_J[F^-]\in\mathfrak C^B(\Xi)$.
\end{proof}

\begin{proposition}
 For $m<0$ if $f\in S^{m}_\rho(\Xi)$ is such that $1+f$ is invertible in
$\mathfrak C^B(\Xi)$, then $(1+f)^--1\in S^m_\rho(\Xi)$.
\end{proposition}

\begin{proof}
 We borrow a simple idea from \cite{LMN}. Choose $f\in S^{m}_\rho(\Xi)$ such that $1+f$ is invertible in
$\mathfrak C^B(\Xi)$. Then $(1+f)^-\in S^0_\rho(\Xi)$. Consequently, by an obvious identity and by the magnetic symbolic calculus
$$
(1+f)^--1=-f+(1+f)^-\dB f\dB f\in S^m_\rho(\Xi).
$$
\end{proof}

\begin{proposition}\label{symb-inv}
Let $m>0$ and $\rho\in[0,1]$.
If $G\in S^m_\rho(\Xi)$ is invertible in $\mathfrak M^B(\Xi)$, with $\mathfrak s_{m}\sharp^B G^{-}\in\mathfrak C^B(\Xi)$,
then $G^-\in S^{-m}_\rho(\Xi)$.
\end{proposition}

\begin{proof}
  First we remark that for any $m\in\mathbb{R}$ we have $\mathfrak{s}_{m}\in S^m_1(\Xi)$.
To see this, one just has to repeat the proof by induction given in Step 6 of the proof of Theorem \ref{main-m} by using symbols
 $\varphi_j\in S^+_1(\Xi)$ in place of the linear symbols used there. Then use Theorem \ref{Bony}.

One has
$$
G\sharp^B\mathfrak s_{-m}\in S^m_\rho(\Xi)
\sharp^B S^{-m}_\rho(\Xi)\subset S^0_\rho(\Xi).
$$
In addition this element is invertible in $\mathfrak C^B(\Xi)$, since we can compute in $\mathfrak M^B(\Xi)$
$$
\left(G\sharp^B\mathfrak s_{-m}\right)^-=\mathfrak s_m\,\sharp^B G^-\in\mathfrak C^B(\Xi).
$$
Then, by Proposition \ref{th}, $\mathfrak s_m\sharp^B G^-\in S^0_\rho(\Xi)$. Consequently
$$
G^{-}=\mathfrak s_{-m}\sharp^B[\mathfrak s_{m}\sharp^B G^{-}]\in S^{-m}_\rho(\Xi)\sharp^B S^0_\rho(\Xi)\subset S^{-m}_\rho(\Xi).
$$

\end{proof}
To verify the boundedness condition in the above Proposition \ref{symb-inv}, the condition of ellipticity is usually needed.

\begin{definition}
 For $m>0$, a symbol $F\in S^m_\rho(\Xi)$ is called {\it elliptic} if there exist two constants $R,C$ for which
$$
|\xi|\geq R\quad\Longrightarrow\quad F(x,\xi)\geq C<\xi>^m.
$$
\end{definition}

We can apply Proposition \ref{symb-inv} to any elliptic symbol of strictly
positive order by using Theorem 4.1. in \cite{IMP}.
In fact that Theorem asserts that
for an elliptic symbol $F\in S^m_\rho(\Xi)$ with $m>0$ and for any vector potential vector $A$ with $B=dA$,
we get a self-adjoint operator $\mathfrak{Op}^A$ having a spectrum $\sigma[F]$ that does not
depend on the choice of the representation (by gauge covariance). Thus for any $z\notin\sigma[F]$
the operator
$\mathfrak{Op}^A(F)-z1=\mathfrak{Op}^A(F-z)$ is invertible with bounded inverse. This
means that the inverse $(F-z1)^-$ exists in $\mathfrak{M}^B(\Xi)$ and that it belongs to
$\mathfrak{C}^B(\Xi)$. Moreover, the Theorem 4.1 in \cite{IMP} implies that
$\mathfrak{Op}^A\big[(F-z)\dB\mathfrak{s}_m^-\big]$ is a bijection
on $L^2(\mathcal{X})$, and thus
$\mathfrak{s}_m\dB(F-z)^-=\big[(F-z)\dB\mathfrak{s}_m^-\big]^-\in\mathfrak{C}^B(\Xi)$. This allows us to use
Proposition \ref{symb-inv} and prove the following statement.

\begin{proposition}\label{ell-inv}
Given a real elliptic symbol $F\in S^m_\rho(\Xi)$, for any $z\notin\sigma[F]$ the inverse $(F-z)^-$ exists and is a symbol
of class $S^{-m}_\rho(\Xi)$.
\end{proposition}

\subsection{Functional calculus}

Relying on Propositions \ref{th} and \ref{ell-inv}, we can obtain results concerning the functional
calculus of elliptic magnetic self-adjoint operators. In any given Hilbert space
representation associated to a vector potential $A$ we have
$$
\Phi\left(\Op^A[f]\right)=:\Op^A\left[\Phi^B(f)\right]
$$
and this gives an intrinsic meaning to the functional calculus for Borel functions $\Phi$.

We recall (cf. \cite{LMN} and references therein) that a $\Psi^*$-{\it algebra}
is a Fr\'echet $^*$-algebra continuously embedded in a $C^*$-algebra, which is
spectrally invariant (i.e. stable under inversion). Our Proposition \ref{th}
says that $S^0_\rho(\Xi)$ is a $\Psi^*$-algebra in the $C^*$-algebra $\mathfrak
C^B(\Xi)$. But $\Psi^*$-algebras are stable under the holomorphic functional
calculus, so we can state:

\begin{proposition}\label{zbauer}
If $f\in S^0_\rho(\Xi)$ and $\Phi$ is a function holomorphic on
some neighborhood of the spectrum of $f$, then $\Phi^B(f)\in S^0_\rho(\Xi)$.
\end{proposition}

If $\Phi\in C_0^\infty(\R)$ (and in many other situations), $\Phi^B(f)$ can be written using the
Helffer-Sj\"ostrand formula
\begin{equation}\label{hesj}
\Phi^B(f)=\frac{1}{\pi}\int_\mathbb C dz\,\partial_{\overline z}\widetilde{\Phi}(z)(f-z)^-,
\end{equation}
$\widetilde{\Phi}$ being a quasi-analytic extension of $\Phi$ (cf. \cite{HS}, \cite{Da1}, \cite{Da2}).

\begin{proposition}\label{func-calc}
If $\Phi\in C^\infty_0(\mathbb{R}), f\in S^m_\rho (\Xi), m\leq 0, f$,elliptic if $m>0$ then $\Phi^B(f)\in S^{-m}_{\rho}(\Xi)$.
\end{proposition}

\begin{proof}
 Let $\widetilde{\Phi}$ be a quasi-analytic extension of $\Phi\in C^\infty_0(\mathbb{R})$, i.e.
 $\widetilde{\Phi}\in C^\infty_0(\mathbb{C})$, $\widetilde{\Phi}|_\mathbb R=\Phi$
 and for any $M\in\mathbb{N}$ there exists $C_M>0$ such that
\begin{equation}\label{est-2}
\left|\frac{\partial\widetilde{\Phi}}{\partial\overline{z}}(\lambda+i\mu)\right|\leq
C_M\left|\mu\right|^M,\quad\text{for }\lambda+i\mu\in\mathbb C.
\end{equation}
Then using formula (\ref{hesj}) we get
$$
\mathfrak s_m\sharp^B\ad^B_{\varphi_1}\dots\ad^B_{\varphi_N}\big[\Phi^B(f)\big]=
\frac{1}{\pi}\int_{\Omega}dz\,\left(\frac{\partial
\widetilde{\Phi}}{\partial\overline{z}}\right)\mathfrak s_m\sharp^B\ad^B_{\varphi_1}\dots\ad^B_{\varphi_N}\big[(f-z)^-\big],
$$
with $\Omega\subset\mathbb{C}$ a bounded domain strictly containing the support of $\widetilde{\Phi}$.

To understand the behavior in
$z=\lambda+i\mu$ of the magnetic derivatives of $(f-z)^-$, we adapt some ideas of \cite{ABG}, section 6.2.
Since $\widetilde\Phi$ has compact support, we concentrate only on the
divergence when $\mu\rightarrow 0$.

Let $z,z_0\notin \sigma(f)$ and $\varphi\in S^+_\rho(\Xi)$.
$$
\ad^B_\varphi\left[(f-z)^{-}\right]=-(f-z)^{-}\dB\ad^B_\varphi(f-z)\dB(f-z)^{-}=
$$
$$
=-(f-z)^{-}\dB(f-z_0)\dB(f-z_0)^-\dB\ad^B_\varphi(f-z_0)\dB(f-z_0)^-\dB(f-z_0)\dB(f-z)^{-}=
$$
$$
=g(z)\dB\ad^B_\varphi\left[(f-z_0)^-\right]\dB g(z),
$$
where
$$
g(z):=(f-z)^{-}\dB(f-z_0)=1+(z-z_0)(f-z)^-.
$$
Then, by reccurence, we prove that $\ad^B_{\varphi_1}\dots\ad^B_{\varphi_N}\left[(f-z)^-\right]$ is a finite linear
combination of terms of the form
$$
(z-z_0)^{m-1}g(z)\dB\mathfrak D^B_{J_1}\left[(f-z_0)^-\right]\dB g(z)\dB\dots\dB g(z)\dB\mathfrak D^B_{J_m}
\left[(f-z_0)^-\right]\dB g(z),
$$
with $(J_1,\dots,J_m)$ a partition of $\{1,\dots,N\}$ and $\mathfrak{D}^B_K:=\ad^B_{\varphi_{k_1}}\dots\ad^B_{\varphi_{k_p}}$
for the ordered set $K:=\{k_1,\dots,k_p\}$. There are at most $N+1$ $g$'s.

Under our assumption on $f$, all the factors $\mathfrak{D}^B_{J_l}\left[(f-z_0)^-\right]$
are elements of $\mathfrak C^B(\Xi)$ with $z$-independent norms. One also has $\parallel g(z)\parallel_{\mathfrak C^B}
\le C|\mu|^{-1}$ on ${\rm supp}\mu$.

The factor $\mathfrak s_m\sharp^B g(z)\sharp^B\mathfrak D_{J_1}[(f-z)^{-}]$
needs a special study.

The first (constant) term composing $g$ is trivial. For the second we write $$\mathfrak s_m\sharp^B [(z-z_0)(f-z)^{-}]
\sharp^B\mathfrak D_{J_1}[(f-z)^{-}]=(z-z_0)\mathfrak s_m\sharp^B (f-z_0)^{-}\sharp^B [(f-z_0)\sharp^B (f-z)^{-}]
\sharp^B\mathfrak D_{J_1}[(f-z)^{-}]$$
and once again we are safe, because $\mathfrak s_m\sharp^B (f-z_0)^{-}\in S^0_\rho (\Xi)\subset \mathcal C^B (\Xi)$.

So finally, putting everything together, we get
\begin{equation}\label{est-1}
\parallel\ad^B_{\varphi_1}\dots\ad^B_{\varphi_N}\left[(f-z)^-\right]\parallel_{\mathfrak C^B}\,\le
\,C\frac{<\lambda>^{N}}{|\mu|^{N+1}}.
\end{equation}

Using now the estimations (\ref{est-1}) and (\ref{est-2}) and Theorem \ref{Bony} we get the stated result.
\end{proof}

\subsection{Fractional powers}

Choosing a vector potential $A$ for the magnetic field $B$ and
considering the associated Schr\"{o}dinger representation on $\mathcal{H}=L^2(\mathcal{X})$, one proves

\medskip
\begin{theorem}\label{codasa}
 Given a lower bounded $F\in S^m_\rho(\Xi)$ with $m\geq0$, elliptic if $m>0$, let $\mathfrak{Op}^A[F]$ be the
 associated self-adjoint, semi-bounded operator on $\mathcal{H}$ given by Theorem 4.1. and Corollary 4.4 in \cite{IMP} and
 let $t_0\in\mathbb{R_+}$ such that for $F_0:=F+t_01$ the operator $\mathfrak{Op}^A[F_0]$ is strictly positive. Then
 for any $s\in\mathbb{R}$ the power $s$ of $\mathfrak{Op}^A[F_0]$ is a magnetic pseudodifferential operator
 with symbol $F_0^{[s]_B}\in S^{sm}_\rho(\Xi)$, i.e. $\left(\mathfrak{Op}^A[F_0]\right)^s=
 \mathfrak{Op}^A\big[F_0^{[s]_B}\big]$.
\end{theorem}

\begin{proof}
Due to (\ref{hor}) and the above Proposition \ref{ell-inv},
the statement is valid for any $s\in\mathbb{Z}$. Once again by (\ref{hor}) it is enough to solve the case $s\in(-1,0)$.

In this case the equality
$$
\left(\mathfrak{Op}^A[F_0]\right)^s=-\frac{1}{2\pi
i}\int_{-i\infty}^{i\infty}\,z^s\left(\mathfrak{Op}^A[F_0]-z\right)^{-1}dz=
-\frac{1}{2\pi i}\mathfrak{Op}^A\left(\int_{-i\infty}^{i\infty}\,z^s(F_0-z)^-dz\right)
$$
may be proved by approximation. First we restrict to vectors $u\in E_{_{\mathfrak{Op}^A[F_0]}}([-N,N])L^2(\mathcal{X})$
and use the Cauchy formula for the analytic function
$z^s$ on the domain $\{\Re z\geq0\}\cap\{0<\epsilon\leq|z|\leq 2N\}$. Here $\epsilon<\inf\Op^A(F_0)$ and $E_T$ denoted the
spectral measure of the self-adjoint operator $T$. Then one lets $\epsilon\searrow0$ and
$N\nearrow\infty$.
Thus we only have to prove that
$$
\int_{\mathbb{R}}\,t^s(F_0+it)^-dt=\int_{\mathbb{R}}\,t^s(F_0+it)^{-1}G_tdt\in S^{ms}_\rho(\Xi),
$$
where $G_t:=(F_0+it)(F_0+it)^-$. This will follow easily from (i) and (ii)
below and the behavior on pointwise products of the seminorms defining the topology of $S^m_{\rho}(\Xi)$
$$
|f|_{m,p,q}:=\underset{(x,\xi)\in\Xi}{\sup}\;\underset{|\alpha|\leq
p}{\max}\;\underset{|a|\leq q}{\max}
\left|<\xi>^{-m+\rho|\alpha|}\left(\partial^\alpha_\xi\partial^a_xf\right)(x,\xi)\right|.
$$

(i) $G_t$ belongs to $S^0_\rho(\Xi)$ uniformly in $t\in\mathbb R$.

Let us notice that $G_t:=(F_0+it)(F_0+it)^-$ is a symbol of type $S^0_\rho(\Xi)$ due to our
previous Proposition \ref{ell-inv} and poinwise multiplication of symbols. It will be enough to prove that
$F_0(F_0+it)^-$ and $t(F_0+it)^-$ are symbols of class $S^0_\rho(\Xi)$ uniformly with respect to $t\in\mathbb{R}_+$.

The results of
Section 4 of \cite{IMP} show that $\mathfrak{Op}^A\left[\mathfrak{s}_m\right]\mathfrak{Op}^A\left[(F_0+it)^-\right]$ is a
bounded operator uniformly with respect to $t\in\mathbb{R}_+$. Let us choose $N\in\mathbb{N}$ and a family
$\{\varphi_1,\ldots,\varphi_N\}\subset S^+_\rho(\Xi)$. Using
(\ref{mult-ad-inv}) one gets
$$
\|\mathfrak{s}_m\dB\left(\ad^B_{\varphi_1}\cdot\ldots\cdot\ad^B_{\varphi_N}
[(F_0+it)^-]\right)\|_{\mathfrak C^B}\leq C
$$
with $C$ independent of $t\in\mathbb R_+$, and this implies immediately  the assertion for $F_0(F_0+it)^-$.

Now use once again (\ref{mult-ad-inv}) and the fact that $F_0\in S^m_\rho(\Xi)$. The results of Section 4 of
\cite{IMP} show that $t\mathfrak{Op}^A\left[(F_0+it)^-\right]$ is a bounded operator uniformly
with respect to $t\in\mathbb{R}_+$, so one gets
$$
\|t\ad^B_{\varphi_1}\cdot\ldots\cdot\ad^B_{\varphi_N}
[(F_0+it)^-]\|_{\mathfrak C^B}\leq C
$$
with $C$ independent of $t\in\mathbb{R}_+$ and this implies the result.

(ii) One has
$$
\int_\mathbb R\,t^s(F_0+it)^{-1}dt\in S^{ms}_\rho(\Xi).
$$
This can be proved by writing
$$
\partial_x^a\partial_\xi^\alpha\left[(F_0+t)^{-1}\right]=
\underset{\underset{|a_1|+\dots+|a_k|=|a|}{|\alpha_1|+\dots+|\alpha_k|=|\alpha|}}
{\sum} C^{a_1,\dots,a_k}_{\alpha_1,\dots,\alpha_k}
\left(\partial_x^{a_1}\partial_\xi^{\alpha_1}F_0\right)\dots
\left(\partial_x^{a_k}\partial_\xi^{\alpha_k}F_0\right)(F_0+t)^{-(k+1)}.
$$
So that we are reduced to evaluating
$$
\int_\mathbb R\,t^s\left|(F_0+it)^{-(k+1)}\right|dt.
$$
But since $s\in(-1,0)$, one has
$$
\int_\mathbb R\,t^s\left|(F_0+it)^{-(k+1)}\right|dt=F_0^{s-k}\int_\mathbb R\,\left(\frac{t}{F_0}\right)^{s-k}
\left|\left(\frac{F_0}{t}+i\right)^{-(k+1)}\right|\frac{dt}{t}\le
$$
$$
\le C<\xi>^{m(s-k)}\int_\mathbb R\tau^{k-s}\left|(\tau+i)^{-(k+1)}\right|\frac{d\tau}{\tau}\le C'<\xi>^{m(s-k)},
$$
and this finishes the proof.
\end{proof}

\section{Magnetic Fourier Integral Operators}

In this section we consider a definition given by J.-M. Bony for an operator to be a Fourier Integral Operator and using our Bony type criterion (Theorem \ref{Bony}) as a starting point, modify it in what we shall call Magnetic Fourier Integral Operators.

Let us consider one-to one mappings $V,W,\dots:\Xi\rightarrow\Xi$.  For complex functions $\varphi$ defined in  phase space, we
introduce formally {\it twisted magnetic commutators}, generalizing our previous commutators $\ad^B_\varphi$:
\begin{equation}\label{tmc}
\ad^{B,V}_\varphi[f]:=\varphi\dB f-f\dB(\varphi\circ V).
\end{equation}
They satisfy simple algebraic properties, that will be basic in the sequel:
\begin{equation}\label{cra}
\ad^{B,V}_\varphi[\lambda f+\mu g]=\lambda\ad^{B,V}_\varphi[f]+\mu\ad^{B,V}_{\varphi}[g],
\end{equation}
\begin{equation}
\ad^{B,V\circ W}_\varphi[f\dB g]=\ad^{B,V}_\varphi[f]\dB g+f\dB\ad^{B,W}_{\varphi\circ V}[g],
\end{equation}
\begin{equation}\label{crb}
\ad^{B,V}_\varphi[f^*]=-\ad^{B,V^{-1}}_{\varphi^*\circ V}[f]^*.
\end{equation}
\begin{definition}\label{noua}
Let $V:\Xi\rightarrow\Xi$ be given and let $\mathcal R$ be a vector subspace of $\mathfrak M^B(\Xi)$, supposed (for simplicity) closed under complex
conjugation and such that $\mathcal{R}\circ V\subset\mathfrak{M}^B(\Xi)$. We set
$$
S(B,V;\mathcal R):=\left\{f\in\mathfrak M^B(\Xi)\mid\ad^{B,V}_{\varphi_1}\dots\ad^{B,V}_{\varphi_N}[f]
\in\mathfrak C^B(\Xi),\ \ \forall N\in N,\ \ \forall \varphi_1,\dots,\varphi_N\in \mathcal R\right\}.
$$
\end{definition}

Clearly $S(B,{\sf id};S^+_\rho(\Xi))=S^0_\rho(\Xi)$, by Theorem \ref{Bony}. In the framework of \cite{Bo3} (cf. Definitions
4.3 and 3.1), for $B=0$ and under some assumptions connecting the diffeomorphism $V$ and the metrics $g_1,g_2$, one has
$$
\Op\left[S(0,V;S^+(1,g_2))\right]={\sf FIO}(V;g_1,g_2)\ \ \ {\rm and}\ \ \ S(0,{\sf id};S^+(1;g)=S(1;g)).
$$

We shall consider a class of diffeomorphisms $\Phi:\Xi\rightarrow\Xi$ and we shall denote $\Phi(X)\equiv\big(y(X),\eta(X)\big)$ (for any $X=x,\xi)\in\Xi$). We shall also consider on $\Xi$ the metric $\mathfrak{g}_{(x,\xi)}(y,\eta):=|y|^2+<\xi>^{-2}|\eta|^2$. We shall supose that $\Phi$ satisfies the following conditions:
\begin{hypothesis}\label{Phi}
 \begin{enumerate}
  \item $\Phi:\Xi\rightarrow\Xi$ is of class $C^\infty$ and symplectic for the canonical symplectic form on $\Xi$;
\item there exists $C>0$ such that
$$
\left(\frac{<\eta(x,\xi)>}{<\xi>}\right)^{\pm1}\leq C;
$$
\item the derivatives of order higher then 1 of $\Phi$ and $\Phi^{-1}$ are bounded with respect to the metric $\mathfrak{g}$ introduced above.
 \end{enumerate}
\end{hypothesis}

One can easily prove that under our Hypothesis \ref{Phi} the class of symbols $S^+_1(\Xi)$ is stable for the composition with $\Phi$. Thus we can define the class $S(B,\Phi;S^+_1(\Xi))$ as above.

For a magnetic field $B$ with components of class $BC^\infty(\mathcal{X})$ we shall consider the class of symbols $S(B,\Phi;S^+_1(\Xi))$. Moreover, chosing a vector potential $A$ for $B$ having components of class $C^\infty_{pol}(\mathcal{X})$ we shall consider the class of \textit{'magnetic Fourier integral operators'}, in the Schr\"{o}dinger representation associated to $A$:
$$
{\sf FIO}^A(\Phi):=\mathfrak{Op}^A\left[S(B,\Phi;S^+_1(\Xi))\right].
$$
In fact we shall prove that for a class of hamiltonians, the unitary evolution group they generate are of class ${\sf FIO}^A(\Phi)$ for a diffeomorphism $\Phi$ given by a Hamiltonian flow.

\subsection{The symbol of the evolution group}

Given any Hamiltonian described by a symbol $h\in S^m_1(\Xi)$ we shall define its associated flow $\Phi_t:\Xi\rightarrow\Xi$, that we shall also denote by $Y(t;X)\equiv\Phi_t(X)$, that is defined by the Cauchy problem:
\begin{equation}\label{Y}
\dot{Y}(t;X)=\mathfrak{X}_h\left[Y(t;X)\right],\qquad Y(0;X)=X,
\end{equation}
with $\mathfrak{X}_h$ the Hamiltonian field associated to $h$ with respect to the canonical simplectic form $\sigma$ on $\Xi$. Explicitely we have
$$
\mathfrak{X}_h:=\left(\partial_\xi h,-\partial_x h\right).
$$
\begin{hypothesis}\label{h}
 Suppose $h\in S^m_1(\Xi)$ is real elliptic and $0< m\leq 1$. 
\end{hypothesis}

\begin{lemma}
Under the above Hypothesis \ref{h} for the Hamiltonian $h$ we have:
\begin{enumerate}
 \item the Cauhy problem (\ref{Y}) has a unique solution $Y(t;X)$ and the map
$$
\mathbb{R}\times\Xi\ni(t,X)\mapsto Y(t:X)\in\Xi
$$
is of class $C^\infty$.
\item for any given $t\in\mathbb{R}$ the flow $\Phi_t$ satisfies the Hypothesis \ref{Phi}.
\end{enumerate}

\end{lemma}
\begin{proof}
 The first conclusion results from the fact that the Hamiltonian has at most a lineara growth. Let us turn now to the proof of the 3 conditions in Hypothesis \ref{Phi}.
The first one is a classic property of Hamiltonian systems. The second one can be easily verified using the ellipticity of $h$ and the conservation of $h$ along the flow it generates. For the third condition we have to integrate the components of the Hamiltonian field (\ref{Y})
\begin{equation}\label{H-syst-int}
\left\{
\begin{array}{lll}
 y_j(t;x,\xi)=&x_j+\int_0^tds\,\big(\partial_{\eta_j}h\big)(y(s;x,\xi),\eta(s;x,\xi))&\\
&&\\
 \eta_j(t;x,\xi)=&\xi_j-\int_0^tds\,\big(\partial_{y_j}h\big)(y(s;x,\xi),\eta(s;x,\xi)).
&\end{array}
\right.
\end{equation}
Now let us compute
$$
\big(\partial_{x_k}y_j\big)(t;x,\xi)=\delta_{jk}+\sum\limits_{1\leq l\leq n}\int_0^tds\,\left[\big(\partial_{y_l}\partial_{\eta_j}h\big)(y(s;x,\xi),\eta(s;x,\xi))\big(\partial_{x_k}y_l\big)(s;x,\xi)+ \right.
$$
$$
+\left.\big(\partial_{\eta_l}\partial_{\eta_j}h\big)(y(s;x,\xi),\eta(s;x,\xi))\big(\partial_{x_k}\eta_l\big)(s;x,\xi) \right],
$$
$$
\big(\partial_{x_k}\eta_j\big)(t;X)=-\sum\limits_{1\leq l\leq n}\int_0^tds\,\left[\big(\partial_{y_l}\partial_{y_j}h\big)(y(s;X),\eta(s;X))\big(\partial_{x_k}y_l\big)(s;X)+ \right.
$$
$$
+\left.\big(\partial_{\eta_l}\partial_{y_j}h\big)(y(s;X),\eta(s;X))\big(\partial_{x_k}\eta_l\big)(s;X)\right].
$$
Thus we have the estimations:
$$
\underset{1\leq j\leq n}{\max}\left|\big(\partial_{x_k}y_j\big)(t;X)\right|\leq 1+C\int_0^tds\,\left[\underset{1\leq l\leq n}{\max}\left|\big(\partial_{x_k}y_l\big)(s;X)\right|+\left< \eta(s;X)\right>^{-1}\left(\underset{1\leq l\leq n}{\max}\left|\big(\partial_{x_k}\eta_l\big)(s;X)\right| \right)\right] ,
$$
$$
\underset{1\leq j\leq n}{\max}\left|\big(\partial_{x_k}\eta_j\big)(t;X)\right|\leq C\int_0^tds\,\left[\left< \eta(s;X)\right>^{m}\left(\underset{1\leq l\leq n}{\max}\left|\big(\partial_{x_k}y_l\big)(s;X)\right| \right)+ \left(\underset{1\leq l\leq n}{\max}\left|\big(\partial_{x_k}\eta_l\big)(s;X)\right| \right)\right].
$$
Due to the ellipticity condition (and $m>0$) and the conservation of the Hamiltonian along the flow we easily obtain that there exists a finite constant $c$ such that $c^{-1}<\xi>\leq<\eta(t;X)>\leq c<\xi>$ for any $t\in\mathbb{R}$. Thus if we denote for $|a|=1$
$$
\mathcal{E}_{a,0}(t;X):=\left\{\underset{1\leq j\leq n}{\max}\left|\big(\partial^a_x y_j\big)(t;X)\right|+<\xi>^{-1}\underset{1\leq j\leq n}{\max}\left|\big(\partial^a_x\eta_j\big)(t;X)\right| \right\},
$$
it defines a positive function for which we have proved the following estimation
$$
\mathcal{E}_{a,0}(t;X)\leq 1+C\int_0^tds\,\mathcal{E}_{a,0}(s;X),
$$
so that by the Gronwall Lemma we conclude that
$$
\mathcal{E}_{a,0}(t;X)\leq e^{Ct},\quad\forall t\in\mathbb{R},\quad\forall X\in\Xi.
$$

Let us consider now the derivations with respect to the $\mathcal{X}^\prime$-variables.
$$
\big(\partial_{\xi_k}y_j\big)(t;X)=\sum\limits_{1\leq l\leq n}\int_0^tds\,\left[\big(\partial_{y_l}\partial_{\eta_j}h\big)(y(s;X),\eta(s;X))\big(\partial_{\xi_k}y_l\big)(s;X)+ \right.
$$
$$
+\left.\big(\partial_{\eta_l}\partial_{\eta_j}h\big)(y(s;X),\eta(s;X))\big(\partial_{\xi_k}\eta_l\big)(s;X) \right],
$$
$$
\big(\partial_{\xi_k}\eta_j\big)(t;X)=\delta_{jk}-\sum\limits_{1\leq l\leq n}\int_0^tds\,\left[\big(\partial_{y_l}\partial_{y_j}h\big)(y(s;X),\eta(s;X))\big(\partial_{\xi_k}y_l\big)(s;X)+ \right.
$$
$$
+\left.\big(\partial_{\eta_l}\partial_{y_j}h\big)(y(s;X),\eta(s;X))\big(\partial_{\xi_k}\eta_l\big)(s;X)\right].
$$
Thus we have the estimations:
$$
\underset{1\leq j\leq n}{\max}\left|\big(\partial_{\xi_k}y_j\big)(t;X)\right|\leq C\int_0^tds\,\left[\underset{1\leq l\leq n}{\max}\left|\big(\partial_{\xi_k}y_l\big)(s;X)\right|+\left< \eta(s;X)\right>^{-1}\left(\underset{1\leq l\leq n}{\max}\left|\big(\partial_{\xi_k}\eta_l\big)(s;X)\right| \right)\right] ,
$$
$$
\underset{1\leq j\leq n}{\max}\left|\big(\partial_{\xi_k}\eta_j\big)(t;X)\right|\leq 1+C\int_0^tds\,\left[\left< \eta(s;X)\right>^{m}\left(\underset{1\leq l\leq n}{\max}\left|\big(\partial_{\xi_k}y_l\big)(s;X)\right| \right)+ \left(\underset{1\leq l\leq n}{\max}\left|\big(\partial_{\xi_k}\eta_l\big)(s;X)\right| \right)\right].
$$
For $|\alpha|=1$ let us denote
$$
\mathcal{E}_{0,\alpha}(t;X):=\left\{\underset{1\leq j\leq n}{\max}\left|\big(\partial^\alpha_\xi y_j\big)(t;X)\right|+<\xi>^{-1}\underset{1\leq j\leq n}{\max}\left|\big(\partial^\alpha_\xi\eta_j\big)(t;X)\right| \right\},
$$
it defines a positive function for which we have proved the following estimation
$$
\mathcal{E}_{0,\alpha}(t;X)\leq <\xi>^{-1}+C\int_0^tds\,\mathcal{E}_{a,0}(s;X),
$$
so that by the Gronwall Lemma we conclude that
$$
\mathcal{E}_{0,\alpha}(t;X)\leq <\xi>^{-1}e^{Ct},\quad\forall t\in\mathbb{R},\quad\forall X\in\Xi.
$$
We conclude that for $|a|+|\alpha|=1$ we have the estimation:
$$
\mathcal{E}_{a,\alpha}(t;X)\leq <\xi>^{-|\alpha|}e^{Ct},\quad\forall t\in\mathbb{R},\quad\forall X\in\Xi.
$$

Let us suppose now that for some $k\geq 1$ we have proved that for any pair of multi-indices $(a,\alpha)$ with $1\leq|a|+|\alpha|\leq k$ we have the inequality
$$
\mathcal{E}_{a,\alpha}(t;X)\leq C_{t,(a,\alpha)}<\xi>^{-|\alpha|},\quad\forall t\in\mathbb{R},\quad\forall X\in\Xi.
$$
Let us choose then a pair of multi-indices $(b,\beta)$, such that $|b|+|\beta|=k+1$ and let us apply the operator $\partial^b_x\partial^\beta_\xi$ to our system \ref{H-syst-int}. Using our induction hypothesis we obtain
$$
\mathcal{E}_{b,\beta}(t;X)\leq C_{t,(b,\beta)}<\xi>^{-|\beta|}+C\int_0^tds\,\mathcal{E}_{b,\beta}(s;X),\quad\forall t\in\mathbb{R},\quad\forall X\in\Xi.
$$
Thus using once again the Gronwall Lemma we obtain that
$$
\mathcal{E}_{a,\alpha}(t;X)\leq C_{t,(a,\alpha)}<\xi>^{-|\alpha|},\quad\forall t\in\mathbb{R},\forall X\in\Xi,\forall(a,\alpha)\in\mathbb{N}^n\times\mathbb{N}^n.
$$
It is now easy to conclude that our third condition in Hypothesis \ref{Y} is satisfied.
\end{proof}

\begin{theorem}\label{FIO-ev}
 We suppose given a magnetic field with components of class $BC^\infty(\mathcal{X})$ and a Hamiltonian $h$ satisfying Hypothesis \ref{h}. In the Schr\"{o}dinger representation associated to a vector potential $A$ of class $C^\infty_{pol}(\mathcal{X}$ we have that $\mathfrak{Op}^A(h)$ defines a self-adjoint operator and its unitary evolution group $P_t:=\exp\{-it\mathfrak{Op}^A(h)\}$ is of class ${\sf FIO}^A(\Phi_t)$ with $\Phi_t$ the solution of problem (\ref{Y}) associated to $h$.
\end{theorem}

The proof of this Theorem is based on the following two Lemmas.
\begin{lemma}\label{7-4}
 Let $a\in S^m_1(\Xi)$ and $c\in S^+_1(\Xi)$. We consider on $S^+_1(\Xi)$ the natural Frechet topology. We denote by $\{.,.\}$ the Poisson bracket defined by the canonical symplectic form $\sigma$ on $\Xi$. Then we have the following statements.
\begin{enumerate}
 \item For any $t\in\mathbb{R}$ we have that $c\circ\Phi_t\in S^+_1(\Xi)$ and the map 
$$
\mathbb{R}\ni t\mapsto c\circ\Phi_t\in S^+_1(\Xi)
$$
is of class $C^\infty(\mathbb{R})$.
\item We have that
$$
c\sharp^Ba-a\sharp^Bc-i^{-1}\{c,a\}\in S^{m-1}_1(\Xi)
$$
and in particular $c\sharp^Ba-a\sharp^Bc\in S^m_1(\Xi)$.
\item For $m\leq 1$ the map 
$$
(c\circ\Phi_t)\sharp^Ba-a\sharp^B(c\circ\Phi_t)-i^{-1}\{(c\circ\Phi_t),a\}\in S^{0}_1(\Xi)
$$
is of class $C^\infty(\mathbb{R})$.
\end{enumerate}
\end{lemma}
The proof of this Lemma may be obtained in a straightforward way from the arguments given in the first two sections of \cite{IMP}.

\begin{lemma}\label{7-5}
 The unitary evolution group $P_t$ generated by $\mathfrak{Op}^A(h)$ satisfies the following relations.
\begin{enumerate}
 \item For any $f\in\mathcal{S}(\mathcal{X})$ and any $t\in\mathbb{R}$ we have that $P_tf\in\mathcal{S}(\mathcal{X})$ uniformly for $t$ in bounded sets.
\item $P_t\in\mathbb{B}(\mathcal{S}(\mathcal{X}))$ for any $t\in\mathbb{R}$.
\item The map $\mathbb{R}\ni t\mapsto P_t\in\mathbb{B}(\mathcal{S}(\mathcal{X}))$ is differentiable for the strong operatorial topology on $\mathbb{B}(\mathcal{S}(\mathcal{X}))$.
\end{enumerate}
\end{lemma}
\begin{proof}
 Let us denote by $\mathcal{H}^m_A$ the domain of the operator $\mathfrak{Op}^A(h)$ with the graph norm (this is a magnetic Sobolev space \cite{IMP}). For any multi-index $\alpha\in\mathbb{N}^n$ let us denote by $\Pi^\alpha_A:=(\Pi^A_1)^{\alpha_1}\cdot\ldots\cdot(\Pi^A_n)^{\alpha_n}$ and by $f(t):=P_tf$. It is enough to prove by induction on $p+q$ (for $(p,q)\in\mathbb{N}\times\mathbb{N}$) that for any $\alpha\in\mathbb{N}^n$ with $|\alpha|=q$ we have
\begin{equation}\label{ind}
 <x>^p\Pi^\alpha_Af(t)\in\mathcal{H}^m_A,
\end{equation}
uniformly for $t$ in bounded sets.

Let us observe that our Lemma \ref{7-4} implies that for any $d\in S^+_1(\Xi)$ we have $[d,h]_{\sharp^B}\in S^m_1(\Xi)$. This allows us to prove that for any $r\in \mathbb{N}$ we have
\begin{equation}\label{E}
 \left[\mathfrak{Op}^A(d)\right]^r\mathfrak{Op}^A(h)f(t)=\sum_{0\leq k\leq r}C^r_k\left[\mathfrak{ad}_{\mathfrak{Op}^A(d)}^{r-k}\cdot\mathfrak{Op}^A(h)\right]\left[\mathfrak{Op}^A(d)\right]^ku(t).
\end{equation}

Suppose that $\varphi\in C^\infty_0(\mathcal{X})$ is such that $\varphi(x)=0$ for $|x|\geq2$ and $\varphi(x)=1$ for $|x|\leq 1$. Denote by $\theta_j(x):=<x>\varphi(x/j)$ for $j\geq 1$; then $\theta_j\in C^\infty_0(\mathcal{X})$. Let us stil introduce the notations 
$$
v_{j,p,\alpha}(t):=\theta_j^p\Pi^\alpha_Af(t), \quad\forall\alpha\in\mathbb{N}^n,\,|\alpha|=q.
$$
One has
$$
\leq\dot{v}_{j,p,\alpha}(t)=\theta_j^p\Pi^\alpha_A\mathfrak{Op}^A(h)f(t)
$$
and using (\ref{E}) we conclude that 
$$
\frac{d}{dt}\left\|v_{j,p,\alpha}\right\|_{L^2(\mathcal{X})}^2\leq \left\|v_{j,p,\alpha}\right\|_{L^2(\mathcal{X})}^2+C,\quad\text{for }|t|\leq T,
$$
for any $T\geq 0$, with $C$ a constant depending on $T$. Integrating this inequality and using the Fatou Lemma we conclude that $<x>^p\Pi^\alpha_Af(t)\in L^2(\mathcal{X})$ uniformly for $t$ in bounded sets. Using once again (\ref{E}) and some evident commutation properties we also obtain that $\mathfrak{Op}^A(h)\left(<x>^p\Pi^\alpha_Af(t)\right)\in L^2(\mathcal{X})$ uniformly for $t$ in bounded sets. This proves (\ref{ind}).

The second conclusion of the Lemma follows from the Uniform Boundedness Principle. The third conclusion follows directly from the inequality
$$
\left\|<x>^p\Pi^\alpha_A\left(P_tf(t)-f(t)+it\mathfrak{Op}^A(h)f(t)\right)\right\|\leq C_{p,q}(f)|t|,\quad\text{for }|t|\leq1,|\alpha|=q,
$$
that can be obtained by induction on $p+q$ using similar arguments as in the proof of (\ref{ind}) and the explicit form of the derivative of the map
$$
\mathbb{R}\ni t\mapsto P_tf(t)-f(t)+it\mathfrak{Op}^A(h)f(t)\in L^2(\mathcal{X}).
$$
\end{proof}

\begin{proofT}
We introduce some more notations. We consider a fixed sequence
$\{d_k\}_{k\in\mathbb{N}^*}$ from $S^+_\rho(\Xi)$ and multi-indices of various lengths:
$$
\mathcal{N}^*:=\left\{\alpha=(\alpha_1,\ldots,\alpha_j)\in(\mathbb{N}^*)^j,\
j\geq1,\
\alpha_1<\ldots<\alpha_j\right\},\qquad\mathcal{N}:=\mathcal{N}^*\cup\{0\};
$$
$$
\alpha=(\alpha_1,\ldots,\alpha_j)\in\mathcal{N}^*\quad\Rightarrow\quad\parallel\!\!\!\!=\alpha:=j;\qquad
\parallel\!\!\!\!=0=0.
$$
For two multi-indices $\alpha$ and $\beta$ from $\mathcal{N}$ we say that $\beta\subset\alpha$ when $\parallel\!\!\!\!=\beta<\
\parallel\!\!\!\!=\alpha$ and $\beta_l\in\{\alpha_1,\ldots\alpha_{\parallel\!\!\!\!=\alpha}\}$.
Then we set 
$$
\mathbb{K}^t_d[T]:=\mathfrak{Op}^A(d\circ\Phi_{-t})\cdot T-T\cdot\mathfrak{Op}^A(d),\quad\text{for any }T\in\mathbb{B}(L^2(\mathcal{X}))
$$
and for $j=\,\parallel\!\!\!\!=\alpha$
$$
\boldsymbol{Q}_\alpha(t):=\mathbb{K}^t_{ d_{\alpha_j}}\ldots\mathbb{K}^t_{ d_{\alpha_1}}
\big[P_t\big]\in\mathbb{B}(\mathcal{S}(\mathcal{X})),\qquad\boldsymbol{Q}_0(t):=P_t\in\mathfrak{C}^B(\Xi).
$$
Due to our previous Lemma the maps $\mathbb{R}\ni t\mapsto \boldsymbol{Q}_\alpha(t)\in\mathbb{B}(\mathcal{S}(\mathcal{X}))$ are differentiable. Using induction on $\parallel\!\!\!\!=\alpha$ we shall now prove that
\begin{equation}\label{Qalpha}
\left\{
\begin{array}{l}
i\partial_t\boldsymbol{Q}_\alpha(t)=\mathfrak{Op}^A(h)\cdot\boldsymbol{Q}_\alpha(t)+\sum\limits_{\beta\in\mathcal{N},\beta\subsetneq\alpha}
\mathfrak{Op}^A(\mathfrak{r}_{\alpha\beta})\cdot\boldsymbol{Q}_\beta(t),\\
\\
\text{$\mathfrak{r}_{\alpha\beta}\in S^0_1(\Xi)$ and depend continuously on $t\in\mathbb{R}$.}
\end{array}
\right.
\end{equation}

For $\alpha=0\in\mathcal{N}$ (\ref{Qalpha}) is true with $\mathfrak{r}_0=0$, being the definition of $P_t$.

Now suppose that we have proved (\ref{Qalpha}) for any $\alpha\in\mathcal{N}$ with $\parallel\!\!\!\!=\alpha\leq s\in\mathbb{N}$ and let us choose
$\alpha_{s+1}\in\mathbb{N}^*$ such that $\alpha_{s+1}\geq1$ for $s=0$ or $\alpha_{s+1}>\alpha_s$ for $s>0$ and denote by
$\widetilde{\alpha}:=(\alpha_1,\ldots,\alpha_s,\alpha_{s+1})$. We differentiate the equality
$$
\boldsymbol{Q}_{\widetilde{\alpha}}(t)=\mathbb{K}^t_{ d_{\alpha_{s+1}}}\left[\boldsymbol{Q}_\alpha(t)\right]=
\left( \mathfrak{Op}^A(d_{\alpha_{s+1}}\circ
P_(-t))\right)\cdot\boldsymbol{Q}_\alpha(t)-\boldsymbol{Q}_\alpha(t)\cdot\mathfrak{Op}^A(d_{\alpha_{s+1}})
$$
and use (\ref{Qalpha}), Lemma \ref{7-4} and the equalities
$$
\partial_t(d_{\alpha_{s+1}}\circ P_-t)=-\left\{h,d_{\alpha_{s+1}}\right\}\circ\Phi_{-t}=\left\{\left(d_{\alpha_{s+1}}\circ\Phi_{-t}\right),h\right\}.
$$
 
In order to finish our proof it is sufficient to prove by induction on $\parallel\!\!\!\!=\alpha$ with $\alpha\in\mathcal{N}$ the fact that uniformly for $t$ in bounded sets we have
\begin{equation}\label{concl}
 \boldsymbol{Q}_\alpha(t)\in \mathbb{B}(L^2(\mathcal{X})).
\end{equation}
The case $\parallel\!\!\!\!=\alpha=0$ is evident. Let $\alpha\in\mathcal{N}*$ and suppose that (\ref{concl}) is true for any $\beta\in\mathcal{N}$ with $\beta\subset\alpha$. Let us denote by $\mathfrak{R}_\alpha(t)$ the sum appearing in the right-hand side of (\ref{Qalpha}). The map
$$
\mathbb{R}\ni t\mapsto\mathfrak{R}_\alpha(t)\in\mathbb{B}(\mathcal{S}(\mathcal{X}))
$$
is continuous and $\mathfrak{R}_\alpha(t)\in\mathbb{B}(L^2(\mathcal{X}))$ uniformly for $t$ in bounded sets. AA direct computation using (\ref{Qalpha}) and the fact that $\Phi_0$ is the identity on $\Xi$ shows that
$$
\frac{d}{dt}\left(P_{-t}\boldsymbol{Q}_\alpha(t)\right)=P_{-t}\mathfrak{R}_\alpha(t),\qquad\boldsymbol(Q)_\alpha(0)=0.
$$
We conclude that
$$
\boldsymbol{Q}_\alpha(t)=\int_0^t P_{t-s}\mathfrak{R}_\alpha(s)\,ds
$$
and thus we obtain (\ref{concl}).
\end{proofT}

\section{Appendix}

\subsection{Regularization procedure}\label{reg-proc}
\begin{lemma}\label{conv-marg}
Let $\mathcal{Y}$ be a finite dimensional real space. Suppose given
$p\in\mathbb{R}$ and $q\in\mathbb{N}$, then we consider the weight
$$
\left\{F\in\mathcal{S^\prime}(\mathcal{Y})\left|\partial^\alpha F\in L^1_{loc}(\mathcal{Y}),\;|\alpha|\leq q\right.\right\}\ni
F\mapsto\nu(F):=\sum_{|\alpha|\leq
q}\sup_{y\in\mathcal{Y}}<y>^p\left|(\partial^\alpha
F)(y)\right|\in\mathbb{R}_+
$$
and the linear spaces:
$$
\mathfrak{L}_\nu:=\left\{F\in\mathcal{S}(\mathcal{Y})^\prime\left|\partial^\alpha F\in L^1_{loc}(\mathcal{Y}),\;|\alpha|\leq q\right.,\;\nu(F)<\infty\right\}
$$
$$
\mathfrak{B}_\nu:=\left\{F\in\mathcal{S^\prime}(\mathcal{Y})\mid\exists
\{\phi_m\}_{m\in\mathbb{N}}\subset\mathcal{S}(\mathcal{Y}),{\rm s.th.}(\phi_m,\varphi)\underset{m\rightarrow
\infty}{\longrightarrow}(F,\varphi),
\forall\varphi\in\mathcal{S}(\mathcal{Y}),\nu(\phi_m)\leq C,\forall
m\in\mathbb{N} \right\}.
$$
Then we have: $\mathfrak{L}_\nu=\mathfrak{B}_\nu$.
\end{lemma}
\begin{proof}
Let $n$ be the dimension of $\mathcal{Y}$. If
$F\in\left\{F\in\mathcal{S^\prime}(\mathcal{Y})\left|\partial^\alpha F\in L^1_{loc}(\mathcal{Y}),\;|\alpha|\leq q\right.\right\}$ is such that
$\nu(F)<\infty$, we choose the cut-off function
$\chi\in\mathcal{D}(\mathcal{Y})$ with $\chi(0)=1$ and define
$\chi_m(y):=\chi(m^{-1}y)$, and we also choose the regularizing function
$\theta\in\mathcal{S}(\mathcal{Y})$ with
$\int_\mathcal{Y}dy\,\theta(y)=1$, and define $\theta_m(y):=m^n\theta(my)$ and
$\phi_m:=\chi_m(\theta_m*F)\in\mathcal{S}(\mathcal{Y})$. Then it is
straightforward to verify that given $\epsilon>0$
$$
(\phi_m,\varphi)\underset{m\rightarrow\infty}{\longrightarrow}(F,\varphi),
\forall\varphi\in\mathcal{S}(\mathcal{Y}),
$$
$$
\nu(\phi_m)=\sum_{|\alpha|\leq
q}\sup_{y\in\mathcal{Y}}<y>^p\left|(\partial^\alpha
\chi_m(\theta_m*F))(y)\right|\leq
$$
$$
\leq\sum_{|\alpha|\leq
q}\sum_{\beta\leq\alpha}C_{\beta}^{\alpha}\sup_{y\in\mathcal{Y}}<y>^p
m^{-|\alpha-\beta|}\left|(\partial^{\alpha-\beta}\chi)_m\left(\theta_m*(\partial^{\beta}F)\right)\right|\leq
$$
$$
\leq C\sum_{|\alpha|\leq q}\sum_{|\beta|\leq
q-|\alpha|}\sup_{y\in\mathcal{Y}}<y>^p
m^{-|\beta|}\left|(\partial^{\beta}\chi)_m\left(\theta_m*(\partial^{\alpha
}F)\right)\right|\leq C^\prime\nu(F),
$$
for $m$ large enough. Thus $\mathfrak{L}_\nu\subset\mathfrak{B}_\nu$\\
For the reversed inclusion suppose we are given
$F\in\mathcal{S^\prime}(\mathcal{Y})$ such that there exists an
approximating sequence
$\{\phi_m\}_{m\in\mathbb{N}}\subset\mathcal{S}(\mathcal{Y})$ as in
the definition of the space $\mathfrak{B}_\nu$. By the usual
properties of tempered distributions it follows that
$\partial^\alpha\phi_m$ converges to $\partial^\alpha F$ in the
sense of distributions for any multiindex $\alpha$ and also that the
product $<y>^p\partial^\alpha\phi_m$ converges in the sense of
distributions to $<y>^p\partial^\alpha F$ that is a well defined
distribution for any $p\in\mathbb{R}$. As $\mathcal{S}(\mathcal{Y})$
is dense in $L^1(\mathcal{Y})$ we conclude from the definition of
$\mathfrak{B}_\nu$ that the sequences
$\{<y>^p\partial^\alpha\phi_m\}_{m\in\mathbb{N}}$, for any $|\alpha|\leq q$, belong to a
finite ball of the space $L^\infty(\mathcal{Y})$ that is the dual
of $L^1(\mathcal{Y})$ and thus the sequence
$\{<y>^p\partial^\alpha\phi_m\}_{m\in\mathbb{N}}$ has accumulation
points in the ball $\|.\|_\infty\leq(1+\epsilon)\nu(F)$, and due to
the weak convergence to $<y>^p\partial^\alpha F$, this distribution
must belong to the above ball $\|.\|_\infty\leq(1+\epsilon)\nu(F)$.
We conclude thus that $\mathfrak{B}_\nu\subset\mathfrak{L}_\nu$.
\end{proof}

\subsection{Some composition formulae}

\begin{lemma}\label{phie}
 For $\varphi\in BC(\mathcal{X})$ and $U\in\Xi$, we have
$$
(\varphi\dB\e U)(Z)=\varphi(z-u/2)\e U(Z),\qquad (\e U\dB\varphi)(Z)=\varphi(z+u/2)\e U(Z).
$$
\end{lemma}
\begin{proof}
By direct computation we get
$$
(\varphi\dB\e U)(Z)=\pi^{-2n}\int_{\Xi}dZ_1\int_{\Xi}dZ_2\,e^{-2i\sigma(Z-Z_1,Z-Z_2)}\Omega^B[\mathcal{T}
(z,z_1,z_2)]\varphi(z_1)e^{-i\sigma(U,Z_2)}=
$$
$$
=\pi^{-2n}\int_{\Xi}dZ_1\int_{\Xi}dZ_2\,e^{-2i\{<\zeta-\zeta_1,z-z_2>-<\zeta-\zeta_2,z-z_1>+<\mu,z_2/2>-
<\zeta_2/2,u>\}}\varphi(z_1)\Omega^B[\mathcal{T}(z,z_1,z_2)]=
$$
$$
=\pi^{-2n}\int_{\Xi}dZ_1\int_{\Xi}dZ_2\,e^{2i\{<\zeta_1,z-z_2>-<\zeta_2,z-z_1-u/2>-<\zeta,z-z_2>+<\zeta,z-z_1>-
<\mu,z_2/2>\}}\varphi(z_1)\Omega^B[\mathcal{T}(z,z_1,z_2)].
$$
Integration in $\zeta_1$ and usual inverse Fourier formula for the Dirac mass gives $z_2=z$;
similarly integration in $\zeta_2$ implies $z_1=z-u/2$. Thus, taking into account that the triangle
$\mathcal{T}(z,z,z-u/2)$ is degenerate, we get:
$$
(\varphi\dB\e U)(Z)=\varphi(z-u/2)e^{-i\sigma(U,Z)}\Omega^B[\mathcal{T}(z,z,z-u/2)]=\varphi(z-u/2)\e U(Z).
$$
Similarly:
$$
(\e U\dB\varphi)(Z)=\pi^{-2n}\int_{\Xi}dZ_1\int_{\Xi}dZ_2\,e^{-2i\sigma(Z-Z_1,Z-Z_2)}\Omega^B[\mathcal{T}
(z,z_1,z_2)]e^{-i\sigma(U,Z_1)}\varphi(z_2)=
$$
$$
=\pi^{-2n}\int_{\Xi}dZ_1\int_{\Xi}dZ_2\,e^{-2i\{<\zeta-\zeta_1,z-z_2>-<\zeta-\zeta_2,z-z_1>+<\mu,z_1/2>-
<\zeta_1/2,u>\}}\varphi(z_2)\Omega^B[\mathcal{T}(z,z_1,z_2)]=
$$
$$
=\pi^{-2n}\int_{\Xi}dZ_1\int_{\Xi}dZ_2\,e^{2i\{<\zeta_1,z-z_2+u/2>-<\zeta_2,z-z_1>-<\zeta,z-z_2>+<\zeta,z-z_1>-
<\mu,z_1/2>\}}\varphi(z_2)\Omega^B[\mathcal{T}(z,z_1,z_2)]=
$$
$$
=\varphi(z+u/2)e^{-i\sigma(U,Z)}\Omega^B[\mathcal{T}(z,z+u/2,z)]=\varphi(z+u/2)\e U(Z).
$$
\end{proof}

\paragraph{Proof of Lemma \ref{chiau}}
By direct computation we obtain
$$
(\e X\dB\e Y)(Z)=\pi^{-2n}\int_{\Xi}dU_1\int_{\Xi}dU_2\,e^{-2i\sigma(Z-U_1,Z-U_2)}
\Omega^B[\mathcal{T}(z,u_1,u_2)]\,e^{-i\sigma(X,U_1)}\,e^{-i\sigma(Y,U_2)}=
$$
$$
=\pi^{-2n}\int_{\Xi}dU_1\int_{\Xi}dU_2\,e^{-2i\mathfrak{s}(X,Y;Z,U_1,U_2)}\Omega^B[\mathcal{T}(z,u_1,u_2)],
$$
where we have introduced the shorthand notation
$$
\mathfrak{s}(X,Y;Z,U_1,U_2):=<\zeta-\mu_1,z-u_2>-<\zeta-\mu_2,z-u_1>+
$$
$$
+<\xi,u_1/2>-<\mu_1/2,x>+<\eta,u_2/2>-<\mu_2/2,y>=
$$
$$
=<\zeta,z-u_2>-<\zeta,z-u_1>+<\xi,u_1/2>+<\eta,u_2/2>-
$$
$$
-<\mu_1,z+x/2-u_2>+<\mu_2,z-y/2-u_1>.
$$
Integration with respect to $\mu_1$ and $\mu_2$ implies $u_2=z+x/2$ and $u_1=z-y/2$ and the phase of the exponential factor will be:
$$
-<\zeta,x/2>+<\zeta,y/2>+<\xi,u_1/2>+<\eta,u_2/2>=
$$
$$
=-<\zeta,x/2>-<\zeta,y/2>+<\xi,z/2>-<\xi,y/4>+<\eta,z/2>+<\eta,x/4>=
$$
$$
=-(1/2)\sigma(Z,X+Y)+(1/4)\sigma(Y,X).
$$
Therefore
$$
(\e X\dB\e Y)(Z)=e^{-i\sigma(X+Y,Z)}\,e^{(i/2)\sigma(X,Y)}\Omega^B[\mathcal{T}(z,z-y/2,z+x/2)]=
$$
$$
=e^{(i/2)\sigma(X,Y)}\Omega^B[\mathcal{T}(z,z-y/2,z+x/2)]\e{X+Y}(Z),
$$
and the first equality is obtained.
The second and the third follow immediately from Lemma \ref{phie}.

\subsection{Proof of Proposition \ref{L-inf-p}}

1) For $(f,g)\in\left[BC(\mathcal{X};L^1(\mathcal{X}^\prime))\right]^2$ and for any $x\in\mathcal{X}$
we can define $f(x,.)\star g(x,.)\in L^1(\mathcal{X}^\prime)$ depending continuously on $(f,g)\in\left[BC(\mathcal{X};L^1(\mathcal{X}^\prime))\right]^2$.
We can verify that
$$
\underset{x\in\mathcal{X}}{\sup}\left\|f(x,.)\star g(x,.)\right\|_1\,\leq\,\|f\|_{\infty,1}\,\|g\|_{\infty,1}.
$$

\noindent
2) In a similar way, using the Hausdorff-Young inequality, for $(f,F)\in BC(\mathcal X;L^1(\mathcal X^\prime))
\times BC(\mathcal X;L^p(\mathcal X^\prime))$ and for any $x\in\mathcal{X}$ we can define $f(x,.)\star g(x,.)\in L^p(\mathcal{X}^\prime)$
and prove that
$$
\underset{x\in\mathcal{X}}{\sup}\left\|f(x,.)\star g(x,.)\right\|_p\,\leq\,\|f\|_{\infty,1}\,\|g\|_{\infty,p}.
$$
For the case $(f,g)\in BC(\mathcal X;L^1(\mathcal X^\prime))\times L^p(\Xi)$ let us first suppose that $p<\infty$
and $g\in\mathcal{S}(\Xi)$. Then for any $x\in\mathcal{X}$ we can define
$$
(f\star g)(x,\xi)\,=\,\int_{\mathcal{X}^\prime}d\eta\,f(x,\xi-\eta)g(x,\eta),
$$
and we remark that
$$
\left|(f\star g)(x,\xi)\right|^p\,\leq\,\|f(x,\cdot)\|_1^{p-1}\int_{\mathcal{X}^\prime}d\eta\,|f(x,\xi-\eta)|
\,|g(x,\eta)|^p\,\leq\,\|f\|_{\infty,1}^{p-1}\int_{\mathcal{X}^\prime}d\eta\,|f(x,\xi-\eta)|\,|g(x,\eta)|^p,
$$
$$
\int_{\mathcal{X}}dx\int_{\mathcal{X}^\prime}d\xi\,\left|(f\star g)(x,\xi)\right|^p\,\leq\,\|f\|_{\infty,1}^{p-1}
\int_{\mathcal{X}}dx\int_{\mathcal{X}^\prime}d\xi\int_{\mathcal{X}^\prime}d\eta\,|f(x,\xi-\eta)|\,|g(x,\eta)|^p\,=
$$
$$
=\|f\|_{\infty,1}^{p-1}\int_{\mathcal{X}^\prime}d\eta\int_{\mathcal{X}}dx\int_{\mathcal{X}^\prime}d\xi\,|f(x,\xi-\eta)|
\,|g(x,\eta)|^p\,\leq
$$
$$
\leq\,\|f\|_{\infty,1}^{p-1}\int_{\mathcal{X}^\prime}d\eta\left(\underset{x\in\mathcal{X}}
{\sup}\|f(x,.)\|_1\right)\int_{\mathcal{X}}dx\,|g(x,\eta)|^p\,=\,\|f\|_{\infty,1}^{p}\,\|g\|_p^p.
$$
The case $g\in L^p(\Xi)$ is now obtained using the density of $\mathcal{S}(\Xi)$ in $L^p(\Xi)$. For $p=\infty$
we simply observe that for any $x\in\mathcal{X}$:
$$
|f(x,.)\star g(x,.)|\,\leq\,\|g\|_\infty\|f(x,\cdot)\|_1\,\leq\,\|g\|_\infty\|f\|_{\infty,1}.
$$

\noindent
3) We observe that
$$
<\xi>^{-m+|\alpha|\rho}|(\partial^a_x\partial^\alpha_\xi(f\star g))(X)|\,\leq
$$
$$
\leq\,\sum_{|b|\leq|a|}c_{a,b}\left|\int_{\mathcal{X}^\prime}d\eta\,<\eta>^{|m-\rho|\alpha||}
\left[\partial^b_x f(x,\eta)\right]\,<\xi-\eta>^{-m+\rho|\alpha|}\left[\partial^{a-b}_x\partial^\alpha_\xi
g(x,\xi-\eta)\right] \right|\,\leq
$$
$$
\leq\,C\max_{|b|\leq|a|}\|\mathfrak{p}_{m}\partial^b_x f\|_{\infty,1}\max_{|b|\leq|a|}\max_{|\beta|\leq|\alpha|}
\sup_{(x,\xi)\in\Xi}|<\xi>^{-m+\rho|\beta|}\partial^b_x\partial^\beta_\xi g(x,\xi)|.
$$
\noindent
4) Evident.

\subsection{Multiple magnetic derivatives}

Using  Corollary \ref{cor-alg-coef} and the formulae in Section \ref{mag-deriv} we obtain
\begin{proposition}\label{est-deriv}
For $B$ with components of class $BC^\infty(\mathcal{X})$ and for any $f\in\mathcal{S}(\Xi)$ we have
\begin{enumerate}
 \item $
\|D_x^a D_\xi^\alpha f\|_\infty\leq C_{|a|}\sum_{b\leq
a}\,\sum_{|\beta|\leq
(|a|-|b|)(2[n/2]+3)}\|(\ad^B_{\boldsymbol{e}})^b(\ad^B_{\boldsymbol{\epsilon}})^{\alpha+\beta}f\|_\infty,
       $
 \item $
\|(\ad^B_{\boldsymbol{e}})^a(\ad^B_{\boldsymbol{\epsilon}})^\alpha
f\|_\infty\leq C_{|a|}\sum_{b\leq a}\,\sum_{|\beta|\leq
(|a|-|b|)(2[n/2]+3)}\|D_x^bD_\xi^{\alpha+\beta}f\|_\infty,
       $
\end{enumerate}
where $\boldsymbol{e}:=\{e_1,\ldots,e_n\}$ and $\boldsymbol{\epsilon}:=\{\epsilon_1,\ldots,\epsilon_n\}$ are the
canonical basis in $\mathcal{X}$ and $\mathcal{X}^\prime$, respectively.
\end{proposition}
\begin{proof}
We first remark that
$$
D_{\xi_j}f=\epsilon_j\cdot D_\xi f=\ad^B_{\epsilon_j}f;\qquad
D_{x_j}f=e_j\cdot D_x f=\ad^B_{e_j}f+\delta^B_j f,
$$
so that
$$
D_\xi^\alpha f=(\epsilon_1\cdot D_\xi)^{\alpha_1}\cdots(\epsilon_n\cdot
D_\xi)^{\alpha_n}f=(\ad^B_{\epsilon_1})^{\alpha_1}\cdots(\ad^B_{\epsilon_n})^{\alpha_n}f\equiv
(\ad^B_{\boldsymbol{\epsilon}})^\alpha f.
$$
Next we use Proposition \ref{L-inf-p} and Corollary \ref{cor-alg-coef} in order to prove that
$$
\|D_{x_j}f\|_\infty=\|e_j\cdot D_x f\|_\infty\leq\|\ad^B_{e_j} f\|_\infty+\|\delta^B_j f\|_\infty\leq
\|\ad^B_{e_j}f\|_\infty+C_\infty\sum_{|\alpha|\leq (2[n/2]+3)}\|(\ad^B_{\boldsymbol{\epsilon}})^\alpha f\|_\infty,
$$
and similarly
\begin{equation}\label{D-x}
\|\ad^B_{e_j}f\|_\infty\leq
\|D_{x_j}f\|_\infty+C_\infty\sum_{|\alpha|\leq
(2[n/2]+3)}\|D_\xi^\alpha f\|_\infty.
\end{equation}
We continue by recurrence using Remark \ref{x-der-c} and Proposition \ref{alg-coef}.
$$
D_x^a
D_{x_j}f=D_x^a(\ad^B_{e_j}f)+D_x^a(\delta_j^Bf)=D_x^a(\ad^B_{e_j}f)+\sum\limits_{1\leq|\alpha|\leq(2[n/2]+3)}\
\sum_{b\leq a}C_b^a(D_x^{a-b}c^B_{j\alpha})\star(D_x^b\partial^\alpha_\xi
f)=
$$
$$
=D_x^a(\ad^B_{e_j}f)+\sum\limits_{1\leq|\alpha|\leq(2[n/2]+3)}\ \sum_{b\leq
a}C_b^a(c^{D_x^{a-b}B}_{j\alpha})\star(D_x^b\partial^\alpha_\xi f).
$$

Let us suppose that for $|a|\leq p$ we have proved that there
exists some finite positive constant $C_p$ such that the following estimation holds
\begin{equation}\label{ind-hyp}
\|D_x^a f\|_\infty\leq C_p \sum_{b\leq a}\,\sum_{|\beta|\leq
(|a|-|b|)(2[n/2]+3)}\|(\ad^B_{\boldsymbol{e}})^b(\ad^B_{\boldsymbol{\epsilon}})^\beta f\|_\infty.
\end{equation}
Then the previous equality (\ref{ind-hyp}) implies that, for any $j\in\{1,\ldots,n\}$, (with $(\theta_j)_k:=\delta_{jk}$),
$$
\|D_x^a D_{x_j}f\|_\infty\leq C_p \sum_{b\leq a}\,\sum_{|\beta|\leq
(|a|-|b|)(2[n/2]+3)}\|(\ad^B_{\boldsymbol{e}})^b(\ad^B_{\boldsymbol{\epsilon}})^\beta(\ad^B_{e_j}f)\|_\infty\,+\,
$$
$$
+c\left[\underset{|a|\leq p}{\max}\ \underset{|\alpha|\leq
(2[n/2]+3)}{\max}\left\|c^{(D_x^{a}B)}_{j\alpha}
\right\|_{\infty,1}\right]\ \sum_{b\leq a}\ \sum_{|\beta|\leq
(2[n/2]+3)}\|D_x^b (\ad^B_{\boldsymbol{\epsilon}})^\beta f\|_\infty\,\leq
$$
\begin{equation}\label{xxx}
\leq\,C_p\,\sum_{b\leq a}\,\sum_{|\beta|\leq
(|a|-|b|)(2[n/2]+3)}\|(\ad^B_{\boldsymbol{e}})^{(b+\theta_j)}(\ad^B_{\boldsymbol{\epsilon}})^\beta
f\|_\infty+
\end{equation}
$$
+\,C_p^\prime C_p\,\sum_{b\leq a}\ \sum_{1\leq|\beta|\leq
(2[n/2]+3)}\sum_{c\leq b}\
\sum_{|\gamma|\leq(|b|-|c|)(2[n/2]+3)}\|(\ad^B_{\boldsymbol{e}})^c(\ad^B_{\boldsymbol{\epsilon}})^\gamma
(\ad^B_{\boldsymbol{\epsilon}})^\beta f\|_\infty.
$$
One can find a constant $K_p$, depending only on $p\in\mathbb{N}$
and on the dimension $n$ of $\mathcal{X}$, such that the expression in (\ref{xxx}) may be estimated by
$$
C_p\,\sum_{b\leq a}\,\sum_{|\beta|\leq
(|a|-|b|)(2[n/2]+3)}\|(\ad^B_{\boldsymbol{e}})^{(b+\theta_j)}(\ad^B_{\boldsymbol{\epsilon}})^\beta
f\|_\infty+\,K_pC_p^\prime C_p\,\sum_{b\leq a}
\sum_{|\beta|\leq(1+|a|-|b|)(2[n/2]+3)}\|(\ad^B_{\boldsymbol{e}})^b
(\ad^B_{\boldsymbol{\epsilon}})^\beta f\|_\infty\leq
$$
$$
\leq C_p\left(1+ K_pC_p^\prime\right)\sum_{b'\leq a'}\,\sum_{|\beta|\leq
(|a'|-|b'|)(2[n/2]+3)}\|(\ad^B_{\boldsymbol{e}})^{(b')}(\ad^B_{\boldsymbol{\epsilon}})^\beta f\|_\infty,
$$
where $a':=a+\theta_j$, so that $|a'|=|a|+1$. Thus, taking
$C_{p+1}=C_p\left(1+ K_pC_p^\prime\right)$, we obtain the condition
(\ref{ind-hyp}) for $|a|=p+1$ and the statement of point 1 of the
Proposition for $|\alpha|=0$. Replacing then $f$ with $\partial^\alpha_\xi
f=(\mathfrak{ad}^B_{\boldsymbol{\epsilon}})^\alpha f$, we completely
prove the assertion of point 1 of the Proposition. Replacing the
norms $\|.\|_\infty$ with the norms $\|.\|_2$ makes no changes, so we also have proved point 2 of the Proposition.

Let us observe now that we can also write
$$
(\ad^B_{\boldsymbol{e}})^a \ad^B_{e_j}f=(\ad^B_{\boldsymbol{e}})^a
(D_{x_j}f)+(\ad^B_{\boldsymbol{e}})^a
(\delta_j^Bf)=(\ad^B_{\boldsymbol{e}})^a
(D_{x_j}f)+\sum\limits_{|\alpha|\leq(2[n/2]+3)}(\ad^B_{\boldsymbol{e}})^a
\left(c_{j\alpha}^B\star \partial^\alpha_\xi f\right),
$$
and a similar induction procedure allows us to prove points 3 and 4 of the Proposition.
\end{proof}

\subsection{Composition of symbols}\label{s-comp-symb}

In order to estimate commutators of symbols we shall need to control the rest in the Taylor series and we consider for $\phi\in C^\infty(\Xi)$
$$
\phi_{s}(X,Y):=\phi(X+s(Y-X)),\qquad\text{for}\ s\in[0,1].
$$
Let us suppose that $\phi\in S^m_\rho(\Xi)$ for some $m\in\mathbb{R}$. Then
$$
|\phi_s(X,Y)|\leq C<\xi+s(\eta-\xi)>^m.
$$

\begin{proposition}\label{comp-symb}
 Suppose we are given $\phi\in S^m_\rho(\Xi)$, $\psi\in S^p_\rho(\Xi)$ and $\theta\in BC^\infty(\mathcal{X};C^\infty_{\text{\sf pol}}(\mathcal{X}^2))$ (the bounded smooth functions on $\mathcal{X}$ with values in the space of smooth functions on $\mathcal{X}^2$ with polynomial growth together with their derivatives). Then
$$
\mathfrak{L}_\sigma(\theta;\phi,\psi;s)(X):=\int_\Xi dY\int_\Xi dZ e^{-2i\sigma(X-Y,X-Z)}\theta(x,y-x,z-x)\phi_s(X,Y)\psi(Z)
$$
defines a symbol of class $S^{m+p}_\rho(\Xi)$ for any $s\in[0,1]$ and the mapping
$$
S^m_\rho(\Xi)\times S^p_\rho(\Xi)\ni(\phi,\psi)\mapsto\mathfrak{L}_\sigma(\theta;\phi,\psi;s)\in S^{m+p}_\rho(\Xi)
$$
is continuous; all is uniform with respect to $s\in[0,1]$.
\end{proposition}
\begin{proof}
 We use integration by parts observing once again that
$$
(y_j-x_j)\,e^{-2i\sigma(X-Y,X-Z)}=\frac{1}{2i}\,\partial_{\zeta_j}e^{-2i\sigma(X-Y,X-Z)},
$$
$$
(\eta_j-\xi_j)\,e^{-2i\sigma(X-Y,X-Z)}=-\frac{1}{2i}\,\partial_{z_j}e^{-2i\sigma(X-Y,X-Z)},
$$
$$
(z_j-x_j)\,e^{-2i\sigma(X-Y,X-Z)}=-\frac{1}{2i}\,\partial_{\eta_j}e^{-2i\sigma(X-Y,X-Z)},
$$
$$
(\zeta_j-\xi_j)\,e^{-2i\sigma(X-Y,X-Z)}=\frac{1}{2i}\,\partial_{y_j}e^{-2i\sigma(X-Y,X-Z)},
$$
so that we have the identity
\begin{equation}\label{int-osc}
e^{-2i\sigma(X-Y,X-Z)}=
\left(\frac{1-i<(\xi-\zeta),\partial_y>}{1+2|\xi-\zeta|^2}\right)^{N_2}
\left(\frac{1+i<(\xi-\eta),\partial_z>}{1+2|\xi-\eta|^2}\right)^{N_1}\times
\end{equation}
$$
\times
\left(\frac{1+i<(x-z),\partial_\eta>}{1+2|x-z|^2}\right)^{M_2}\left(\frac{1-i<(x-y),\partial_\zeta>}
{1+2|x-y|^2}\right)^{M_1}\,e^{-2i\sigma(X-Y,X-Z)}
$$
for any exponents $N_1$, $N_2$, $M_1$, $M_2$. Then we start by considering $\phi$ and $\psi$ as test functions and we integrate by parts. Due to our hypothesis we easily obtain the estimation
$$
\left|\mathfrak{L}_\sigma(\theta;\phi,\psi;s)(X)\right|\leq
$$
$$
\leq C\left(\int_\Xi dY  <\xi-\eta>^{-N_1}<\xi+s(\eta-\xi)>^m<x-y>^{r_1(N_1,N_2)-M_1}\right)\times
$$
$$
\times\left(\int_\Xi dZ<\xi-\zeta>^{-N_2}<\zeta>^p<x-z>^{r_2(N_1,N_2)-M_2}\right)\leq
$$
$$
\leq C^\prime<\xi>^{m+p},\qquad \text{uniformly in}\ s\in[0,1],
$$
where we choose $N_1>|m|+n$, $N_2>|p|+n$, $M_1>r_1(N_1,N_2)+n$ and $M_2>r_2(N_1,N_2)+n$, with $r_j(N_1,N_2)$ the powers dominating $\partial_z^{N_1}\partial_y^{N_2}\theta(x,y-x,z-x)$.
Let us compute now the $\xi$-derivative of $\mathfrak{L}_\sigma(\theta;\phi,\psi)$
$$
\left(\partial_{\xi_j}\mathfrak{L}_\sigma(\theta;\phi,\psi;s)\right)(X)=
$$
$$
=-\int_\Xi dY\int_\Xi dZ \left[(\partial_{\eta_j}+\partial_{\zeta_j})e^{-2i\sigma(X-Y,X-Z)}\right]\theta(x,y-x,z-x)\phi_s(X,Y)\psi(Z)+
$$
$$
+\int_\Xi dY\int_\Xi dZ e^{-2i\sigma(X-Y,X-Z)}\theta(x,y-x,z-x)\left[(1-s)\left(\partial_{\xi_j}\phi\right)_s(X,Y)\right]\psi(Z)=
$$
$$
=\int_\Xi dY\int_\Xi dZ e^{-2i\sigma(X-Y,X-Z)}\theta(x,y-x,z-x)\left[s\left(\partial_{\xi_j}\phi\right)_s(X,Y)\right]\psi(Z)+
$$
$$
+\int_\Xi dY\int_\Xi dZ e^{-2i\sigma(X-Y,X-Z)}\theta(x,y-x,z-x)\phi_s(X,Y)\left[\left(\partial_{\zeta_j}\psi\right)(Z)\right]+
$$
$$
+\int_\Xi dY\int_\Xi dZ e^{-2i\sigma(X-Y,X-Z)}\theta(x,y-x,z-x)\left[(1-s)\left(\partial_{\xi_j}\phi\right)_s(X,Y)\right]\psi(Z)=
$$
$$
=\mathfrak{L}_\sigma(\theta;(\partial_{\xi_j}\phi),\psi;s)(X)+\mathfrak{L}_\sigma(\theta;\phi,(\partial_{\xi_j}\psi);s)(X).
$$

Considering the $x$-derivative we obtain in a similar way that
$$
\left(\partial_{x_j}\mathfrak{L}_\sigma(\theta;\phi,\psi;s)\right)(X)=
$$
$$
=\mathfrak{L}_\sigma(\theta;(\partial_{x_j}\phi),\psi;s)+\mathfrak{L}_\sigma(\theta;\phi,(\partial_{x_j}\psi);s)+\mathfrak{L}_\sigma(\tilde{\theta};\phi,\psi;s)(X)
$$
where $\tilde{\theta}(x,y-x,z-x):=\partial_{x_j}\theta(x,y-x,z-x)$.
\end{proof}

\subsection{Some confinement results}
\label{confinement}

In this Appendix we include some technical results inspired by \cite{BC}. In fact, as we only need some very particular case of the results in \cite{BC}, we prefered to include here some complete proofs for these simpler Lemmas.

\begin{lemma}\label{confinare}
Suppose given the family $\{F_{(x,y)}\}_{_{(x,y)\in\mathcal{X}\times\mathcal{X}}}\subset \mathcal{S}^m_0(\Xi)$ for some $m\in\mathbb{R}$, uniformly for $(x,y)\in\mathcal{X}\times\mathcal{X}$ and $\chi\in C^\infty_0(\mathcal{X})$ with
$\text{\rm supp}\chi\subset B_R(0)$. For $|x-y|>2R$ and for any $N\in\mathbb{N}$ we have
$$
<x-y>^N\tau_{x}[\chi]\sharp F_{(x,y)}\sharp\tau_{y}[\chi]\in S^m_0(\Xi)
$$
uniformly in $x$ and $y$ in the given domain of $\mathcal{X}\times\mathcal{X}$.
\end{lemma}
\begin{proof}
By the Theorem on composition of symbols $\tau_{x}[\chi]\sharp F_{(x,y)}\sharp\tau_{y}[\chi]$ is a symbol of
type $S^m_0(\Xi)$ and we have
$$
\left\{\tau_{x}[\chi]\sharp
F_{(x,y)}\sharp\tau_{y}[\chi]\right\}(Z)=\pi^{-2n}\int_{\Xi}dZ_1\int_{\Xi}dZ_2\,
e^{-2i\sigma(Z-Z_1,Z-Z_2)}\chi(z_1-x)\left\{F_{(x,y)}\sharp\tau_{y}[\chi]\right\}(Z_2)=
$$
$$
=\pi^{-3n}\int_{\mathcal{X}}dz_1\int_{\mathcal{X}^\prime}d\zeta_2\int_{\Xi}dZ_3\int_{\Xi}dZ_4\,
\left.e^{2i(z-z_1,\zeta-\zeta_2)}e^{-2i\sigma(Z_2-Z_3,Z_2-Z_4)}\right|_{z_2=z}\chi(z_1-x)\chi(z_4-y)F_{(x,y)}(Z_3)=
$$
$$
=\pi^{-2n}\int_{\mathcal{X}}dz_1\int_{\mathcal{X}^\prime}d\zeta_2\int_{\mathcal{X}}dz_4\int_{\mathcal{X}^\prime}d\zeta_3
\,e^{2i(z-z_1,\zeta-\zeta_2)}e^{-2i(z-z_4,\zeta_2-\zeta_3)}\chi(z_1-x)\chi(z_4-y)F_{(x,y)}(z,\zeta_3)=
$$
$$
=\pi^{-2n}\int_{\mathcal{X}}dz_1\int_{\mathcal{X}^\prime}d\zeta_2\int_{\mathcal{X}}dz_4\int_{\mathcal{X}^\prime}d\zeta_3
\,e^{2i(z_1,\zeta-\zeta_2)}e^{-2i(z_4,\zeta_2-\zeta_3)}\chi(z-z_1-x)\chi(z-z_4-y)F_{(x,y)}(z,\zeta_3)=
$$
$$
=\pi^{-n}\int_{\mathcal{X}}dz_1\int_{\mathcal{X}^\prime}d\zeta_3\,
e^{2i(z_1,\zeta-\zeta_3)}\chi(z-z_1-x)\chi(z+z_1-y)F_{(x,y)}(z,\zeta_3)=
$$
\begin{equation}\label{quasi-loc}
=\pi^{-n}\int_{\mathcal{X}}du\,
e^{i(u,\zeta)}\chi(z-u/2-x)\chi(z+u/2-y)\,\mathcal{F}_2[F_{(x,y)}(z,\cdot)](-u),
\end{equation}
where $\mathcal{F}_2$ is the Fourier transform with respect to the second variable.
The hypothesis $F_{(x,y)}\in \mathcal{S}^m_0(\Xi)$ implies that for fixed $z\in\mathcal{X}$,
$\mathcal{F}_2[F_{(x,y)}(z,\cdot)]$ is a tempered distribution having rapid decay (i.e. extending to a continuous linear functional on $C^\infty_{pol}(\mathcal{X})$) and such
that for any $\varphi\in C^\infty_{pol}(\mathcal{X})$ the map $\mathcal{X}\ni z\mapsto
<\mathcal{F}_2[F_{(x,y)}(z,\cdot)],\varphi>\in\mathbb{C}$ is of class $BC^\infty(\mathcal{X})$ uniformly for $(x,y)\in\mathcal{X}\times\mathcal{X}$.
If $\text{\rm supp}\chi\subset B_R(0)$ then the integral in (\ref{quasi-loc}) is to be taken only on the domain
$B_{2R}(y-x)$ and
$
\tau_{x}[\chi]\sharp F_{(x,y)}\sharp\tau_{y}[\chi](z,\zeta)=0
$
for $z\notin B_{2R}((x+y)/2)$.
For any $N\in\mathbb{N}$ we have
$$
<x-y>^{2N}\left\{\tau_{x}[\chi]\sharp
F_{(x,y)}\sharp\tau_{y}[\chi]\right\}(Z)=\pi^{-n}<x-y>^{2N}\int_{\mathcal{X}}du\,
e^{i(u,\zeta)}\chi(z-u/2-x)\chi(z+u/2-y)\,\mathcal{F}_2[F_{(x,y)}(z,\cdot)](u)=
$$
$$
=\pi^{-n}\int_{B_{2R}(y-x)}du\,
e^{i(u,\zeta)}\chi(z-u/2-x)\chi(z+u/2-y)\,\left(\frac{<x-y>}{<u>}\right)^{2N}\mathcal{F}_2[(1-\Delta)^NF_{(x,y)}(z,\cdot)](u)=
$$
$$
=\pi^{-n}\left\{\left(\mathcal{F}[\check{\Phi}_{x,y,z}]\right)*\left[(1-\Delta)^NF_{(x,y)}(z,\cdot)\right]\right\}(\zeta).
$$
Thus
$$
\underset{(z,\zeta)}{\sup}<\zeta>^{-m}\left|<x-y>^{2N}\left\{\tau_{x}[\chi]\sharp
F_{(x,y)}\sharp\tau_{y}[\chi]\right\}(z,\zeta)\right|\leq
$$
$$
\leq\pi^{-n}\underset{(z,\zeta)}{\sup}<\zeta>^{-m}\left|\left\{\left(\mathcal{F}[\check{\Phi}_{x,y,z}]\right)*\left[(1-\Delta)^NF_{(x,y)}(z,\cdot)\right]\right\}(\zeta)\right|\leq
$$
$$
\leq\pi^{-n}\underset{z}{\sup}\|<.>^{|m|}\Phi_{x,y,z}\|_{_1}\ \underset{(z,\zeta)}{\sup}<\zeta>^{-m}\left|\left[(1-\Delta)^NF_{(x,y)}(z,\zeta)\right]\right|\leq C
$$
uniformly in $x$ and $y$ with $|x-y|>2R$. In fact, here we have denoted by
$$
\Phi_{x,y,z}(u):=\chi(z-x-u/2)\chi(z-y+u/2)\left(\frac{<x-y>}{<u>}\right)^{2N}
$$
that is of class $C^\infty_0(\mathcal{X})$ and such that together with all its derivatives, they have bounded $L^2$-norms uniformly with respect to $x$, $y$ and $z$.
The derivatives $\partial_z^a\partial_\zeta^\alpha\left(\tau_{x}[\chi]\sharp
F_{(x,y)}\sharp\tau_{y}[\chi]\right)(Z)$ are clearly handled by completely similar arguments.
\end{proof}

\begin{lemma}\label{CS-confinare}
Assume that $\{G_x\}_{x\in\mathcal{X}}\subset S^m_0(\Xi)$ are uniformly bounded with respect to $x\in\mathcal{X}$ for the topology of $S^m_0(\Xi)$ and
$\mathfrak{v}_j$ (with j=1,2) are symbols of class $S^0_1(\Xi)$
with rapid decay in the $\mathcal{X}$-variable. Then, by denoting
$$
\mathfrak{A}_x:=\mathfrak T^B_{(x,0)}[\mathfrak{v}_1]\,\sharp^B\,
G_x\,\sharp^B\,\mathfrak T^B_{(x,0)}[\mathfrak{v}_2],
$$
the family of symbols $\mathfrak{s}^-_m\sharp^B\mathfrak{A}_x$ indexed by $x\in\mathcal{X}$, defines in any Schr\"{o}dinger
representation a family of operators $\{A_x\}_{x\in\mathcal{X}}$
that satisfies the hypothesis of the Proposition \ref{Cotlar-Stein}.
\end{lemma}
\begin{proof}
First we observe that
$$
\mathfrak{s}_m^-\sharp^B\mathfrak{A}_x=\mathfrak{s}_m^-\sharp^B\,\mathfrak T^B_{(x,0)}[\mathfrak{v}_1]\,
\sharp^B G_x\sharp^B\,\mathfrak T^B_{(x,0)}[\mathfrak{v}_2]=
$$
$$
=\mathfrak T^B_{(x,0)}\left[\mathfrak{s}_m^-\sharp^B\mathfrak{v}_1\sharp^B\mathfrak{s}_m\right]\,
\sharp^B\left[\mathfrak{s}_m^-\sharp^B G_x\right]\sharp^B
\left[\mathfrak T^B_{(x,0)}[\mathfrak{v}_2]\right],
$$
and a similar formula is valid for
$\mathfrak{A}_x\sharp^B\mathfrak{s}_m^-$. 

Remark that if $\mathfrak{v}_j$ (with j=1,2) is a symbol of class $S^0_1(\Xi)$
with rapid decay in the $\mathcal{X}$-variable, then $\mathfrak{s}_m^-\sharp^B\mathfrak{v}_1\sharp^B\mathfrak{s}_m$ is also a symbol of class $S^0_1(\Xi)$ (Theorem on composition of symbols) and has also rapid decay in the $\mathcal{X}$-variable because:
$$
<x>^p\big[\mathfrak{s}_m\sharp^B\mathfrak{w}\big](X)=
$$
$$
=\pi^{-2n}\,\int_{\Xi}\int_{\Xi}\,dY\,dZ\,\left[<x-z>^pe^{-2i\sigma(X-Y,X-Z)}\right]\Omega^B[\mathcal{T}(x,y,z)]\mathfrak{s}_m(Y)\big[<z>^p\mathfrak{w}(Z)\big]\left[\frac{<x>^p}{<x-z>^p<z>^p}\right]
$$
and we apply the usual integration by parts technique to control the growing factor $<x-z>^p$.

So we can easily reduce the proof of the Lemma to the case $m=0$. Let us compute
$$
\overline{\mathfrak{A}_y}\sharp^B\mathfrak{A}_z=\overline{\mathfrak T^B_{(y,0)}[\mathfrak{v}_2]}\,\sharp^B
\overline{G_y}\,\sharp^B\overline{\mathfrak T^B_{(y,0)}[\mathfrak{v}_1]}\,\sharp^B\,\mathfrak T^B_{(z,0)}
[\mathfrak{v}_1]\,\sharp^B G_z\,\sharp^B\,\mathfrak T^B_{(z,0)}[\mathfrak{v}_2].
$$
We have then:
$$
\underset{y\in\mathcal{X}}{\sup}\int_{\mathcal{X}}dz\left\|\overline{\mathfrak{A}_y}\sharp^B
\mathfrak{A}_z\right\|_{\mathfrak C^B}^{1/2}\leq
\underset{x\in\mathcal{X}}{\sup}\left\|G_x\sharp^B\mathfrak T^B_{(x,0)}[\mathfrak{v}_2] \right\|_{\mathfrak C^B}\underset{y\in\mathcal{X}}{\sup}\int_{\mathcal{X}}dz\left\|
\overline{\mathfrak T^B_{(y,0)}[\mathfrak{v}_1]}\,\sharp^B\,\mathfrak T^B_{(z,0)}
[\mathfrak{v}_1]\right\|_{\mathfrak C^B}^{1/2}.
$$
But, extending formula (\ref{aut}) to functions in $S^0_1(\Xi)$ with rapid decay in the $\mathcal{X}$-variable, 
$$
\left(\overline{\mathfrak T^B_{(y,0)}[\mathfrak{v}_1]}\,\sharp^B\,\mathfrak T^B_{(z,0)}
[\mathfrak{v}_1]\right)(X)=
$$
$$
=\int_{\Xi}\int_{\Xi}dX_1\,dX_2\,e^{-2i\sigma(X-X_1,X-X_2)}\Omega^B\big(\mathcal{T}(x,x_1,x_2)\big)
\overline{\left(\widetilde{\Omega^B_\mathcal{P}}[y]\star\big(\mathfrak T_{(y,0)}\mathfrak{v}_1\big)\right)(X_1)}
\left(\widetilde{\Omega^B_\mathcal{P}}[z]\star\big(\mathfrak T_{(z,0)}\mathfrak{v}_1\big)\right)(X_2)
$$
so that we can write
$$
<y-z>^N\left|\left(\overline{\mathfrak T^B_{(y,0)}[\mathfrak{v}_1]}\,\sharp^B\,\mathfrak T^B_{(z,0)}
[\mathfrak{v}_1]\right)(X)\right|=
$$
$$
=<y-z>^N\left|\int_{\Xi}\int_{\Xi}dX_1\,dX_2\,e^{-2i\sigma(X-X_1,X-X_2)}\Omega^B\big(\mathcal{T}(x,x_1,x_2)\big)<x-x_1>^{-p}<x-x_2>^{-p}
\times\right.
$$
$$
\times\overline{\left[\left(\frac{1-(i/2)<x-x_2,\partial_{\xi_1}>}{<x-x_2>}\right)^p
\left(\widetilde{\Omega^B_\mathcal{P}}[y]\star\big(\mathfrak T_{(y,0)}\mathfrak{v}_1\big)\right)\right](X_1)}\times
$$
$$
\left.\times\left[\left(\frac{1-(i/2)<x-x_1,\partial_{\xi_2}>}{<x-x_1>}\right)^p
\left(\widetilde{\Omega^B_\mathcal{P}}[z]\star\big(\mathfrak T_{(z,0)}\mathfrak{v}_1\big)\right)\right](X_2)
\right|\leq
$$
$$
\leq C_{N,n}(B)\sum_{|a|\leq q}\sum_{|b|\leq
q}\int_{\Xi}\int_{\Xi}dX_1\,dX_2\,<\xi-\xi_1>^{-q}<\xi-\xi_2>^{-q}<x-x_1>^{N+r(q)-p}<x-x_2>^{N+r(q)-p}\times
$$
$$
\times<y-x_1>^N\partial^a_{x_1}\overline{\left[\left(\frac{1-(i/2)<x-x_2,\partial_{\xi_1}>}{<x-x_2>}\right)^p
\left(\widetilde{\Omega^B_\mathcal{P}}[y]\star\big(\tau_{(y,0)}\mathfrak{v}_1\big)\right)\right](X_1)}\times
$$
$$
\times<z-x_2>^N\partial^b_{x_2}\left[\left(\frac{1-(i/2)<x-x_1,\partial_{\xi_2}>}{<x-x_1>}\right)^p
\left(\widetilde{\Omega^B_\mathcal{P}}[z]\star\big(\tau_{(z,0)}\mathfrak{v}_1\big)\right)\right](X_2)
\leq C_{N,n}(B)
$$
\end{proof}

\begin{lemma}\label{CS-s-confinare}
Assume that $\{G_x\}_{x\in\mathcal{X}}\subset S^m_0(\Xi)$ are uniformly bounded with respect to $x\in\mathcal{X}$ for the topology of $S^m_0(\Xi)$ and
$\mathfrak{v}$ is a symbol of class $S^0_1(\Xi)$
with rapid decay in the $\mathcal{X}$-variable. Then, by denoting
$$
\mathfrak{A}_x:=\mathfrak T^B_{(x,0)}[\mathfrak{v}]\,\sharp^B\,G_x,
$$
the family of symbols $\mathfrak{s}^-_m\sharp^B\mathfrak{A}_x$ indexed by $x\in\mathcal{X}$, defines in any Schr\"{o}dinger
representation a family of operators $\{A_x\}_{x\in\mathcal{X}}$
that satisfies the hypothesis of the Proposition \ref{Cotlar-Stein}.
\end{lemma}
\begin{proof}
 As remarked at the begining of the proof of the previous Lemma \ref{CS-confinare} it is enough to consider the case $m=0$. Evidently the product $\overline{\mathfrak{A}_y}\sharp^B\mathfrak{A}_z$ is treated identically as in the proof of Lemma \ref{CS-confinare}. So let us consider the opposite situation $\mathfrak{A}_y\sharp^B\overline{\mathfrak{A}_z}$ for $m=0$:
$$
\underset{y\in\mathcal{X}}{\sup}\int_{\mathcal{X}}dz\left\|\mathfrak{A}_y\sharp^B\overline{\mathfrak{A}_z}\right\|^{1/2}_{\mathfrak C^B}=\underset{y\in\mathcal{X}}{\sup}\int_{\mathcal{X}}dz\left\|\mathfrak T^B_{(y,0)}[\mathfrak{v}]\sharp^BG_y\sharp^B\overline{G_z}\sharp^B\mathfrak T^B_{(z,0)}[\overline{\mathfrak{v}}]\right\|^{1/2}_{\mathfrak C^B}.
$$
Let us choose now $\varphi\in C^\infty_0(\mathcal{X})$ such that $\int_{\mathcal{X}}\varphi^2(x)\,dx=1$ and thus, using Lemma \ref{confinare} and our $L^2$-continuity result in \cite{IMP}
$$
\mathfrak T^B_{(y,0)}[\mathfrak{v}]\sharp^BG_y\sharp^B\overline{G_z}\sharp^B\mathfrak T^B_{(z,0)}[\overline{\mathfrak{v}}]=\int_{\mathcal{X}}dx\,\mathfrak T^B_{(y,0)}[\mathfrak{v}]\sharp^BG_y\sharp^B\varphi_x\sharp^B\varphi_x\sharp^B\overline{G_z}\sharp^B\mathfrak T^B_{(z,0)}[\overline{\mathfrak{v}}],
$$
$$
\underset{y\in\mathcal{X}}{\sup}\int_{\mathcal{X}}dz\left\|\mathfrak{A}_y\sharp^B\overline{\mathfrak{A}_z}\right\|^{1/2}_{\mathfrak C^B}\leq\underset{y\in\mathcal{X}}{\sup}\int_{\mathcal{X}}dz\int_{\mathcal{X}}dx\,\left\|\mathfrak T^B_{(y,0)}[\mathfrak{v}]\sharp^BG_y\sharp^B\mathfrak{T}^B_{(x,0)}[\varphi]\right\|^{1/2}_{\mathfrak C^B}\,\left\|\mathfrak{T}^B_{(x,0)}[\varphi]\sharp^B\overline{G_z}\sharp^B\mathfrak T^B_{(z,0)}[\overline{\mathfrak{v}}]\right\|^{1/2}_{\mathfrak C^B}.
$$
But
$$
\mathfrak T^B_{(y,0)}[\mathfrak{v}]\sharp^BG_y\sharp^B\mathfrak{T}^B_{(x,0)}[\varphi]=\int_{\mathcal{X}}du\,\mathfrak T^B_{(y,0)}[\mathfrak{v}] \sharp^B\varphi_u\sharp^B\varphi_u\sharp^BG_y\sharp^B\mathfrak{T}^B_{(x,0)}[\varphi]
$$
so that (using the rapid decay of $\mathfrak{v}$)
$$
\underset{y\in\mathcal{X}}{\sup}\int_{\mathcal{X}}dz\left\|\mathfrak{A}_y\sharp^B\overline{\mathfrak{A}_z}\right\|^{1/2}_{\mathfrak C^B}\leq C_N\underset{y\in\mathcal{X}}{\sup}\int_{\mathcal{X}}dz\int_{\mathcal{X}}dx\int_{\mathcal{X}}du\,<y-u>^{-N}<x-u>^{-N}<x-z>^{-N}
$$
(where for any $N\in\mathbb{N}$ there exists a finite positive constant $C_N$ such that the inequality is true).
\end{proof}

\noindent{\bf Acknowledgements: }IV and RP acknowledge partial support from the ANCS Contract No. 2 CEx06-11-18/2006 and MM acknowledge partial support from the Fondecyt Grant No.1085162 and from the N\'ucleo Cientifico ICM P07-027-F "Mathematical
 Theory of Quantum and Classical Systems". RP acknowledges partial support from the SCOPES Programme and thanks EPF Lausanne where part of this paper has been completed and Tudor Ratiu for their hospitality.

E-mail:Viorel.Iftimie@imar.ro, Marius.Mantoiu@imar.ro, Radu.Purice@imar.ro

\end{document}